\documentclass[fleqn,usenatbib]{mnras}
\defcitealias{SSL17}{SSL17}

\usepackage{newtxtext,newtxmath}

\usepackage[T1]{fontenc}

\DeclareRobustCommand{\VAN}[3]{#2}
\let\VANthebibliography\thebibliography
\def\thebibliography{\DeclareRobustCommand{\VAN}[3]{##3}\VANthebibliography}

\usepackage{graphicx}	
\usepackage{amsmath}	
\usepackage{bm}
\usepackage{comment}
\usepackage{xspace}
\usepackage{gensymb}
\usepackage[normalem]{ulem}
\usepackage[dvipsnames]{xcolor}
\usepackage{threeparttable}
\usepackage{multirow}
\usepackage{makecell}
\usepackage{array}
\usepackage{array}
\newcolumntype{C}[1]{>{\centering\arraybackslash}p{#1}}

\newcommand{\emissionline}[2]{#1\,{\footnotesize #2}}

\newcommand{\mstar}[1][]{\ensuremath{M_{*}^{#1}}\xspace}
\newcommand{\mbh}[1][]{\ensuremath{M_{\mathrm{BH}^{#1}}}\xspace}
\newcommand{\mbulge}[1][]{\ensuremath{M_{\mathrm{*,bulge}}^{#1}}\xspace}

\newcommand{\logmstar}[1][]{\ensuremath{\log(M_{*})^{#1}}\xspace}
\newcommand{\logmbh}[1][]{\ensuremath{\log(M_{\mathrm{BH}})^{#1}}\xspace}
\newcommand{\logmbulge}[1][]{\ensuremath{\log(M_{\mathrm{*,bulge}})^{#1}}\xspace}

\newcommand{\disk}{\textsc{DESI-disk}\xspace}
\newcommand{\bulgeless}{\textsc{DESI-bulgeless}\xspace}
\newcommand{\somebulge}{\textsc{DESI-some-bulge}\xspace}
\newcommand{\control}{\textsc{control}\xspace}

\title[Secular galaxy-SMBH co-evolution in DESI]{Galaxy-SMBH co-evolution in 2,435 DESI disk galaxies and merger-free BH growth in bulgeless disks} 

\author[Sophie M. Jewell et al.]{
Sophie M. Jewell,$^{1}$\thanks{E-mail: sophie.jewell@physics.ox.ac.uk (SMJ)}
Rebecca J. Smethurst,$^{1}$
Chris J. Lintott$,^{1}$
Brooke D. Simmons,$^{2}$
Ricarda S. Beckmann,$^{3}$
\newauthor 
Izzy L. Garland,$^{4,2}$
Tobias G\'eron,$^{5}$
Benne W. Holwerda,$^{6}$
Karen L. Masters,$^{7}$
Ragadeepika Pucha,$^{8}$
\newauthor
and Ashley Spindler$^{9}$\\
$^{1}$Oxford Astrophysics, Department of Physics, University of Oxford, Denys Wilkinson Building, Keble Road, Oxford OX1 3RH, UK\\
$^{2}$Physics Department, Lancaster University, Lancaster LA1 4YB, UK\\
$^{3}$Institute for Astronomy, University of Edinburgh, Royal Observatory, Edinburgh EH9 3HJ, UK\\
$^{4}$Department of Theoretical Physics and Astrophysics, Faculty of Science, Masaryk University, Kotl\'{a}\v{r}sk\'{a} 2, Brno, 611 37, Czech Republic\\ 
$^{5}$Dunlap Institute for Astronomy and Astrophysics, University of Toronto, 50 St. George Street, Toronto, ON M5S 3H4, Canada\\
$^{6}$Department of Physics and Astronomy, University of Louisville, Natural Science Building 102, 40292 KY Louisville, USA\\
$^{7}$Department of Physics and Astronomy, Haverford College, Lancaster Avenue, Ardmore, PA 19041 USA\\
$^{8}$Department of Physics and Astronomy, University of Utah, 115 South 1400 East, Salt Lake City, UT 84112, USA\\
$^{9}$ Centre for Astrophysics Research, Department of Physics, Astronomy and Mathematics, University of Hertfordshire, Hatfield, Hertfordshire, AL10 9AB, UK\\
}

\date{Accepted XXX. Received YYY; in original form ZZZ}

\pubyear{\the\year{}}

\begin{document}
\label{firstpage}
\pagerange{\pageref{firstpage}--\pageref{lastpage}}
\maketitle

\begin{abstract} 

    We analyse black hole scaling relations for disk-dominated galaxies, and present evidence for secularly driven galaxy-black hole co-evolution. We do this observationally by looking at ``bulgeless'' galaxies, where the lack of a stellar bulge suggests a quiet merger history. We first compose a sample of 2,435 disk-dominated galaxies hosting optical broad line active galactic nuclei (AGN) from DESI data, and find a \mbh--\mstar scaling relation of \logmbh$= - 3.7^{+0.3}_{-0.3} + 1.00^{+0.03}_{-0.03} \log(M_*)$. From these, we identify 546 bulgeless galaxies hosting AGN, a factor of $\gtrsim5$ larger than pre-existing samples, and find they follow the same \mbh-\mstar relation as the general disk galaxy sample. Comparing the bulgeless galaxies to those with some bulge component, we find good agreement in the \mbh--\mstar relation (within $3\sigma$), but disagreement in the \mbh--\mbulge relation (at $>7\sigma$). This is the opposite of what is expected from merger-driven co-evolution, suggesting secular growth is significant in establishing observed \mbh--\mstar scaling relations. Furthermore, in a control sample of early-type galaxies we find tentative evidence of increased AGN luminosity compared to the bulgeless sample, but no statistical difference in black hole mass. This suggests the assumed merger-dominated evolutionary history of early-type galaxies does not lead to significantly increased long-term BH growth. These results support numerous recent observational and theoretical studies which suggest secular processes fuel significant supermassive black hole growth and galaxy-black hole co-evolution.
    
\end{abstract}




\section{Introduction} \label{sec:intro}

The relationship between supermassive black holes (SMBHs) and their host galaxies is of prime importance to modern astrophysics. A co-evolutionary relationship between the two is seen in multiple scaling relations between black hole (BH) mass (\mbh) and host galaxy properties such as velocity dispersion \citep[e.g.][]{Magorrian1998, Ferrarese2000, Kormendy2001, McConnell2013, Baldassare2020}, bulge stellar mass \citep[\mbulge; e.g.][]{McLure2002, Marconi2003, Haring2004, Saglia2016, Sahu2019} and total stellar mass \citep[\mstar; e.g.][]{Cisternas2011, Reines2015, SSL17, Davis2018, Sahu2019, Harikane2023, Pucha2025, Pucha2026}.

However, there remain many unknowns regarding this co-evolution and the physical mechanisms driving it. This has recently been re-emphasised by JWST observations. One of the most challenging findings from JWST is that BHs at $z\gtrsim4$ are over-massive with respect to their host galaxies, relative to local-universe scaling relations \citep[e.g.][]{Goulding2023, Harikane2023, Kokorev2023, Ubler2023, Furtak2024, Maiolino2024}. In addition, JWST studies of galaxy morphology have  revealed that disk galaxies are much more prevalent at $z\gtrsim1$ than predicted, a  result which holds out to $z\sim7-8$ \citep{Ferreira2022, Suess2022, Kartaltepe2023, Huertas-Company2024, Tohill2024, Smethurst2025}. Secular mechanisms play an important role in the evolution of disk galaxies \citep[e.g.][]{Kormendy2004, Debattista2006, Sellwood2014}, where ``secular'' refers to processes occurring over long timescales in the absence of significant galaxy merger activity. Secular evolution is driven by both internal structures (bars, spiral arms, dark matter halos, central SMBHs etc.) and environment (infalling gas, minor mergers etc.) \citep[see][]{Kormendy2004}. Given their significance in the evolution of disk galaxies, it is important to consider the role of secular mechanisms in BH growth within these systems, and whether such secular processes could account for the over-massive BHs at high-redshift found by JWST. However, our current understanding of secular BH growth and co-evolution is limited and work to grow this understanding must begin at low-redshifts, where disks and their sub-structures are better resolved.

Over the last decade, there has been increasing evidence at low-redshifts that secular growth plays a significant role in driving galaxy-SMBH co-evolution. Some of this evidence comes from simulations: \citet{Martin2018} investigated the role of galaxy mergers over cosmic time using the \textsc{Horizon-AGN} cosmological hydrodynamical simulation \citep{Dubois2014}. They found that $\sim65\%$ of BH growth since $z\sim3$ is attributed to secular mechanisms, and the position of local, massive galaxies on the \mbh-\mstar relation is unaffected by their merger history. In a later study using the Evolution and Assembly of GaLaxies and their Environment Simulation \citep[\textsc{eagle};][]{Crain2015, Schaye2015}, \citet{McAlpine2020} find that, although galaxy mergers cause periods of enhanced BH growth, this accounts for $\lesssim15\%$ of an individual BH's mass at the final redshift of the simulation. The results of these two studies suggest that galaxy mergers do not dominate population-level or individual BH growth on cosmological timescales.

However, this is difficult to test with observations as the merger history of a galaxy cannot be fully known from observational data. One approach has been to study ``bulgeless'' galaxies, which have little-to-no stellar bulge component (typically with bulge-to-total ratio $B/T<0.1$). Such galaxies are expected to have had quiet merger histories. This is supported by simulation studies \citep[e.g.][]{Merloni2010, Martin2018}, which find highly disk-dominated galaxies have not experienced major merger activity since $z\gtrsim2$. \citet{Simmons2013} identified 13 active galactic nuclei (AGN) in massive bulgeless disk galaxies, demonstrating that secularly evolving systems can host growing BHs. Of these AGN, 2 were broad-line AGN (BL-AGN) for which they calculate virial \mbh estimates and find they are higher than predicted by pre-established \mbh--\mbulge relations.

This was followed up by \citet{SSL17} (herein \citetalias{SSL17}), who visually identify bulgeless galaxies to compose a sample of 101 BL-AGN in disk-dominated galaxies in the Sloan Digital Sky Survey (SDSS). They find a \mbh--\mstar relation that is in good agreement with the \citet{Haring2004} scaling relation for early-type galaxies. This demonstrated that co-evolution is occurring effectively in bulgeless galaxies. Placing upper limits on bulge mass, they find no statistically significant correlation between bulge mass and BH mass for their bulgeless sample, with the majority of the BHs being over-massive relative to the \citet{Haring2004} relation. Under the assumption that bulgeless galaxies have quiet merger histories, these results suggest that merger-free processes are important in driving co-evolution.

However, these studies have so far been limited by sample size due to the rarity of BL-AGN and the challenge of identifying bulgeless galaxies. Studies of bulgeless SMBH hosts require large surveys with complementary photometry and spectroscopy. The Dark Energy Spectroscopic Instrument (DESI) survey provides an opportunity for this work, with its large scale optical spectroscopic dataset, accompanying optical photometry from the DESI Legacy Surveys (LS).

In this paper, we use DESI spectroscopy and LS photometry to compile a sample of 546 bulgeless BL-AGN host galaxies at $0.0172 \leq z \leq 0.2504$: a factor of $\gtrsim5$ increase in sample size from \citetalias{SSL17}. With morphological classifications from Galaxy Zoo (GZ) DESI \citep{Walmsley2023} and \textsc{EmFit} emission line models \citep{Pucha2025, Pucha2026}, we compose a sample of 2,435 disk galaxies hosting optical BL-AGN. We find a correlation in BH mass and total stellar mass for this sample, and fit a linear relation. We fit multi-component surface brightness profiles to $r$-band imaging with \textsc{galfitm} \citep{Peng2002, Peng2010, Vika2013}. From the resulting models we identify the bulgeless galaxies. With this large bulgeless AGN sample, we aim to study the \mbh--\mstar and \mbh--\mbulge scaling relations, and investigate the role of secular processes in galaxy-SMBH co-evolution.

The structure of this paper is as follows. In Section \ref{sec:data}, we introduce the DESI data, before describing the selection of the disk-morphology BL-AGN host galaxies in Section \ref{sec:sample}, and the selection of a comparison sample of early-type galaxies, weighted to match the disk sample in redshift and stellar mass, in Section \ref{subsec:control}. In Section \ref{sec:analysis} we describe the \mbh estimates used in the paper (\ref{subsec:mbh}), and the process of bulge-disk decomposition using \textsc{galfitm} and subsequent identification of bulgeless disk galaxies (\ref{subsec:galfitm}). We present our results in Section \ref{sec:results}, and discuss the possible underlying physics and potential implications in Section \ref{sec:discussion}. We assume a flat $\rm \Lambda CDM$ cosmology with $H_0=67.66$ and $\Omega_M=0.31$ throughout \citep{Planck2020}.


\section{Data} \label{sec:data}

This paper is based on spectroscopic observations from the DESI\footnote{Dark Energy Spectroscopic Instrument} survey \citep{DESI2025} and accompanying Legacy Survey \citep[LS;][]{Dey2019} photometry. DESI is an optical, multi-object spectrograph aboard the Mayall 4m telescope at the Kitt Peak Observatory, USA \citep{DESI2016b}. The unique arrangement of 5020 fibres allows DESI to observe 5000 targets at a time \citep{DESI2016b, Silber2023}. The primary aim of the DESI survey is to map the large-scale structure of the Universe across a large sky area and redshift range, covering $\sim17,000 \text{ deg}^2$ across eight years of observation.

\begin{table*}
    \centering
    \caption{Selection of BL-AGN in host galaxies with disk morphologies and a control sample of BL-AGN in non-merging galaxies of early-type morphology.}
    \label{tab:selection}

    \begin{threeparttable}
    
        \begin{tabular}{| 
                        C{0.12\textwidth} |
                        C{0.21\textwidth} |
                        C{0.12\textwidth} |
                        C{0.11\textwidth} |
                        C{0.1\textwidth} |
                        C{0.14\textwidth} |}
            \hline
            \textbf{Population} & \textbf{Catalogue} & \textbf{Variable} & \textbf{Cut} & \textbf{Sample Size} & \textbf{Control Sample Size} \\
            \hline

            Visibly & \multirow{5}{0.21\textwidth}{\centering GZ DESI \citep{Walmsley2023}} & $p_{\rm asked-merging}$ & $\geq 0.5$ & \multicolumn{2}{c|}{\multirow{2}{0.25\textwidth}{\centering 2,081,790} }\\
            non-merging & & $\zeta_{\rm merger}$ & $\leq 0.3$ & \multicolumn{2}{c|}{} \\
            \cline{1-1} \cline{3-6}
             
            Disk galaxies & & $f_{\rm features-or-disk}$ & $\geq0.3$ & 385,417 & -- \\
            \cline{1-1} \cline{3-6}
            Smooth featureless & & $f_{\rm smooth}$ & $\geq0.7$ & \multirow{2}{*}{\centering --} & \multirow{2}{*}{\centering 1,527,198} \\
            galaxies & & $p_\text{asked-smooth}$ & $\geq0.5$ & & \\
            \hline

             & & \multirow{2}{*}{--} & Included in \textsc{EmFit} catalogue & \multirow{2}{*}{378,663} & \multirow{2}{*}{1,490,685} \\
             \cline{3-6}
             Broad H$\alpha$ & \textsc{EmFit} & FWHM$_{\rm H\alpha,broad}$ & $\geq 300\text{ km s}^{-1}$ & & \\
             Emitters & (\citeauthor{Pucha2025} \citeyear{Pucha2025}, & SNR$(F_{\rm H\alpha,broad})$ & $\geq 3$ & & \\
             & \citealp{Pucha2026}) & SNR$(\sigma_{\rm H\alpha,broad})$ & $\geq 3$ & 4,314 & 9,603 \\
             & & $\rm ANR_{H\alpha,broad}$ & $\geq 2$ & & \\
             & & \textsc{PROB\_BROAD} & $\geq 80\%$ & & \\
            \hline
            
            \multirow{2}{*}{BL-AGN} & \multirow{2}{*}{--} & GMM cluster association & \multirow{2}{*}{$\geq 99\%$} & \multirow{2}{*}{2,619} & \multirow{2}{*}{5,923} \\
            \hline

            Reliable & \multirow{2}{0.21\textwidth}{\centering \citet{Siudek2024}} & $\chi^2_\nu$ & $<17$ & \multirow{2}{0.1\textwidth}{\centering 2,435} & \multirow{2}{0.14\textwidth}{\centering 5,879} \\
            stellar mass & & $M_{*,{\rm best}}/M_{*,{\rm bayes}}$ & $\geq1/5$; $\leq5$ & & \\
            \hline

            Early-type & DESI DR1 photometry catalogue & $n$ & $\geq3$ & -- & 3,226 \\
            \hline
            
        \end{tabular}

    \end{threeparttable}
        
\end{table*}

\subsection{\label{subsec:photo} Photometry}

The DESI LS is a culmination of three surveys: the Dark Energy Camera \citep[DECam;][]{Flaugher2015} Legacy Survey \citep[DECaLS;][]{Dey2019}, the Beijing-Arizona Sky Survey \citep[BASS;][]{Zou2017} and the Myall z-band Legacy Survey \citep[MzLS;][]{Silva2016}. These observations were completed prior to the start of the DESI survey, and were used to select targets for spectroscopic observation.

DECaLS uses DECam, which is mounted on the 4m Blanco Telescope at the Cerro Tololo Inter-American Observatory in Chile. DECaLS consists of imaging covering $\sim9000\text{ deg}^2$ of sky in the southern hemisphere in the $g$, $r$ and $z$ bands. BASS and MzLS cover $\sim5000\text{ deg}^2$ of the northern sky. Both used cameras mounted on telescopes located at Kitt Peak observatory in Arizona, USA: BASS used the 90Prime camera on board the Bok 2.3m Telescope to image in $g$ and $r$ bands, while MzLS used the Mosaic-3 camera at the prime focus of the 4m Myall Telescope to image in the $z$-band. These three surveys compose the LS, providing optical images in $g$, $r$ and $z$.

In this paper, we use photometry from the ninth data release of the LS (DR9)\footnote{DESI LS DR9 information available at \href{https://www.legacysurvey.org/dr9/description/}{www.legacysurvey.org/dr9/}}. This release includes the final full-sky coverage of the survey first published in DR8, with improved data reduction. The images used in this paper are single-band co-added images. The LS pipeline includes single component morphological fitting with \textsc{the Tractor} \citep{Lang2016}. This classifies each object as a point source (PS), round exponential, exponential, de Vaucouleurs \citep{deVaucouleurs1948} and S\'ersic \citep{Sersic1968}. This provides useful approximations of properties such as size, axis ratio and S\'ersic index which we use as initial parameters in our own morphological fitting (see Section \ref{subsec:galfitm}).

\subsubsection{\label{subsec:GZ} Galaxy Zoo DESI}

The Galaxy Zoo (GZ) DESI catalogue \citep{Walmsley2023} contains detailed morphological classifications for 8.67 M galaxies in LS DR8 imaging. These classifications are produced by a deep learning model, \textsc{Zoobot} \citep{Walmsley2022} trained on GZ citizen science vote fractions\footnote{\textsc{Zoobot} code is available at \href{https://github.com/mwalmsley/zoobot?tab=readme-ov-file}{github.com/mwalmsley/zoobot}}. In the traditional citizen science approach, the citizen scientists (CSs) are presented with questions regarding imaging of a galaxy, with a number of possible responses. For example, the question “Is the galaxy simply smooth and rounded, with no sign of a disk?” has three possible responses: “smooth”, “features or disk”, and “star or artifact”. The ``vote fraction'', $f$, for a given response is then the number of CSs who choose this response as a fraction of total responses to the question. \textsc{Zoobot} outputs predicted versions of these vote fractions. The catalogue contains predicted vote fractions for each response to each question included in the training set.

The GZ DESI \citep{Walmsley2023} morphology classifications are used in Section \ref{subsec:disks} for the selection of a disk galaxy sample.

\subsection{\label{subsec:spec} Spectroscopy}

The DESI survey aims to obtain spectra of $\sim60$M galaxies, having exceeded the initial target of $\sim40$M \citep{DESI2016a}. We use spectra from the first data release of DESI DR1 \citep{DESI2025}. Released to the public on 25$^\text{th}$ March 2025, DR1 contains observations from the first 13 months of main survey observations, with additional survey validation data previously released in the Early Data Release \citep[EDR;][]{DESI2024}. This includes the spectra of $\sim 18.7$ M unique targets across $\sim9,700\text{ deg}^2$ of sky. Redshifts and basic spectral type classifications (GALAXY, QSO, STAR) are included in the DESI spectral catalogues, obtained from the \textsc{Redrock} spectral fitting code \citep{Brodzeller2023, Anand2024}. According to these classifications, this includes {$\sim13$ M} galaxy spectra and {$\sim 1.5$ M} quasar spectra.

DESI DR1 also includes a number of value-added catalogues, including the \textsc{EmFit} catalogue of emission line models \citep{Pucha2025, Pucha2026} that we use for the selection of BL-AGN (see Section \ref{subsec:blagn}), and the AGN Host Galaxies Physical Properties catalogue \citep{Siudek2024} from which we obtain total stellar mass measurements.

\subsubsection{Stellar masses} \label{subsec:mstar}

We use stellar masses from the DESI DR1 AGN Host Galaxies Physical Properties catalogue \citep{Siudek2024}, containing the results of Spectral Energy Density (SED) fitting for $\sim 17$M galaxies classified as \texttt{QSO} or \texttt{GALAXY} by \textsc{Redrock} in the DESI pipeline. The SED fitting was performed using \textsc{CIGALE} \citep{Boquien2019} with $g,r,z$ optical photometry from LS and $W1, W2, W3$ and $W4$ mid-IR photometry the Wide-field Infrared Survey Explorer \citep[WISE;][]{Wright2010}. They use the single stellar population (SSP) models of \citet{Bruzual2003} with a \citet{Chabrier2003} initial mass function (IMF), and implement a delayed tau star formation history (SFH) with a possible exponential burst of star-formation \citep[e.g.][]{Ciesla2015} and the \citet{Inoue2011} nebular emission model. For dust, they use the dust emission models of \citet{Dale2014} and a \citet{Calzetti2000} dust attenuation curve. An AGN component is included in the \textsc{CIGALE} SED model, using AGN emission models from \citet{Fritz2006}. being fitted simultaneously. Galaxy and AGN components are fitted simultaneously and the AGN contribution is allowed to go to zero.

Any one of the model choices outlined above can affect the derived stellar mass. For a detailed assessment of biases in the SED fitting, including how these pertain to BL-AGN host galaxies specifically, we refer the reader to \citet{Siudek2024}. The three factors discussed by \citet{Siudek2024} which are the most relevant to our \disk sample are: the IMF, the dust attenuation model, and the inclusion of WISE photometry. The \citet{Chabrier2003} IMF is found to produce stellar masses $\sim0.24\text{ dex}$ smaller than the alternative \citet{Salpeter1955} IMF, which has increased contribution from high-mass stars, in agreement with pre-existing literature \citep[e.g.][]{Ilbert2010}. The \citet{Calzetti2000} dust attenuation model is found to underestimate the stellar mass for AGN and composite galaxies by $\sim 0.1 \text{ dex}$ compared to the \citet{Charlot2000} model which treats young and old stellar populations separately. While the inclusion of the four WISE bands has little impact on stellar mass for the general population, it increases the stellar mass by $\sim 0.2\text{ dex}$ for BL-AGN host galaxies, compared to using $grz$ alone. 69\% of the \disk sample have photometry with $S/N>3$ in all 4 WISE bands, and $97\%$ have $S/N>3$ in at least 3 WISE bands. Therefore the vast majority of the sample benefits from the inclusion of the WISE photometry, in regard to the serived stellar mass.

\citet{Siudek2024} also find the inclusion of the AGN component in the their \textsc{CIGALE} fitting model has a significant impact on stellar mass for BL-AGN host galaxies, decreasing it by $\sim 0.26\text{ dex}$. This can largely be understood physically: the AGN will contribute significantly to the optical and IR emission in the case of a BL-AGN, therefore including the AGN component re-attributes light to the AGN that would otherwise be falsely attributed to stars, reducing the inferred stellar mass.

We select confident stellar mass estimates from the SED fitting using the recommended quality cuts:
\begin{align}
    &\chi^2_\nu \leq 17 \\
    &1/5 \leq M_{*,{\rm best}}/M_{*,{\rm bayes}} \leq 5
\end{align}
where $M_{*,\rm{best}}$ and $M_{*,\rm{bayes}}$ are CIGALE estimates from the best model and the marginalised probability density function (PDF) respectively \citep{Siudek2024}.


\section{Sample Selection}  \label{sec:sample}

\subsection{Selection of disk galaxies} \label{subsec:disks}

We begin by selecting a sample of disk galaxies which do not visibly appear to be undergoing a galaxy merger. For this, we use the GZ DESI morphology catalogue described in Section \ref{subsec:GZ}. Disk galaxies were selected by implementing a threshold cut on the predicted votes fraction for ``features or disk'' in response to the question ``Is the galaxy simply smooth and rounded, with no sign of a disk?''. A cut of $f_{\rm features-or-disk}\geq0.3$ is used, as suggested in \citet{Walmsley2022} based on GZ DECaLS vote fractions.

Galaxies which show signs of recent galaxy merger activity in their imaging were discarded from this sample by defining a ``merger prominence'' parameter, as implemented by \citet{Garland2024}:
\begin{equation}
    \zeta_{\rm merger} = 0.2f_\text{minor} + 0.8f_\text{major} + f_\text{merger}
\end{equation}
where $f_\text{minor}$, $f_\text{major}$ and $f_\text{merging}$ are the fraction of votes for ``minor disturbance'', ``major disturbance'' and ``merging'' respectively, in response to this question of merger activity. We select galaxies without visible signs of merger activity by applying a conservative cut, requiring $\zeta_{\rm merger} \leq 0.3$. For consistency, we implement the same definition and cut as \citet{Garland2024}, and confirm the suitability of this threshold by visual inspection of randomly selected galaxy images in bins of $\zeta_{\rm merger}$.

It is typical when using citizen science vote fractions to implement a minimum number of responses to a question before making a selection based on the vote fractions \citep[e.g.][]{Land2008}. However, this information is not available for \textsc{Zoobot} classifications. Instead, the model includes the probability that any random CS would be asked this question \citep{Walmsley2023}. Therefore, we require that a given CS is more like than not be asked if the galaxy is ``merging or disturbed'' (i.e. $p_{\rm asked-merging}\geq0.5$).

The above selections give in a sample of 385,417 visibly non-merging disk galaxies from GZ DESI ($4\%$ of the full GZ DESI catalogue).

\subsection{Selection of BL-AGN} \label{subsec:blagn}

\begin{figure}
    \centering
    \includegraphics[width=\linewidth]{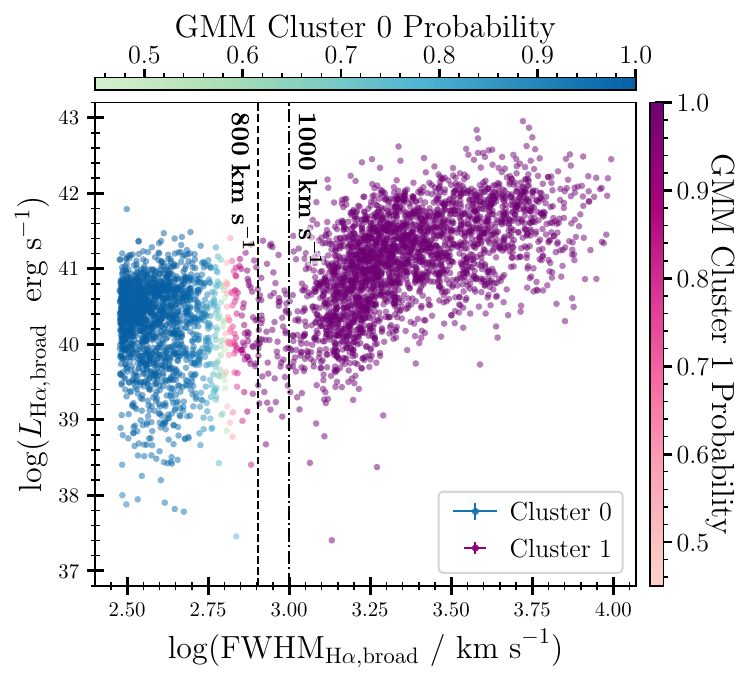}
    \caption{Luminosity against $\rm FWHM$ of broad $\rm H\alpha$ emission for disk galaxies, identified as displaying broad $\rm H\alpha$ emission. Two distinct underlying sub-populations are distinguished using a Gaussian Mixture Model (GMM), and are plotted with different colour maps. In each instance, the colour corresponds to the probability of the given data point belonging to its assigned GMM group. The vertical lines indicate $\rm{ FWHM_{H\alpha, broad}}=800\text{ km s}^{-1}$ ($--$) and $1000\text{ km s}^{-1}$ ($\bm{-\cdot-}$) respectively.}
    \label{fig:lha_fwhm_gmm}
\end{figure}

As we aim to study BH scaling relations, we require BH mass measurements. For this reason, we select BL-AGN for which we can estimate the virial BH mass from the broad line emission (see Section \ref{subsec:mbh}).

We select optical BL-AGN within our disk galaxy sample using emission line models in the DESI DR1 \textsc{EmFit} catalogue \citep{Pucha2025, Pucha2026}. \textsc{EmFit} is a Python-based emission line fitting code, written primarily for use on DESI spectra\footnote{The \textsc{EmFit} code is available at \href{https://github.com/Ragadeepika-Pucha/EmFit}{github.com/Ragadeepika-Pucha/EmFit}.}. The DR1 catalogue includes H$\beta$, [\emissionline{O}{III}$\lambda4959,5007$], [\emissionline{N}{II}$\lambda6548,6583$], H$\alpha$ and [\emissionline{S}{II}$\lambda6716,6731$] line models for all DESI DR1 sources at $z\leq0.45$ assigned the GALAXY or QSO spectrum type in the DESI pipeline. 

\textsc{EmFit} uses \textsc{FastSpecFit} \citep{Moustakas2023} to model and subtract the stellar continuum from the DESI spectrum, before fitting multi-component models to the emission lines. In addition to the narrow line components, broad line components are included in the models of H$\alpha$ and H$\beta$, and outflow components are allowed for all emission lines. All individual components are modelled as Gaussians. 

Of the 385,417 disk galaxies from GZ DESI, we find 378,663 in the \textsc{EmFit} catalogue. This loss is likely due to a lack of DESI DR1 spectroscopic observations for some sources in LS DR9, as DESI DR1 is not complete over the full DESI footprint. Following \citet{Pucha2025}, we select broad $\rm H\alpha$ emitters by making cuts based on the signal-to-noise ratio (SNR), amplitude-to-noise ratio (ANR) and SNR in line width, of the broad $\rm H\alpha$ model component. These cuts are detailed in \autoref{tab:selection}, and lead to a selection of 4,314 broad $\rm H\alpha$ emitters.

We then look for the BL-AGN amongst the broad $\rm H\alpha$ emitters. A common approach used to select BL-AGN in large optical datasets is to take a velocity cut in the full-width-half-maximum of the broad $H\alpha$ emission, $\rm{ FWHM_{H\alpha, broad}}$. Most commonly, $\rm{ FWHM_{H\alpha, broad}}\geq1000\text{ km s}^{-1}$ is used \citep[e.g.][]{VandenBerk2006, Schneider2010, Stern2012, Stern2013, Harikane2023}, though a number of previous studies have implemented a lower threshold of $800 \text{ km s}^{-1}$ to probe the lower-luminosity BL-AGN population \citep[e.g.][]{Oh2015, Liu2019}. Stellar processes such as star formation outflows and supernovae can also produce broad $\rm H\alpha$ emission \citep[see e.g.][]{Smith2011, Gutierrez2017, Lopez-Coba2020, Dessart2025} that can lead to minor contamination in $\rm FWHM_{H\alpha, broad}$-selected BL-AGN samples \citep[e.g.][]{Oh2015}. Alternatively, emission line ratio diagrams such as BPT \citep[][]{Baldwin1981} may be used to identify AGN however, BL-AGN are widely distributed on such diagrams, and traditional cuts often exclude many BL-AGN \citet{Greene2007, Stern2013, Birchall2020, Malkan2026}. We therefore choose not to implement such emission line selections, opting for sample completeness over purity.

To identify BL-AGN, we consider the distribution in $\rm FWHM_{H \alpha,broad}$ and $L_{\rm H \alpha,broad}$ shown in \autoref{fig:lha_fwhm_gmm}. The relationship between $\rm{ FWHM_{H\alpha, broad}}$ and $L_{\rm H\alpha, broad}$ depends on the source of the broad $\rm H\alpha$ emission. In a population with a common dominant source of broad $\rm H\alpha$ emission, such as the broad-line emitting region (BLR) of an AGN, we expect to see a population-level correlation between $\rm{ FWHM_{H\alpha, broad}}$ and $L_{\rm H\alpha, broad}$.

Three outliers at ${\rm FWHM_{H \alpha,broad} } > 10^4 \text{ km s}^{-1}$ were identified\footnote{DESI IDs of $\rm FWHM_{H \alpha,broad}$ outliers: 39627761706339881, 39627862663235667, 39628507487145762}. On inspection, they were found to have erroneous \textsc{EmFit} models and were discarded from the sample. Two distinct distributions are visible in \autoref{fig:lha_fwhm_gmm}, one towards lower $\rm FWHM_{H \alpha,broad}$ and one towards higher $\rm FWHM_{H \alpha,broad}$. To distinguish between these populations, we fitted a two-component Gaussian Mixture Model \citep[GMM; see e.g.][]{Bishop2008} to the distribution using \textsc{scikit-learn}\footnote{\href{https://scikit-learn.org/stable/modules/generated/sklearn.mixture.GaussianMixture.html}{scikit-learn.org/sklearn.mixture.GaussianMixture.html}}. A full covariance matrix is used in the fitting, meaning the two Gaussian components are each assigned an independent general covariance matrix. This approach assumes no shared shape, size or orientation between the underlying distributions and allows correlations between parameters to be captured. 

The result of the fitted GMM is two ``clusters'' of associated data points. This is shown in \autoref{fig:lha_fwhm_gmm}, where each point is coloured by the likelihood of its association with the assigned cluster. We refer to the cluster towards lower $\rm FWHM_{H \alpha,broad}$ as Cluster 0, and the one towards higher $\rm FWHM_{H \alpha,broad}$ as Cluster 1. Two vertical lines indicate the ${\rm FWHM_{H\alpha,broad}} = 800 \text{ km s}^{-1}$ and $1000 \text{ km s}^{-1}$ thresholds. Of the 2,727 objects associated with Cluster 1, 96\% have $\rm{ FWHM_{H\alpha, broad}} \geq 800\text{ km s}^{-1}$, and $94\%$ have $\rm{ FWHM_{H\alpha, broad}} \geq 1000 \text{ km s}^{-1}$. Conversely, none of the 1,584 sources associated with Cluster 0 of the GMM have $\rm{ FWHM_{H\alpha, broad}} \geq 800 \text{ km s}^{-1}$. Therefore the two clusters assigned with GMM are largely similar to velocity thresholds commonly used in the literature.

\begin{figure}
    \includegraphics[width=\linewidth]{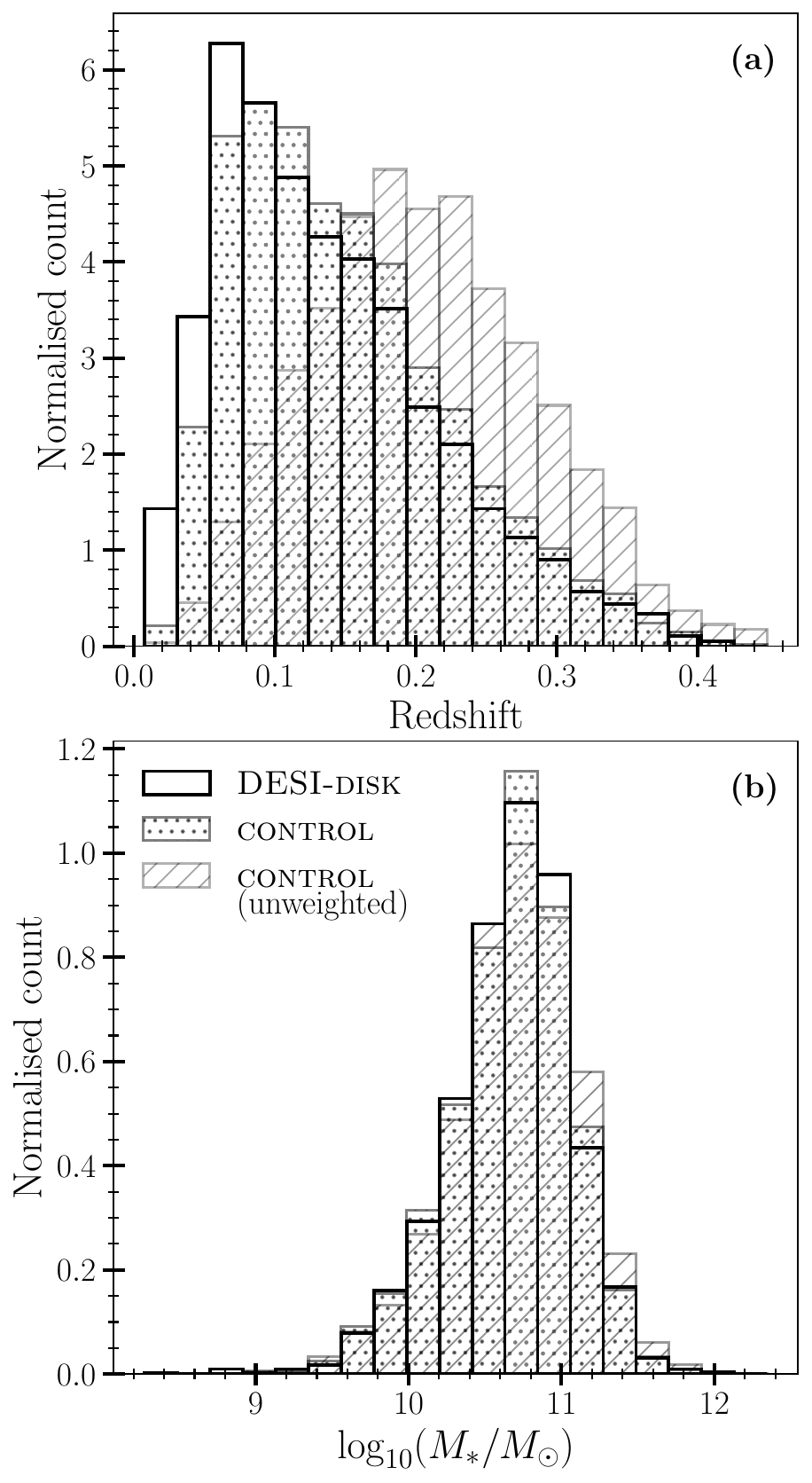}
    \caption{Redshift (left) and total stellar mass (right) distributions of 2,435 \disk galaxies (solid black), and the weighted (dotted) and un-weighted (hatched) distributions of the 3,226 matched \control (dotted grey). The weights are produced to match \control to \disk in redshift-stellar mass space. Redshifts are taken from the DESI DR1 spectral catalogue. Stellar masses are taken from the DESI DR1 AGN host properties catalogue \citep{Siudek2024}. The \disk sample peaks at $0.0424 \leq z \leq 0.0888$ and $10.56 \leq \log(M_*/M_\odot) \leq 11.04$, and spans $0.0075 \leq z \leq 0.4213$ and $8.27 \leq \log(M_*/M_\odot) \leq 12.04$.}
    \label{fig:z_mstar_dist}
\end{figure}

For each cluster, we test for monotonicity between $\log({\rm FWHM_{H \alpha,broad}})$ and $\log({\rm L_{H \alpha,broad}})$ using the Spearman rank-order correlation coefficient. This test indicates that there is no meaningful monotonic relationship in Cluster 0 ($\rho=-0.07$ with 95\% CI $[-0.12, -0.02]$), while there is a strong, positive monotonic relationship in Cluster 1 ($\rho=0.62$ with 95\% CI [0.60, 0.65]). We interpret Cluster 1 as tracing the underlying BL-AGN population, where accretion onto the central BH dominates the broad $\rm H\alpha$ emission. Of the 2,727 sources associated with Cluster 1, $\sim96\%$ have $\geq99\%$ likelihood of belonging to this cluster. These sources are selected as BL-AGN, giving a final sample of 2,619 BL-AGN hosted in non-merging disk galaxies ($0.7\%$ of the disk galaxy sample). Of this final BL-AGN sample, $\geq99.9\%$ and $\geq97.6\%$ have $\rm{ FWHM_{H\alpha, broad}} \geq 800 \text{ km s}^{-1}$ and $\geq 1000 \text{ km s}^{-1}$, respectively. Therefore, this selection is highly analogous to conventional $\rm FWHM$ selection. The redshift distribution of this sample is shown in panel (a) of \autoref{fig:z_mstar_dist}. The redshift range represented in this sample is $0.0075 \leq z \leq 0.4213$, with a peak in the distribution at $z\sim0.0656$.

We obtain stellar masses from the CIGALE catalogue, as described in Section \ref{subsec:mstar}. A single outlier was identified at \logmstar$\sim7.6$.\footnote{DESI TARGETID: 39627896582573918} By inspection of the \textsc{EmFit} emission line model, we determine this is an unreliable broad $\rm H\alpha$ detection and discard it from the sample. These cuts leave 2,435 BL-AGN host galaxies with reliable stellar mass estimates. The distribution of stellar mass is shown in panel (b) of \autoref{fig:z_mstar_dist}, with range $8.3 \leq \log(M_*) \leq 12.0$ and mean $\log(M_*)=10.7$. This sample is herein referred to as the \disk sample. 

\subsection{Selection of early-type galaxy comparison sample} \label{subsec:control}

We also compile a control sample of DESI galaxies with early-type morphologies and hosting BL-AGN, for the purpose of comparison with the \disk sample. As early-type galaxies are expected to have merger-dominated histories, unlike late-type disks \citep[e.g.][]{Toomre1977, Deeley2017, Martin2018b}, this sample provides a useful comparison to bulgeless galaxies with expected quiet merger histories. 

We select visibly-non-merging early-type galaxies using GZ DESI morphologies, in a similar manner to the disk galaxy selection described in Section \ref{subsec:disks}. The same cuts on $\zeta_{\rm merger}$ and $p_{\rm asked-merging}$ as used the \disk sample are applied, however, to select early-type instead of disks, we implement a cut on the predicted vote fraction for ``smooth'' in response to the question ``Is the galaxy simply smooth and rounded, with no sign of a disk?''. We require $f_{\rm smooth}\geq0.7$, using the cut suggested in \citet{Walmsley2022} as before. Visual inspection of randomly selected objects in bins of $f_{\rm smooth}$ confirm this to be a sensible threshold choice.

We follow the same BL-AGN selection criteria used for the \disk sample described in Section \ref{subsec:blagn}. We obtain stellar mass estimates from the \citet{Siudek2024} SED fitting catalogue, making the suggested quality cuts (see Section \ref{subsec:mstar} for full description). These various selections give a sample of 5,879 BL-AGN. Interestingly, $\sim0.4\%$ of the smooth, featureless galaxies are found to host BL-AGN, whereas the BL-AGN fraction for the selected disk galaxies (Section \ref{subsec:disks}) is $\sim0.7\%$. Therefore we find the BL-AGN fraction in disks is higher than in the selected early-type galaxies.

\begin{figure*}
    \includegraphics[width=\textwidth]{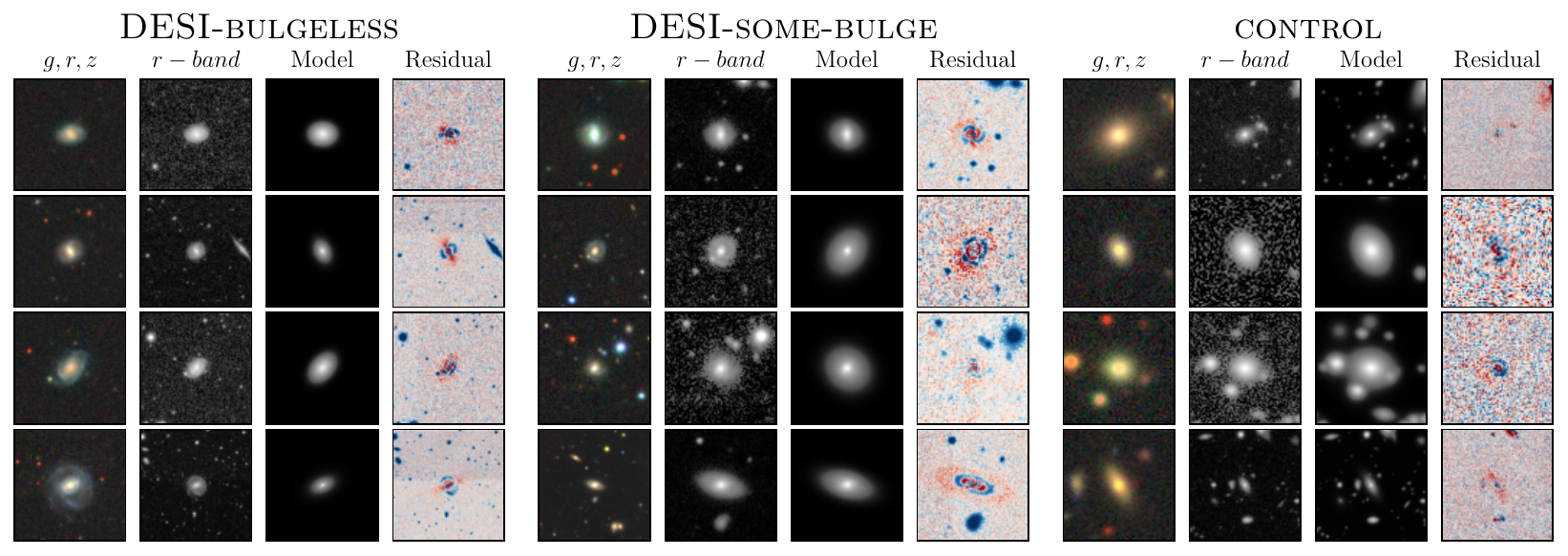}
    \caption{DESI LS photometry and surface brightness profile models and residuals of galaxies in \bulgeless (left), \somebulge (centre) and \control (right) samples. The galaxies shown are randomly selected from each sample. Four columns are shown for each sample, containing (from left to right) $g,r,z$ colour images, $r-$band images, the surface brightness profile models to the $r-$band images, and the residuals. For the \bulgeless and \somebulge samples, the models shown are \textsc{GALFITM} models from this work. For the \control sample, the models are \textsc{TRACTOR} models created in the LS DR9 photometry pipeline.}
    \label{fig:image_grid}
\end{figure*}

The GZ DESI morphology selection may suffer from contamination from featureless disks. Furthermore, the early-type galaxies tend towards higher redshift (see \autoref{fig:z_mstar_dist}) which may cause featured objects to appear smooth in the DESI imaging. To account for this, we use the results of the \textsc{TRACTOR} single-component surface brightness model fits included in the DESI DR1 photometry catalogues (see Section \ref{subsec:photo}). We exclude any objects with PS or exponential light profiles, and any with S\'ersic index $n<3$. These various selections are outlined in \autoref{tab:selection}, and result in a sample of 3,226 early-type galaxies hosting BL-AGN. This then forms our early-type BL-AGN host sample, hereafter referred to as the \control. The DESI colour images of nine randomly selected galaxies in the \control are shown in \autoref{fig:image_grid}, alongside a selection of \disk galaxies.

In order to reliably compare the \control to the \disk sample, we must account for differences in redshift and stellar mass. The largest point of discrepancy in the two samples is at the lowest redshifts ($z\lesssim0.14$), where there are more \disk objects than \control objects. We choose to apply individual weights to the \control objects. This approach allows us to match the \disk sample in redshift and stellar mass, without excluding objects from the sample or artificially supplementing the data.

We calculate the weight of each object according to the difference in the 2D probability density (PD) of the two samples in redshift-\logmstar space using Kernel Density Estimation (KDE). In KDE, a chosen kernel function is attributed to each data point, and the combination of these functions produces a smooth PDF. We choose a 2D Gaussian as the kernel function. The bandwidth of the kernel function, $h$, is a critical parameter in KDE. In the case of a 2D Gaussian kernel, $h$ is the covariance matrix of the Gaussian, relating to the standard deviations in each dimension. 

We use \textsc{scikit-learn} to perform the KDE. First we choose a value of $h$ to apply across both the \disk and \control datasets, allowing consistent comparison of the fitted PDFs. To do this, we combine both samples into one dataset and perform a grid search over a logarithmically spaced set of 30 $h$ values, $-1.5 \leq \log(h) \leq 0.5$, using \textsc{scikit-learn}'s \textsc{GridSearchCV}. With this method, 5-fold cross-validation of the KDE is performed on the combined dataset for value of $h$. In each instance, the KDE is scored by the total log-likelihood under the model, and the average score across the 5-folds is taken for this $h$ value. The $h$ value with the highest average log-likelihood is then taken as the best value of $h$. We find $h=0.213$ is the best choice of bandwidth, and use this for all further KDEs.

Using this $h$ value, we fit a Kernel Density (KD) model to the redshift-\logmstar distributions of the \control and \disk samples using the using \textsc{scikit-learn}'s \textsc{KernelDensity}. We then compute the log-likelihood PD of the \control sample under the \control KD model ($\log({P\!D_{control}})$) and under the \disk KD model ($\log({P\!D_{disk}})$). The normalised weights, $w$, on the \control sample objects are then calculated as:
\begin{equation}
    w_i = \frac{\log(P\!D_{\rm disk})_i}{\log(P\!D_{\rm control})_i} \times \frac{N_{\rm disk}}{N_{\rm control}} 
\end{equation}
where $N_{\rm disk}$ and $N_{\rm control}$ are the sizes of the respective samples.

In order to ensure that a small number of highly weighted objects do not have a disproportionate bearing on the results, we clip the weights at the 99.5\% level. This clipping affects 30 objects with $w_i \gtrsim 5.13$. All the weights are re-normalised after this clipping. 

The redshift and stellar mass weighted and un-weighted distributions of the \control sample are shown in \autoref{fig:z_mstar_dist}. The weighted \control peaks in redshift at $0.0656 \leq z \leq 0.1121$, approximately 0.02 higher than the peak in the \disk distribution. The distributions in total stellar mass of the two samples peak in the same \logmstar bin ($ 10.56 \leq \log(M_*/M_\odot) \leq 11.04$).


\section{Analysis} \label{sec:analysis}

\subsection{Virial estimation of black hole masses} \label{subsec:mbh}

In BL-AGN, emission from the BLR of the AGN is present in the optical spectrum as broadened $\rm H\alpha$ and $\rm H\beta$ emission lines. This broad emission traces the size of the BLR and the $\rm FWHM$ acts as a proxy for gas velocity. Under the assumption that the BLR of the AGN is virialised, an estimate of \mbh can be obtained:
\begin{equation}
    M_{\rm BH} \propto \frac{v^2 R_{\rm BLR}}{G}
    \label{eq:m_vir}
\end{equation}
where $v$ and $R_{\rm BLR}$ are the velocity dispersion and radius of the BLR.

\citet{Greene2005} developed a method to obtain such virial \mbh estimates for BL-AGN from properties of the broad $\rm H\alpha$ emission alone. The \citet{Greene2005} equation was updated by \citet{Reines2013}, by implementing a more recent form of the $R_{\rm BLR}-L_{5100}$ correlation from \citet{Bentz2013}, yielding:
\begin{equation}
    \begin{split}
        \log( \frac{ M_{\rm BH} }{M_\odot} ) = \log(\epsilon) + 0.47\log( \frac{ L_{\rm H\alpha,broad} }{10^{42}\text{ erg s}^{-1}} ) \\ 
        + 2.06\log( \frac{ \rm FWHM_{H\alpha,broad} }{10^3\text{ km s}^{-1}} )
    \end{split}
\end{equation}
where $\epsilon$ is the virial scale factor. This factor depends on the geometry of the BLR and potentially the accretion physics of the BH \citep{Liu2024}, and is one of the primary sources of uncertainty in virial \mbh estimates. Values in the literature vary around 1 \citep[$0.55 - 1.4$, translating to $\sim0.4$ dex difference in BH mass;][]{Onken2004, Greene2007, Grier2013, Woo2015, GRAVITY2024}. We assume $\epsilon=1$, and implement the \citet{Reines2013} equation to obtain dynamical estimates of BH mass, using the $\rm H\alpha$ emission line models.

This method is underpinned by the assumption that the BLR of the AGN is virialised. While there is evidence to support this \citep[e.g.][]{Gaskell1988, Bentz2009b, Grier2017} and the method is widely used in the field \citep[e.g.][]{Greene2005, Reines2015}, potential complications such as the geometry of the BLR remain \citep[e.g.][]{Shapovalova2010}. We also note that this method is dependent on multiple local relations ($R_{\rm BLR}-L_{5100}$ \citet{Bentz2009}; $L_{\rm H\alpha}-L_{5100}$ and $\rm FWHM_{H\alpha}-FWHM_{H\beta}$ \citet{Greene2005}). 

Virial \mbh estimates following this approach show a scatter of $\sim 0.5$ dex \citep[see e.g.][]{Peterson2010, Shen2013}. We therefore propagate a 0.5 dex scatter term in the errors on \logmbh, which significantly outweighs systematic errors from the \textsc{EmFit} emission line models.

\begin{table*}
    \centering
    \caption{Summary of the identification of bulge components in \textsc{galfitm} models.}
    \label{tab:identify_bulges}

    \begin{threeparttable}
    
        \begin{tabular}{| 
                        C{0.12\textwidth} |
                        C{0.07\textwidth} |
                        C{0.07\textwidth} |
                        C{0.07\textwidth} |
                        C{0.2\textwidth} |}
            \hline
            {\vspace{2pt} \large \textbf{Model}} & {\vspace{2pt} \large $\bm{n}$} & {\vspace{2pt} \large $\bm{b/a}$} & {\vspace{2pt} \large $\bm{R_e}$} & {\vspace{2pt} \large \textbf{Bulge / Not Bulge} } \\
            \Xhline{1.2pt}
            
            PS $+$ S\'ersic & - & - & - & - \\
            \hline
            
            \multirow{4}{0.12\textwidth}{\centering PS $+$ Exp. Disk \newline $+$ S\'ersic} 
                & $\leq0.5$ & $\leq 0.85$ & - & not bulge \\
            \cline{2-5}
                & $\leq 0.5$ & $> 0.85$ & - & bulge \\
            \cline{2-5}
                & $> 0.5$ & $\leq 0.3$ & - & not bulge \\
            \cline{2-5}
                & $> 0.5$ & $> 0.3$ & - & bulge \\
            \hline
            
            \multirow{8}{0.12\textwidth}{\centering PS $+$ Exp. Disk \newline $+$ S\'ersic $+$ S\'ersic} 
                & \multicolumn{3}{C{0.26\textwidth}|}{ S\'ersic with min. $b/a$ and S\'ersic with} & min. $b/a$ = not bulge; \\
            & \multicolumn{3}{C{0.26\textwidth}|}{min. $R_e$ are different} & min. $R_e$ = bulge \\
            \cline{2-5}
                & \multicolumn{3}{C{0.25\textwidth}|}{Single S\'ersic has min. $b/a$ \newline and min. $R_e$}
                & Apply single-S\'ersic criteria to both components \\
            \cline{2-5}
                & \multicolumn{3}{C{0.25\textwidth}|}{One bulge, one not bulge by single-S\'ersic criteria}
                & One bulge, one not bulge \\
            \cline{2-5}
                & \multicolumn{3}{C{0.25\textwidth}|}{Both S\'ersics satisfy bulge criteria}
                & Largest $n$ = bulge; \newline other = not a bulge \\
            \cline{2-5}
                & \multicolumn{3}{C{0.25\textwidth}|}{Both satisfy not-bulge criteria}
                & Smallest $b/a$ = not bulge; \newline other = not a bulge \\
            \hline

        \end{tabular}

    \end{threeparttable}

\end{table*}

\subsection{Bulge-Disk decomposition} \label{subsec:galfitm}

\subsubsection{Modelling surface brightness profiles} \label{subsubsec:galfitm:modelling}
In order to quantify the contribution of a bulge component to each galaxy's light profile and identify bulgeless disks in the DESI-BL-AGN sample, we model the surface brightness distribution using \textsc{galfitm}\footnote{\textsc{galfitm} is available at \url{https://www.nottingham.ac.uk/astronomy/megamorph/}}, developed from \textsc{galfit3.0} \citep{Peng2002, Peng2010} as part of the \textsc{MegaMorph} project \citep{Bamford2011, Haussler2013, Vika2013}. The surface brightness profile can be measured parametrically using functions such as the S\'ersic profile, describing the radial intensity profile \citep{Sersic1968}:
\begin{equation}
    I(R) = I_e \exp\{ -b_n [ ( \frac{R}{R_e} ) ^{1/n} - 1 ] \}
    \label{eq:sersic}
\end{equation}
where $b(n)$ is a constant, defined such that $R_e$ is the half-light radius, containing half of the total flux of the object, $I_e$ is the intensity at $R_e$ and $n$ is the S\'ersic index. Setting $n=4$ yields a \citet{deVaucouleurs1948} profile, and $n=1$ gives an exponential profile.

The primary aim is to perform bulge-disk decompositions of the light profile in an automated manner. We do this by fitting multiple multi-component surface brightness models to $r$-band imaging of the \disk sample, and selecting the best fitting model.

Before fitting, we create a mask image for each galaxy using \textsc{photutils}\footnote{\href{https://photutils.readthedocs.io/en/stable/}{photutils.readthedocs.io}} segmentation mapping with a $3\sigma$ detection threshold. This mask is used to ensure secondary sources do not affect the galaxy model fitting process. To capture extended emission, the segmentation is performed on a smoothed $r$-band image, made by convolving the image with a 2D Gaussian kernel. The central segmentation area is associated with the target galaxy.

We retrieve DESI $r-$band photometry information for all sources in the image, including their position and morphology\footnote{The morphology included in the DESI photometry catalogue comes from the \textsc{TRACTOR} software, which is included in the DESI LS pipeline. See \href{https://thetractor.readthedocs.io/en/latest/}{thetractor.readthedocs.io}.}. Each source is linked to an area of the segmentation map based on its distance from the segment centroid. We create an ellipsoidal mask for the target galaxy using the centroid, orientation, and semimajor axis from the segmentation. We increase the semimajor axis by a factor of 5 to capture faint disk emission. This is the target object footprint.

For sources with segmentation areas overlapping with the target footprint, we perform aperture photometry to compare the brightness of the source to its local background. To avoid masking bright spots, we do not mask overlapping sources with flux $<5$ times the background level. Some overlapping sources need modelling with \textsc{galfitm} to account for their emission. Overlapping PSs with aperture flux $>100$ times their local background are modelled as PSs. Overlapping extended sources with a segmentation area greater than $>1/3$ of the target object, and $r<20$ in the DESI catalogue are modelled as S\'ercics. We use DESI photometry for initial parameters and hold all parameters except position angle fixed. These modelled sources are not masked.

This gives the final source mask that is passed to \textsc{galfitm}: the segmentation map excluding the target object segment and any bright overlapping sources to be modelled.

To estimate the sky background, a more conservative mask is required. We invert the segmentation image to mask background pixels. We then dilate the background-masked image using the DESI model PSF image. We create a new mask from this by masking pixels with flux $>0.1\%$ of the peak flux in the masked image. This expands the masked areas of the original segmentation map. We add the elliptical target footprint to this mask. Applying this mask to the science image isolates the sky background emission. We use the median of this background emission as the sky background in \textsc{galfitm} modelling, holding it fixed throughout the fitting.

In modelling surface brightness profiles of BL-AGN host galaxies, the exclusion of point source component in the model leads to an over estimation of the contribution from the bulge component \citep[e.g.][]{Kim2008a, Simmons2008, Gabor2009, Bruce2016, Fahey2025}. We therefore include a point source in each model. In \textsc{galfitm}, the point source is modelled using a user-supplied PSF image, for which we use PSF model images from DESI LS DR9\footnote{LS DR9 PSF information available at \href{https://www.legacysurvey.org/dr9/psf/}{www.legacysurvey.org/dr9/psf/}}. The PSF model used is an extended PSF model composed of two components: a flexible inner PSF and a fixed outer PSF. The PSF of each CCD is first estimated using \textsc{PSFEx} \citep{Bertin2011}. The inner PSF is then a \citet{Moffat1969} profile fitted to the \textsc{PSFEx} for the CCD. The outer PSF is a power law for the $g-$ and $r-$bands, and a \citet{Moffat1969} profile for the $z$-band.

The modelling process is as follows. First we fit the simplest model: a two-component model consisting of a single S\'ersic and a point source. We use the DESI LS DR9 photometry to inform the input magnitudes of the two components. We perform an initial fit of this simple model allowing all component parameters to vary. The central coordinate of the fitted S\'ersic is then used to fix the central coordinates of all components in the proceeding models. We re-fit the two component model with the positions fixed, using the results of the initial fit as the input parameters. 

Secondly, we fit a three-component model, composed of an exponential disk, a S\'ersic and a point source. \textsc{galfitm}'s exponential disk profile is equivalent to the S\'ersic profile in \autoref{eq:sersic} with a fixed index of $n=1$. The input parameters of the S\'ersic are chosen such that it reflects a bulge component: $R_e = 0.1 R_{e,{\rm disk}}$; $n=2$ \citep[the border between classical and pseudo-bulges according to][]{Fisher2008}; $b/a=0.8$. However, none of these parameters are fixed, so the final S\'ersic component in the fitted model may not represent a stellar bulge and we must check (see Section \ref{subsec:galfit:bls_sb}). The input magnitudes of the exponential disk and S\'ersic components are set to 80\% and 20\% of the magnitude of the S\'ersic in the fitted two-component model, respectively.

Many of the galaxies in the \disk sample are barred (see Section \ref{subsubsec:disc:bars_and_spirals} in the discussion), therefore we fit a final four-component model by adding an additional S\'ersic component. This allows both the bar and bulge to be modelled where both are present. Excluding a bar component in barred galaxies leads to significant under-estimation of the bulge contribution \citep[e.g.][]{Gadotti2008} The input parameters of the two S\'ersic components were chosen such that one was more representative of a bulge ($R_e = 0.25 R_{e,\text{prev. sersic}}$; $n=2$; $b/a=0.8$), and the other of a bar ($R_e = 0.5 R_{e,\text{disk}}$; $n=0.7$; $b/a=0.2$). While bars have been modelled using both S\'ersic and \citet{Ferrers1877} profile, it has been demonstrated that the two functions yield similar results \citep{Kim2015}.

Throughout the fitting, we constrain $R_e$ and $n$ S\'ersic to $0.5<R_e<700\text{ pixels}$ and $0<n<8$. The same constraint is places on $R_e$ of the exponential disk. Furthermore, the magnitude of each component is constrained to be within $6\text{ dex}$ of the input value. These soft constraints are based on the method of \citet{Kruk2018}, and are in place to prevent convergence to extremely small or large, unphysical parameter values.

Having performed two- three- and four-component model fits, the best-fitting model was determined as that with the reduced $\chi^2$ ($\chi_\nu^2$) value closest to 1. For each galaxy, a final fit of the best-fitting model was performed, using the output of the best fit as the input parameters. In this final fit, the central coordinates were no longer held fixed. However, for four-component models, the centres of the two sub-disk S\'ersic components were bound together (as done by \citealp{Kruk2018}).

Note, we have not explicitly excluded edge-on or nearly edge-on galaxies from the fitting. Visual inspection of the $r$-band images galaxies with GZ DESI vote fraction $f_{\rm edge-on-yes}>0.8$ for ``yes'' in answer to the question ``Could this be a disk viewed edge-on?'', showed a number of these galaxies have extended emission visible in the $r$-band which is less apparent in the $r,g,b$ colour imaging. Therefore we include these objects in the fitting process for completeness.

Of the 2,435 \disk galaxies fitted with \textsc{galfitm}, 1,709 ($\sim70\%$) successfully reached the end of the full process outlined above. Of the remaining 726, at least one \textsc{galfitm} model was successfully fitted for 448. For these objects, we identify a best fitting model by comparing $\chi^2_\nu$ of the model successful fitted by \textsc{galfitm}. This gives 1,927 galaxies with \textsc{galfitm} models ($\sim89\%$ of the \disk sample).

We identify and remove poor-quality fits from the sample by looking for convergence of parameters, poorly constrained radii, and sub-disk components with unphysical radii. \textsc{galfitm} flagged a possible convergence of parameters in the models of 219 objects ($\sim10\%$ of the 2,157 models obtained). We identify components with a poorly constrained effective radius, $R_e$, if the output error on the radius, $\sigma_{R_e}$, is $\geq0.5 R_e$. This effects 189 objects ($\sim8\%$). We also look for sub-disk components (e.g. bulges and bars) with $R_e$ larger than that of the disk. Such unphysical component radii were identified in 429 models ($\sim20\%$). With all of these criteria combined, a total of 391 models were flagged as poor quality ($\sim18\%$), leaving 1,766 objects.

We make a further selection on $\chi^2_\nu$, excluding models with $\chi^2_\nu\geq2$. Thus, 1,433 objects in our \disk sample were found to have good-quality \textsc{galfitm} models.\footnote{The smallest axis ratio ($b/a$) of the disk is $0.106$, equivalent to an inclination angle $i\sim84^\degree$ for a thin, circular disk.}

\subsubsection{Identification of bulgeless galaxies} \label{subsec:galfit:bls_sb}

For best fitting models composed of two components (a PS and a S\'ersic) there is no bulge component. However, models with exponential disks and additional S\'ersic components may contain a bulge component. Our priority in identifying bulge components is the purity of the bulgeless sample. The criteria detailed below were chosen primarily to distinguish between bulges and bars, while prioritising the purity of the bulgeless sample. We classify components simply as ``bulge'' or ``not bulge'', and do not discriminate between pseudo- and classical bulges. The criteria applied are summarised in Table \ref{tab:identify_bulges}. 

For models with three components (PS + exponential disk + S\'ersic), the S\'ersic is classified as bulge or not bulge based on $n$ and $b/a$. If $b/a>0.85$, it is classified as a bulge irrespective of the value of $n$ as we consider this too round a feature to be classified as a bar. If $n>0.5$ and $b/a>0.3$, it is classified as a bulge. Pseudobulges are thought to have $n=0.5-2$ \citep{Fisher2008}, while a S\'ersic profile with $n\lesssim0.5$ is similar in shape to the \citet{Ferrers1877} profile, commonly used to model bars \citep{Peng2010}. We consider profiles with $b/a\leq0.3$ too elongated to be considered a bulge. Otherwise, the component is classified as not a bulge.

The case in models with four components (PS +exponential disk + S\'ersic + S\'ersic) is more complex. We first compare $R_e$ and $b/a$ of the S\'ersics: the one with the smallest radius is taken to be the bulge while the most elongated (smallest $b/a$) is taken to be the bar and, therefore, not a bulge. If a single S\'ersic is found to be both the smallest and most elongated of the two, we apply the single-S\'ersic criteria used for the three component model to both S\'ersic components. In most instances, one S\'ersic is classified as a bulge and the other as not a bulge under these criteria. However, rare instances do arise where both S\'ersics are both are classified as a bulge, or both as not a bulge, under the single S\'ersic criteria. If both are classified as a bulge, the S\'ersic with the largest $n$ is taken to be the true bulge. In this case, the second the flatter profile is more likely to be an inner disk.

Selecting those galaxies with no identified bulge in their \textsc{galfitm} model, we obtain a sample of 1,233 bulgeless galaxies. Similarly, selecting those with an identified bulge component, we have a sample of 278 galaxies with some bulge component.

Our selection of bulgeless galaxies as those without a bulge in the best-fitting model differs from the traditional classification of bulgeless galaxies as those with bulge-to-total mass ratio $B/T<0.1$. Under this traditional classification, some galaxies with a bulge component in the model may be classified as bulgeless (see panel (d) of \autoref{fig:mbh_mstar_mbulge_bt_hists}). However, given the limited resolution of the ground-based DESI LS imaging, the classification based on the presence/absence of a bulge component in the model is more conservative.

\subsubsection{Completeness and purity measured via injection tests}\label{subsec:galfit:inject}

The ability to reliably de-compose the bulge and disk components in a galaxy's surface brightness profile depends on the resolution of the imaging, and the physical size of the bulge relative to the $\rm FWHM$ of the PSF. Poor spatial resolution or a small bulge relative to the $\rm FWHM$ can mean the bulge is not decomposed in the modelling, and may lead to false identification of ``bulgeless'' galaxies.

To assess the bulge-disk decomposition described above, we perform injection tests; taking a sample of bulgeless galaxies in DESI LS DR9 with no central point source, inject artificial bulge and point source components to the $r-$band imaging, and perform bulge+disk decomposition with \texttt{GALFITM}.

We select a sample of 10 bulgeless galaxies using the GZ DESI morphology catalogue, which do not contain a BL-AGN. Non-merging disk galaxies are selected following the procedure outlined in Section \ref{subsec:disks}. As we are selecting galaxies with no central point source, the GZ DESI vote fractions in response to the question ``Is there a central bulge? If so, what size is it compared with the galaxy?'' can be used for this selection, where it could not be used for the \disk sample. Selecting a vote fraction $>0.6$ for ``no bulge'' in response to this question provides a bulgeless sample. We then selected 10 galaxies from this sample, herein \textsc{test}, with redshift and $r-$band magnitude, ${\rm mag}(r)$, close to the median of the \disk sample ($z\sim0.1234$ and ${\rm mag}(r)\sim16.1$). Prior to the injection of any artificial components, we fit a single S\'ersic model to each of the \textsc{test} galaxies with \texttt{GALFITM}. Output parameters of these fits are then used as input for the disk component in subsequent model fits.

We select a range of 10 linearly-spaced point-source magnitudes [18.60, 19.28, 19.96, 20.63, 21.31, 21.99, 22.67, 23.34, 24.02, 24.70] and 10 linearly-spaced bulge magnitudes [16.60, 17.17, 17.73, 18.30, 18.87, 19.43, 20.00, 20.57, 21.13, 21.70], chosen to encapsulate a $3\sigma$ range around the median values of these properties for the 278 galaxies with a bulge component identified in the \texttt{GALFITM} fitting (Section \ref{subsec:galfit:bls_sb}). Each combination of these point-source and bulge magnitudes, added to each of the \textsc{test} galaxies, forms a unique injection ``experiment''. Additionally, we consider the case where no bulge component is injected, but a point source is injected, giving a total of 1,100 unique experiments on which to test the bulge-disk decomposition with \texttt{GALFITM}.

We form artificial bulge component images using a S\'ersic profile, with $n=2.0$, $R_{e,\text{bulge}}=0.5R_{e,\text{SS}}$ where $R_{e,\text{SS}}$ is taken from the single-S\'ersic fit for the given \textsc{test} galaxy. The centroid, position angle and $b/a$ of the bulge are set to match the single-S\'ersic output. This S\'ersic profile image is then convolved with the DESI LS DR9 PSF image corresponding to the \textsc{test} galaxy. Artificial point source component images are formed by convolving a delta function with the DESI LS DR9 PSF. Each artificial component image is then normalised such that the total flux, $F_{\text{art}}$, is equivalent to the desired injected source magnitude.

We must add noise to these artificial component images, which will also be added to the \textsc{test} galaxy's invariance image. This is somewhat complicated by the use of DESI co-add images, as the exact gain information across the field-of-view of an image is not known. To model the noise, we therefore calculate the effective gain of each \textsc{test} galaxy co-add image from the r-band image and invariance image, using a process similar to that of the \texttt{lsst.source.injection} package, for synthetic source injection in LSST co-add images \citep[][]{VeraRubin2025}. A first degree polynomial is fitted to the \textsc{test} galaxy variance and flux images in the form $v_i=af_i+b$, where $v$ is variance and $f$ is source flux. The effective gain is then given by $g_{\rm eff}=1/a$ (in flux units, i.e. $\text{nMgy}$). An artificial variance image is then produced as $v_{{\rm art},i} = f_{{\rm art},i}/g_{\rm eff}$, which is added to the original LS DR9 variance image linearly. We generate Gaussian noise, using $v_{{\rm art},i}$ as the standard deviation, and add this to the artificial source image such that $f_{\text{art},i}^{\rm noisy}=f_{\rm art}+\mathcal{N}(0, v_{{\rm art},i})$. The artificial components can the be injected into the LS DR9 image by linear addition, $f_{{\rm inj},i}=f_i+f_{\text{art},i}^{\rm noisy}$.

For each of the 1,100 injection experiments, we inject the relevant artificial components as described, and fit two surface brightness models to the injected source images using \textsc{GALFITM}. The first model consists of a single S\'ersic component and a point source. The second model consists of an exponential disk, a S\'ersic component and a point source. The best-fitting of the two models is determined by $\chi_\nu^2$, as done for the \disk sample (Section \ref{subsubsec:galfitm:modelling}). The results of the modelling described below are then in relation to the best-fitting model.

We define four possible outcomes of the \textsc{GALFITM} fitting of each injection experiment: ``true positive'' (TP) where an injected bulge component is successfully recovered; ``true negative'' (TN) where no bulge component was injected and no bulge component was recovered; ``false positive'' (FP) where no bulge was injected but the best fitting model included a bulge component; and ``false negative'' (FN) where an injected bulge component was not recovered in the final model.

Using these definitions, we calculate the completeness ($C$) and purity ($P$) of the bulgeless and some-bulge test samples as:
\begin{equation*}
    C_{\rm some-bulge} = \frac{\rm TP}{\rm TP + FN} \quad\text{and}\quad C_{\rm bulgeless} = \frac{\rm TN}{\rm TN + FP}
\end{equation*}

\begin{equation*}
    P_{\rm some-bulge} = \frac{\rm TP}{\rm TP + FP} \quad\text{and}\quad P_{\rm bulgeless} = \frac{\rm TN}{\rm TN + FN}
\end{equation*}

From the injection test, we initially find $C_{\rm some-bulge}=52\%$, $C_{\rm bulgeless}=62\%$, $P_{\rm some-bulge}=58\%$ and $P_{\rm bulgeless}=56\%$.

To investigate the effect of spatial resolution, $r_{\rm res}$, we calculate the $C$ and $P$ of the injection tests as a function of $r_{\rm res}$. Both $C$ and $P$ are seen to worsen significantly at $r_{\rm res}\gtrsim2.4$. Similarly, to investigate the effect of size of the galaxy relative to the $\rm FWHM$ of the PSF, we calculate $C$ and $P$ as functions of $R_{e,{\rm disk}}/{\rm FWHM}$. Significant improvement in $C$ and $P$ is seen at $R_{e,{\rm disk}}/{\rm FWHM} \gtrsim 2.6$.

Selecting only the injection tests with $r_{\rm res} < 2.4$ and $R_{e,{\rm disk}}/{\rm FWHM} > 2.6$, we find $C_{\rm some-bulge}=48\%$, $C_{\rm bulgeless}=93\%$, $P_{\rm some-bulge}=88\%$ and $P_{\rm bulgeless}=64\%$. We note there is a decrease in $C_{\rm some-bulge}$. However, for the purpose of this paper, we are more concerned with purity than completeness, which shows a significant increase for both bulgeless and some-bulge injection tests.

\begin{figure}
    \centering
    \includegraphics[width=\linewidth]{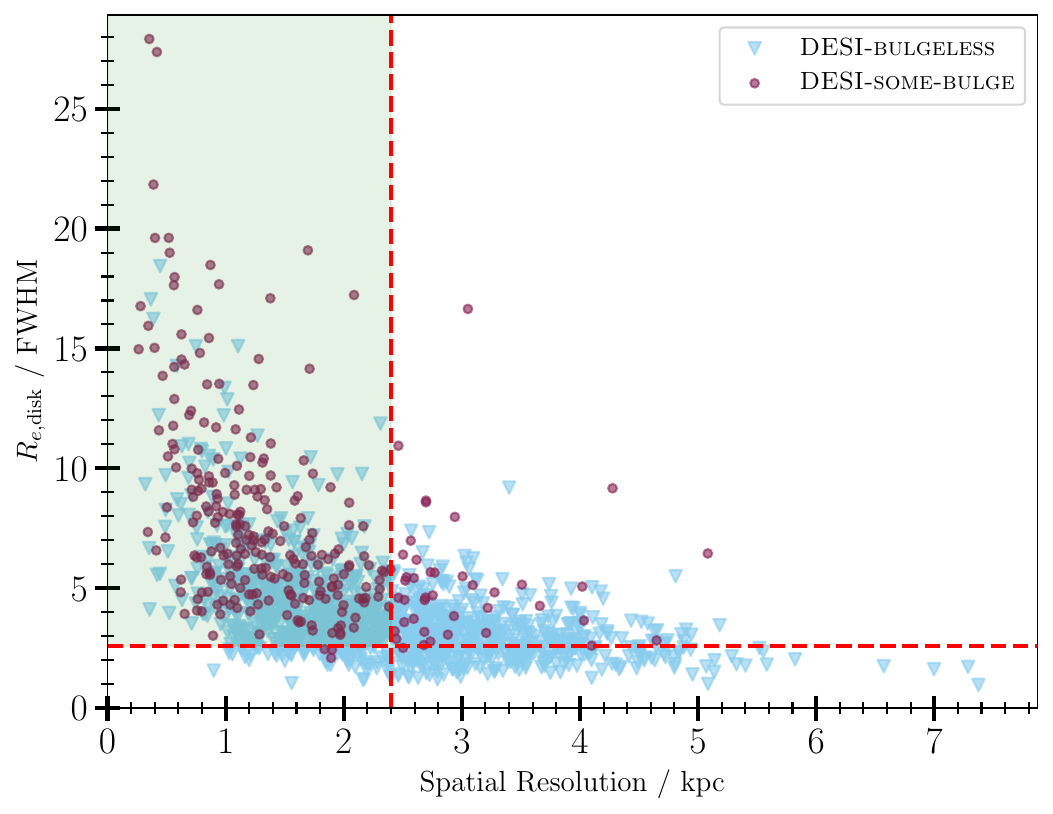}
    \caption{Ratio of disk radius ($R_{e,{\rm disk}}$) to PSF $\rm FWHM$ against spatial resolution  for \bulgeless galaxies (blue triangles) and \somebulge (burgundy circles). The horizontal and vertical red dashed lines indicate $R_{e,{\rm disk}}/{\rm FWHM}=2.6$ and $r_{\rm res}=2.4\text{ kpc}$. The green shaded region shows the selected sample region.}
    \label{fig:Reff_FWHM_v_resolution} 
\end{figure}

Upon initial examination, the purity of the bulgeless sample derived from the injection tests is low. However, we find the false negatives from the injection test tend to result from fainter injected stellar bulge components, with $\sim48\%$ of the false negatives having $B/T<0.1$. Furthermore, the \disk sample probes lower $r_{\rm res}$ than represented in the \textsc{test} sample, therefore the values of completeness and purity reported for the injection tests, after making the above selections, should be thought of as lower limits on the completeness and purity of the science sample, after the same selections are made. The selection thresholds chosen improve $C$ and $P$ of the bulgeless sample sufficiently, while maintaining a good sample size and avoiding over-tuning to the injection test sample over the true science sample.

We make the same cuts on the 1,433 objects in the \disk sample with successful \textsc{GALFITM} models, shown in \autoref{fig:Reff_FWHM_v_resolution}. This reduces the sample to 786 objects with reliable \textsc{GALFITM} models. The number of bulgeless galaxies reduces to 550 (by $\sim56\%$) and the number of galaxies with some bulge component reduces to 236 (by $\sim14\%$). The significantly larger reduction in bulgeless galaxies indicates that many apparent bulgeless galaxies with unreliable decompositions have been removed from our sample by this selection.

We visually inspected the \textsc{GALFITM} models of 23 bulgeless galaxies which have an additional non-bulge component in the final model, and the properties of this component. Of these, 4 of the non-bulge components were re-classified as a bulge, giving final samples of 546 bulgeless galaxies, herein \bulgeless, and 240 galaxies with some bulge component, herein \somebulge.

\begin{figure*}
    \centering
    \includegraphics[width=\linewidth]{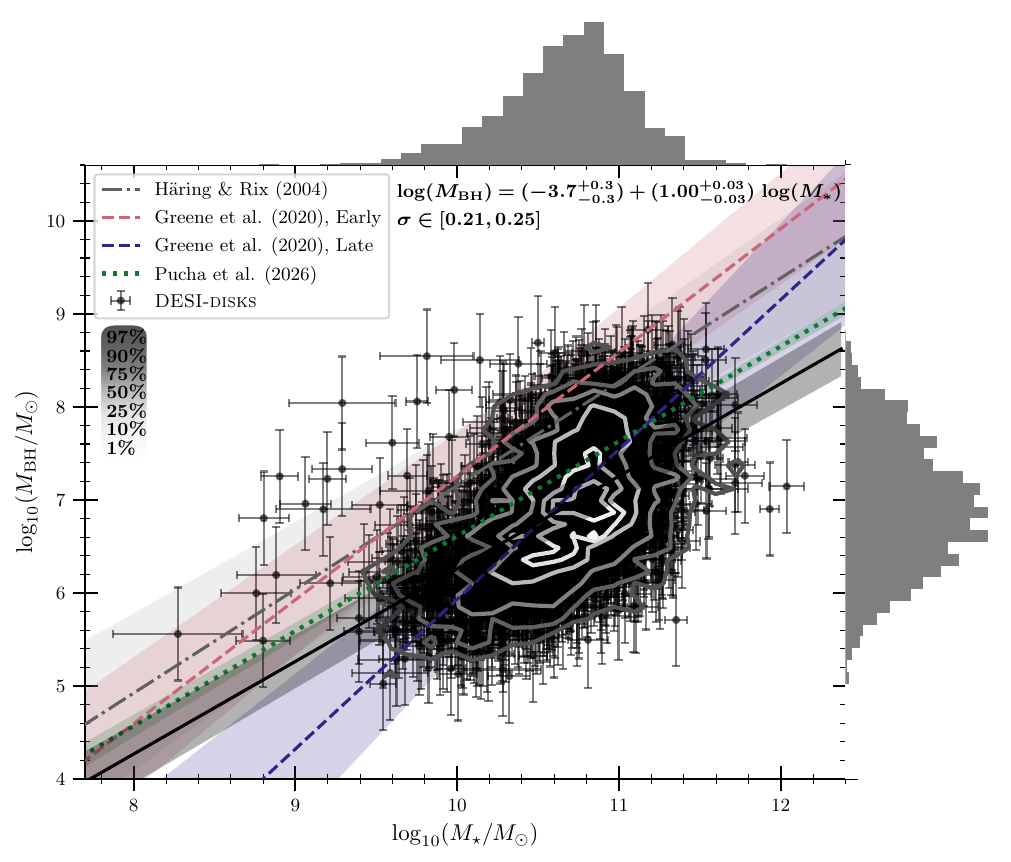}
    \caption{Black hole mass against total stellar mass for the \disk sample of disk galaxies hosting optical BL-AGN. Black hole masses are estimated from broad $\rm H\alpha$ emission using the \citep{Reines2013} relation. Total stellar mass values are taken from \citet{Siudek2024} SED fitting results using DESI LS photometry, performed with \textsc{CIGALE} \citep{Boquien2019}.
    The 1\%, 10\%, 25\%, 50\%, 75\%, 90\% and 97\% contours are plotted (light-to-dark showing the smallest-to-largest percentages). The best-fitting straight-line \mbh--\mstar relation is shown as a black solid line (--). The mathematical form of this relation is shown as text, along with the intrinsic scatter on the relation, $\sigma$. Multiple correlations from the literature are also shown: \citet{Haring2004} for early-type and bulge-dominated galaxies as a light grey, dot-dashed line ($\cdot-\cdot$); \citet{Greene2020} for early-type galaxies and for late-type galaxies as dashed lines ($--$) in red and blue respectively; \citet{Pucha2026} is shown as a green dotted line ($\cdot\cdot\cdot$). For each line plotted, the corresponding shaded region shows the $1\sigma$ uncertainty with intrinsic scatter. The distributions of black hole mass and total stellar mass are shown as histograms across the right-hand and upper axes, respectively.}
    \label{fig:mbh_mstar}
\end{figure*}

\subsubsection{Stellar masses of galactic bulges} \label{subsec:galfit:mbulge}

We estimate the stellar mass of the bulge component, \mbulge, by assuming a constant stellar mass-to-light ratio ($M/L$) across the galaxy. We measure $M/L$ from the total \mstar of the galaxy and the combined flux of all components of the best-fitting \textsc{galfitm} galaxy model. An estimate of \mbulge is then obtained from the flux of the identified bulge component of the model:
\begin{equation}
    M_{*,{\rm bulge}} = (M/L)_{\rm total} \times F_{\rm bulge}
\end{equation}
In reality, $M/L$ will vary between the disk and any central bulge component, as the stellar population of the bulge is found to be older than that of the disk \citep{DeJong1996, Moffett2015, Moffett2016}. However, there is not a well-defined relationship between the two that can be applied across all systems in the sample. Attempting to account for this effect would require further assumptions and introduce further uncertainty in the \mbulge estimates. Using the total $M/L$ across all galaxy components requires fewer assumptions and is therefore preferred in this case.

For \bulgeless galaxies, without any bulge component in the model, we calculate upper limits on \mbulge. This allows us to compare the full \bulgeless sample to \somebulge on the \mbh--\mbulge plane. The upper limit should represent the brightest possible bulge component that could potentially be hidden behind the flux of the central point source. Using the \somebulge sample, we can identify the fitted bulge which is closest in brightness to the point source i.e., with the smallest difference in flux ($\Delta F = F_{\rm PS}-F_{\rm bulge}$). This then represents the minimum difference in brightness which which both point source and bulge are distinguishable. Subtracting this flux difference from the point source flux for \bulgeless galaxies, then provides a physically-motivated upper limit on the flux of the indeterminable bulge component:
\begin{equation}
    F_{\rm bulge} \leq F_{\rm PS} - \min(\Delta F)
\end{equation}
We found $\min(\Delta F)\simeq0.05\text{ dex}$ in magnitude. For 135 \bulgeless galaxies ($25\%$), $F_{\rm PS}<\Delta F$ leading to an upper limit \mbulge$<0$, which is unphysical. For these galaxies, we take $F_{\rm PS}$ as the limiting flux, using this to obtain an upper limit on \mbulge. 

\citet{SSL17} use bulge magnitude from bulge-disk decompositions of \citet{Simard2011}, which do not include any point source component, as conservative upper limits on \mbulge. This approach attributes all the flux from the central BL-AGN to the stellar bulge. We could do something similar by attributing all of the point source flux in the \textsc{GALFITM} model to a stellar bulge. Given all the \disk galaxies host BL-AGN, the presence of a point source in the galaxy is known, therefore this is not a physical limit. However, it is the most conservative possible approach. Calculating them in this way would increase the bulge mass upper limits for the \bulgeless sample by a median value of $0.07\pm0.3\text{ dex}$ (with $>1\text{ dex}$ for only 2\% of the bulgeless sample).


\section{Results} \label{sec:results}

\subsection{Disk galaxies on the BH mass -- stellar mass plane} \label{subsec:res:mbh_mstar}

The full \disk sample is plotted on the stellar mass - black hole mass plane in \autoref{fig:mbh_mstar}. To the best of our knowledge, this is the largest sample of AGN hosts with known late-type morphologies to be plotted on the total stellar mass - black hole mass plane to date. There is a moderate positive correlation between \logmbh and \logmstar for the \disk ($r=0.536^{+0.016}_{-0.016}$) sample\footnote{Due to the large sample size, the null hypothesis of no correlation between \logmbh and \logmstar is rejected at extremely high significance, with $p\ll10^{-10}$.}. While there is a well-defined core cluster of points, there is also some considerable scatter ($\sigma_{x}=0.4$, $\sigma_{y}=0.7$, average distance from the mean $=0.7$). 

Using the \textsc{linmix}\footnote{Python port available here: \href{https://github.com/jmeyers314/linmix?tab=readme-ov-file}{github.com/jmeyers314/linmix}} software \citep{Kelly2007} we fit a straight line with intercept $\beta$ and slope $\alpha$, giving the relation:
\begin{equation}
    \log(M_{\rm BH}/M_\odot) = -3.7^{+0.3}_{-0.3} + 1.00^{+0.03}_{-0.03} \log(M_*/M_\odot)
    \label{eq:mbh_mstar}
\end{equation}
with intrinsic scatter in \logmbh at fixed \logmstar of $\sigma = 0.23^{+0.02}_{-0.02}$ dex. The fitted relation is shown as the solid line in \autoref{fig:mbh_mstar}, and the $1\sigma$ uncertainty on the line (plus intrinsic scatter) is shown as a shaded region.

The distribution of \logmbh is shown on the right-hand axis of \autoref{fig:mbh_mstar}. The peak occurs around \logmbh$\sim6.9$. However, there is a potential second peak visible in the distribution at \logmbh$\sim7.6$, with lower amplitude. This feature persists down to histogram bins of \logmbh$\sim0.25$ dex in width. We compare the fit of a single Gaussian and double Gaussian curve to the \logmbh distribution, and find the inclusion of the second Gaussian improves $\chi^2_\nu$ of the fit by $\sim0.8$. If this observation is real, and reflects some physical trait of or process occurring in the sample, it could indicate some change in physics between two underlying populations. \citet{Pucha2026} also find this feature in their DESI BL-AGN sample, where the host galaxies have not been selected on a particular morphology, suggesting this feature is not driven by morphology. However, further work is required to form any conclusions on the cause of this feature.


\subsection{Comparison of \bulgeless and \somebulge galaxies}
\label{subsec:res:compare}

\begin{figure}
    \includegraphics[width=\linewidth]{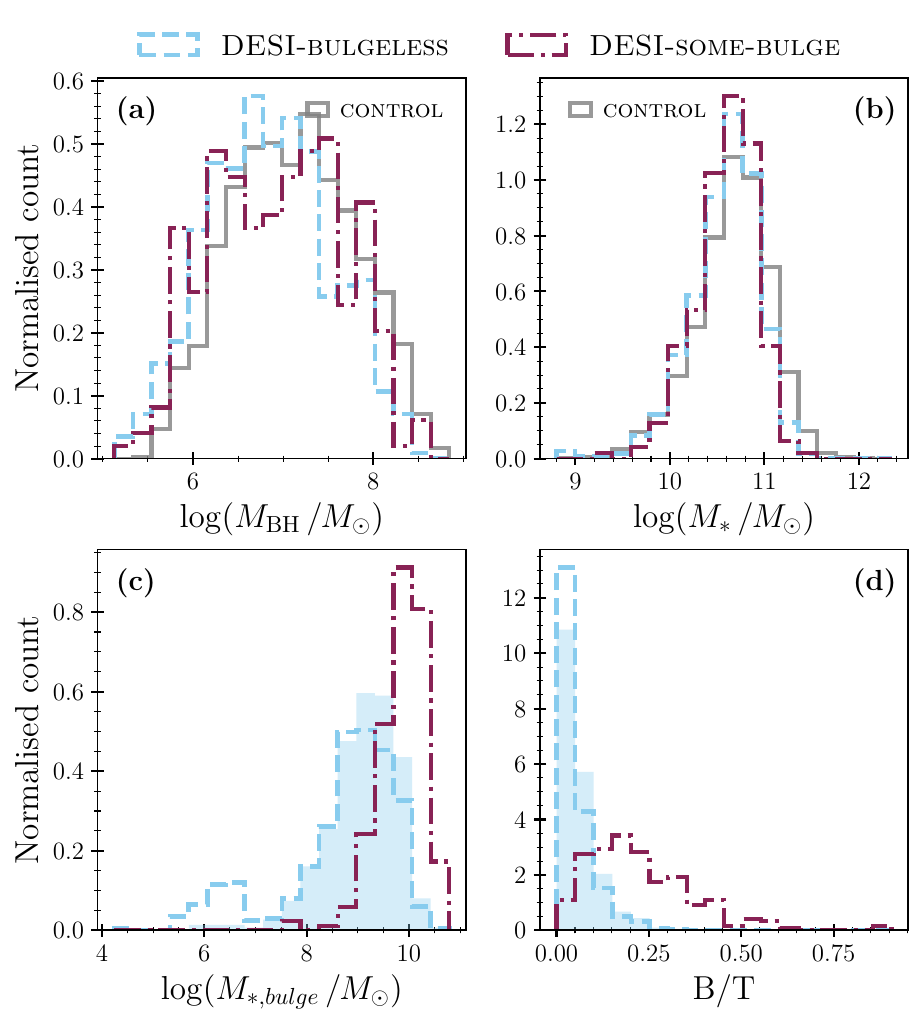}
    \caption{Distributions of black hole mass (\textbf{a}), total stellar mass (\textbf{b}), bulge stellar mass (\textbf{c}) and bulge-to-total ratio (\textbf{d}) for \bulgeless galaxies in blue (dashed line $--$), and \somebulge galaxies in burgundy (dot-dashed line $\cdot-\cdot$). The black hole mass distribution of the \control sample of early-type galaxies, is also included in panels (a) and (b) in grey (solid line --). The \bulgeless sample excluding points-source-limited bulge properties are also shown in panels (c) and (d), as shaded blue histograms.}
    \label{fig:mbh_mstar_mbulge_bt_hists}
\end{figure}

In this section, we compare the physical properties of the \bulgeless and \somebulge populations. Their BH mass, total stellar mass, bulge mass and $B/T$ distributions are shown in \autoref{fig:mbh_mstar_mbulge_bt_hists}, and demonstrate the real difference between the two populations.

The distributions of BH mass for the two populations, shown in panel (a) of \autoref{fig:mbh_mstar_mbulge_bt_hists}, cover a very similar range ($5.1 \lesssim$ \logmbh$\lesssim 8.6$ for \bulgeless; $5.3 \lesssim$ \logmbh$\lesssim 8.6$ for \somebulge). The median \logmbh is $6.8\pm0.5$ for \bulgeless and $7.0\pm0.5$ for \somebulge. However, the \somebulge distribution has a double-peak structure, with apparent peaks at \logmbh$\sim6.4$ and $\sim7.2$. The positions of these peaks differ from the double peak structure in the BH mass distribution of the overall \disk sample noted in Section \ref{subsec:res:mbh_mstar} (see \autoref{fig:mbh_mstar}), being shifted towards lower BH mass by $\sim0.4-0.5$ dex. This could indicate a difference in the emergence of this structure in each sample. However, as noted in Section \ref{subsec:res:mbh_mstar}, this requires further investigation to draw firm conclusions, which is left to future work. An Anderson-Darling \citep[AD;][]{Anderson1954} test gives statistic 1.7 and $p=0.07$, meaning the difference is not significant at $\sim1.8\sigma$. Therefore, we cannot reject the null hypothesis that the BH masses in the \bulgeless and \somebulge samples are drawn from the same distribution. This implies the same mechanism is growing the BHs in both \bulgeless and \somebulge samples. Given the assumed quiet-merger histories of the \bulgeless sample (as demonstrated by their lack of significant stellar bulge), we hypothesise that secular mechanisms are driving the BH growth in both samples, and discuss this further in Section \ref{subsec:disc:secular-growth}.

The total stellar mass distributions of the \bulgeless and \somebulge populations are shown in panel (b) of \autoref{fig:mbh_mstar_mbulge_bt_hists}. Recall, these two populations are not matched in stellar mass in any way. Both populations' distributions peak at $10.47\leq\log(M_*)\leq10.87$, and their median values differ by $\lesssim0.09\sigma$. Both populations cover a similar range of \logmstar, with the \somebulge having a slightly smaller range ($8.8<\log(M_*)<11.5$ for \bulgeless and $9.3<\log(M_*)<11.4$ for \somebulge). An AD test suggests moderate difference in the distributions ($p=0.38$) but with low significance ($\sim0.9\sigma$).

The bulge mass distributions are shown in panel (c) of \autoref{fig:mbh_mstar_mbulge_bt_hists}. Recall that the bulge masses of the \bulgeless sample are upper limits, of which $25\%$ are limited purely by the point source flux (Section \ref{subsec:galfit:mbulge}). By design, we expect the \bulgeless sample to have generally lower \mbulge than \somebulge. Indeed, we see that the \bulgeless population peaks at lower bulge mass (peak \logmbulge$\sim 9.2$ and median $\sim9.0$, compared to peak $\sim 9.9$ and median $\sim9.9$ for \somebulge) and extends to lower bulge mass (\logmbulge$\sim 4.2$ compared to $\sim 7.7$ for \somebulge). Furthermore, in this instance an AD test hits the imposed floor value of $p=0.0001$, meaning the difference is significant at $>3\sigma$. A secondary peak is seen in the \bulgeless bulge mass distribution at \logmbulge$\sim6.6$. This is composed only of objects where the limit on bulge mass is taken from the point source flux, as seen by comparison to the \bulgeless distribution excluding these PS-limited bulge masses shown in \autoref{fig:mbh_mstar_mbulge_bt_hists}.

The $B/T$ distributions are shown in panel (d) of \autoref{fig:mbh_mstar_mbulge_bt_hists}. Although we do not select \bulgeless galaxies using the traditional selection of $B/T<0.1$ (see Section \ref{subsec:galfit:bls_sb}), we note that the \bulgeless sample clearly peaks below this threshold at $B/T\sim0.008$. The \somebulge distribution is spread more widely, and peaks above this threshold at $B/T\sim0.2$.


\begin{figure*}
    \includegraphics[width=\linewidth]{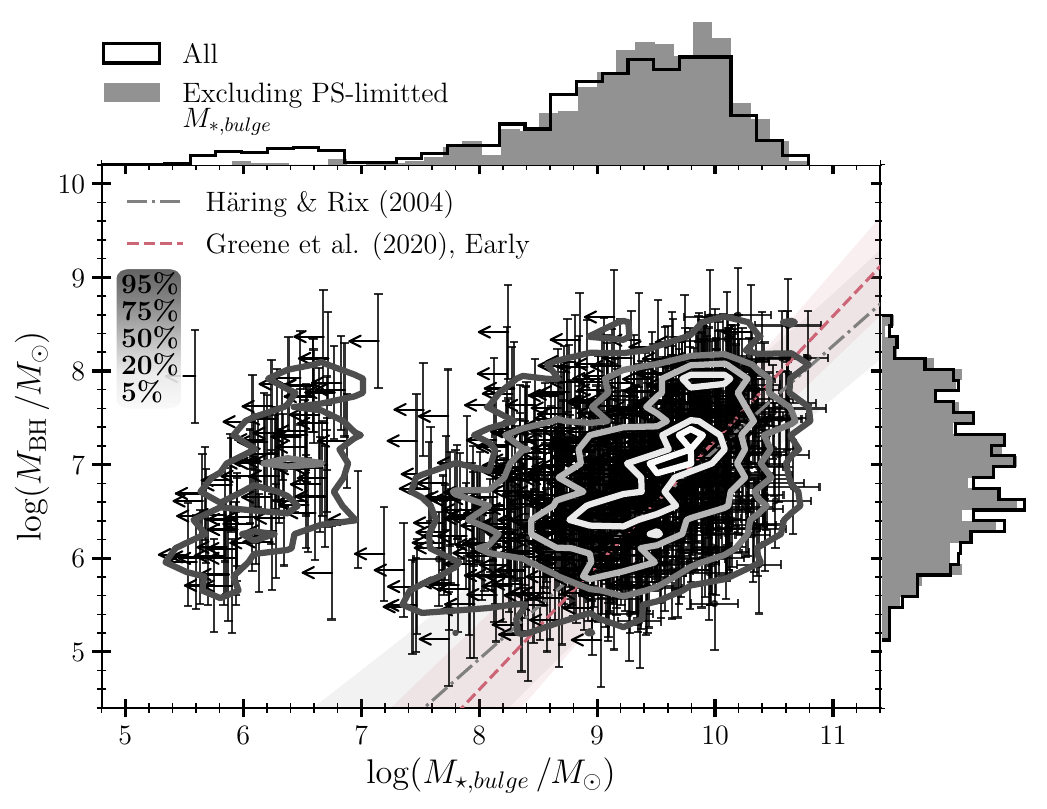}
    \caption{Black hole mass against bulge stellar mass for 778 objects in the \disk sample, with \textsc{galfitm} surface brightness models. The distributions of black hole mass and bulge stellar mass are shown as histograms across the right-hand and upper axes, respectively. For \somebulge with bulge components in the model, bulge masses are found using the total $M/L$ of the galaxy. For the \bulgeless galaxies with no bulge component in the model, upper limits are used and plotted as arrows. Upper limits are found by subtracting the minimum difference in point source and bulge flux in the \somebulge sample from the point source flux of the \bulgeless galaxy. Where this difference is greater than the point source flux of the \bulgeless, the point source flux is taken as the limit. The filled histograms show the distributions excluding the 135 \bulgeless galaxies with point-source-limited bulge mass. Black hole mass values are as in Figure \ref{fig:mbh_mstar}. The \citet{Haring2004} relation for ellipticals and bulge-dominated galaxies is shown in grey. The \citet{Greene2020} for late-type galaxies is shown in red. The filled histograms include those with upper limits on bulge mass, while the black line histograms include only objects with measured bulge mass values. A straight line fit yields a slope of $\sim0.15-0.40$, where the lower and upper bounds are set by including and excluding the point-source-limited bulge mass limits, respectively.}
    \label{fig:mbh_mbulge}
\end{figure*}

\subsection{Black hole scaling relations for bulgeless disk galaxies} \label{subsec:b_v_sb}

The 778 objects with \textsc{galfitm} models are plotted on the \mbh--\mbulge plane in \autoref{fig:mbh_mbulge}. The sample shows a clear deviation from the \citet{Haring2004} relation and \citet{Greene2020} early-type relation, with many black holes being over-massive compared to these relations at \logmbulge$\lesssim 9$. The distribution of bulge mass is shown on the upper axes of the plot. This distribution contains both \somebulge and \bulgeless, and therefore both measured values and upper limits. The median bulge mass across the sample is \logmbulge$\sim9.3$.

\begin{figure*}
    \includegraphics[width=\linewidth]{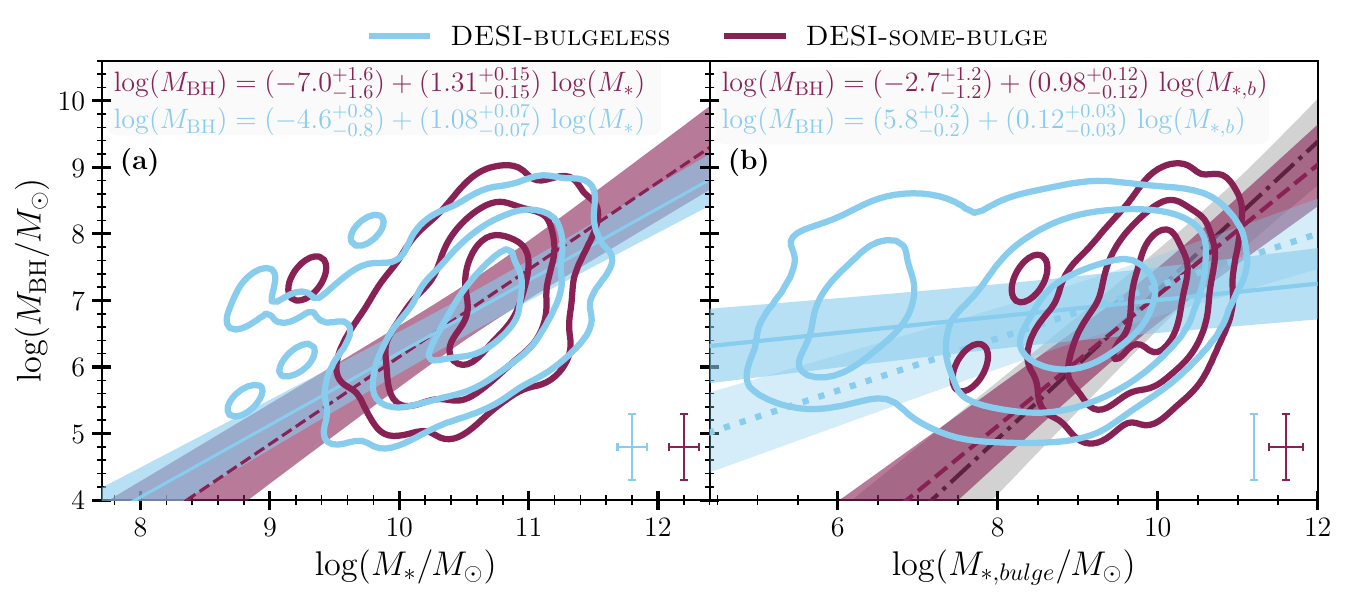}
    \caption{Black hole mass against total stellar mass (left, panel a) and against bulge stellar mass (right, panel b) for 784 galaxies with bulge-disk decompositions from \textsc{galfitm} surface brightness model fitting. The \bulgeless galaxy sample is shown in blue and the \somebulge sample is shown in burgundy. The $1\sigma, 2\sigma$ and $3\sigma$ contours are shown. Black hole masses and total stellar masses are defined as in \autoref{fig:mbh_mstar}. Bulge stellar mass is defined as in \autoref{fig:mbh_mbulge}, with upper limits used for the \bulgeless galaxies. The median best-fitting straight line correlations are plotted as a solid blue line for \bulgeless and a dashed burgundy line (--) for \somebulge, with shaded $1\sigma$ regions. These relations are also shown explicitly as text. The median best-fitting straight line for the \bulgeless sample excluding objects where the bulge mass is limited by the brightness of the central point source, is shown as a blue dotted line ($\cdot\cdot\cdot$). The \citet{Haring2004} relation is shown on panel (b) as a grey dash-dotted line ($\cdot-\cdot$).} 
    \label{fig:mbh_mstar_mbulge_fits}
\end{figure*}

To investigate if the deviation from pre-existing relations exists for both \bulgeless and \somebulge, and explore the difference in their BH scaling relations more generally, we plot the two populations separately on the \mbh--\mstar and \mbh--\mbulge planes in \autoref{fig:mbh_mstar_mbulge_fits}. Relations were fitted using \textsc{linmix}, and are shown. 

Consider first the \mbh-\mstar relation, shown on panel (a) of \autoref{fig:mbh_mstar_mbulge_fits}. For \bulgeless, the fitted relation has $\alpha=-4.6^{+0.8}_{-0.8}$ and $\beta=1.08^{+0.07}_{-0.07}$, with a scatter of $\sigma\in[0.20, 0.29]$. For \somebulge, $\alpha=-7.0^{+1.6}_{-1.6}$ and $\beta=1.31^{+0.15}_{-0.15}$, with a scatter of $\sigma\in[0.29, 0.40]$. The slope of the \somebulge relation is slightly steeper, with a difference of $\Delta\beta=0.23$ ($\sim1.4\sigma$). Therefore we find the \mbh-\mstar relations of the \somebulge and \bulgeless populations are consistent within $3\sigma$. We note the increased scatter for \somebulge: this is likely due to the smaller sample statistics in the line fitting.

Consider now the \mbh--\mbulge relation, shown in panel (b) of \autoref{fig:mbh_mstar_mbulge_fits}. For \bulgeless, treating the upper limits on \logmbulge as measurements, we find $\alpha=5.8^{+0.2}_{-0.2}$ and $\beta=0.12^{+0.03}_{-0.03}$, with a scatter of $\sigma\in[0.41, 0.47]$. For \somebulge, $\alpha=-2.7^{+1.2}_{-1.2}$ and $\beta=0.98^{+0.12}_{-0.12}$, with a scatter of $\sigma\in[0.29, 0.40]$. The difference in slopes of the \mbh--\mbulge relations of the \bulgeless and \somebulge populations is then $\Delta\beta=0.86$, or $\sim7.0\sigma$. Thus, we find the \mbh--\mbulge relations of the \bulgeless and \somebulge populations are statistically distinct.

Excluding the PS-limited upper limits from the \bulgeless sample yields a steeper slope ($\beta=0.39^{+0.04}_{-0.04}$ with $\alpha=3.3^{+0.4}_{-0.4}$) with increased scatter ($\sigma\in[0.29, 0.35]$). The difference in slope compared to \somebulge is then reduced to $\Delta\beta=0.59$ or $4.7\sigma$. Therefore, the relations remain statistically distinct regardless of the inclusion or exclusion of these sources, and this does not change the conclusions of this paper.

The \mbh-\mbulge relation of the \bulgeless sample shows a clear deviation from the early-type relation of \citet{Haring2004}, with BHs that are over-massive relative to their bulge mass at low bulge masses, compared to the predictions of pre-established \mbh--\mbulge relations. This is in agreement with \citetalias{SSL17}, who find 60\% of their sample of 101 disk-dominated BL-AGN hosts in SDSS lie above the \citet{Haring2004} relation. We find $\sim87\%$ of the \bulgeless galaxies lie above the \citet{Haring2004} relation ($\sim 82\%$ excluding PS-limited bulge mass limits), compared to $\sim46\%$ of the \somebulge galaxies. Given the assumed quiet merger history of the \bulgeless galaxies, this suggests secular processes are capable of growing SMBHs to masses consistent with those in disk galaxies with mixed merger histories. This is also reflected in the similarity in BH mass distribution noted in Section \ref{subsec:res:compare}.

\section{Discussion} \label{sec:discussion} 

\subsection{BH--stellar mass relation in disk galaxies} \label{sec:disc:mbh_mstar}

Three pre-existing \mbh--\mstar correlations are plotted alongside the \disk sample in \autoref{fig:mbh_mstar}: the \citet{Haring2004} relation, derived as \mbh--\mbulge for early type galaxies, and the \citet{Greene2020} early-type and late-type relations. While the median on the slope of our relation is closest to \citet{Haring2004} ($\beta=1.2\pm0.06$), it is also consistent with \citet{Greene2020} late-type ($\beta=1.61\pm0.24$) and early-type ($\beta=1.33\pm0.12$) relations within $3\sigma$. This is due to the larger reported uncertainties on the \citet{Greene2020} relations, which may arise from the small samples (37 late-type and 83 early-type).

\citet{Pucha2026} report a \mbh--\mstar relation for a sample of 17,949 BL-AGN in DESI DR1, with virial mass estimates derived from the \textsc{EmFit} emission line information. The reported relation has $\alpha=-3.6\pm0.11$ and $\beta=1.02\pm0.01$ (in the form of Equation \ref{eq:mbh_mstar}). This is in very close agreement with our relation in both slope ($\Delta\beta = 0.02$ being within $1\sigma$) and normalisation ($\Delta\alpha = 0.1$ being within $1\sigma$). \citet{Pucha2025} identify AGN using emission line ratio BPT diagrams. The agreement between the \mbh--\mstar relations of this sample and \disk therefore suggests that AGN populations selected using our BL-AGN selection method (Section \ref{subsec:blagn}) and using BPT are broadly in agreement in the \mbh--\mstar plane.

A trail of points can be seen in \autoref{fig:mbh_mstar}, lying above the general population at higher \mbh, at $\log(M_{*})\lesssim 10.5$. Visual inspection of their photometry and spectra suggests the \logmbh or \logmstar values of these objects are reliable. Approx. 1/3 of these objects are in our \bulgeless population (see panel (a) of \autoref{fig:mbh_mstar_mbulge_fits}). We speculate that this trail could result from a period of significant growth via secular mechanism(s), during which the BH grows without a significant concurrent increase in stellar mass. During such periods, secular gas inflow mechanisms, such as bar or spiral arm in-fuelling (as discussed below in Section \ref{subsubsec:disc:bars_and_spirals}), may continue to provide fuel for the central SMBH while removing gas from the reservoir for star formation.

We note that the trail emerges from the general population at $\log(M_{*})\sim 10^{10.5}$. Stellar masses around this value have been noted as significant transitionary masses \citep[\mstar$\sim2-3\times10^{10}$;][]{Kauffmann2003, Baldry2004} in a large number of properties, including galaxy morphology \citep[e.g.][]{Kauffmann2003, Moffett2016}, gas fraction \citep[e.g.][]{Kannappan2004}, star formation rate \citep[e.g.][]{Brinchmann2004, Kauffmann2004, Salim2007}, metallicity \citep[e.g.][]{Tremonti2004}, galaxy environment \citep[e.g.][]{Baldry2004, Keres2005, Dekel2006}, AGN activity \citep[e.g.][]{Heckman2004, Kewley2006} and feedback \citep[e.g.][]{Baldry2004}. Therefore we speculate that the emergence of the trailing data points may be related to any number of these galaxy and/or AGN properties. $\sim32\%$ of the objects in this trail are also in the \bulgeless sample. Regardless of the mechanism responsible for the growth of these over-massive BHs, this suggests that the secular processes operating in these \bulgeless systems are capable of growing BHs to such large masses.

\subsection{The importance of secular BH growth} \label{subsec:disc:secular-growth}

\begin{figure}
    \centering
    \includegraphics[width=\linewidth]{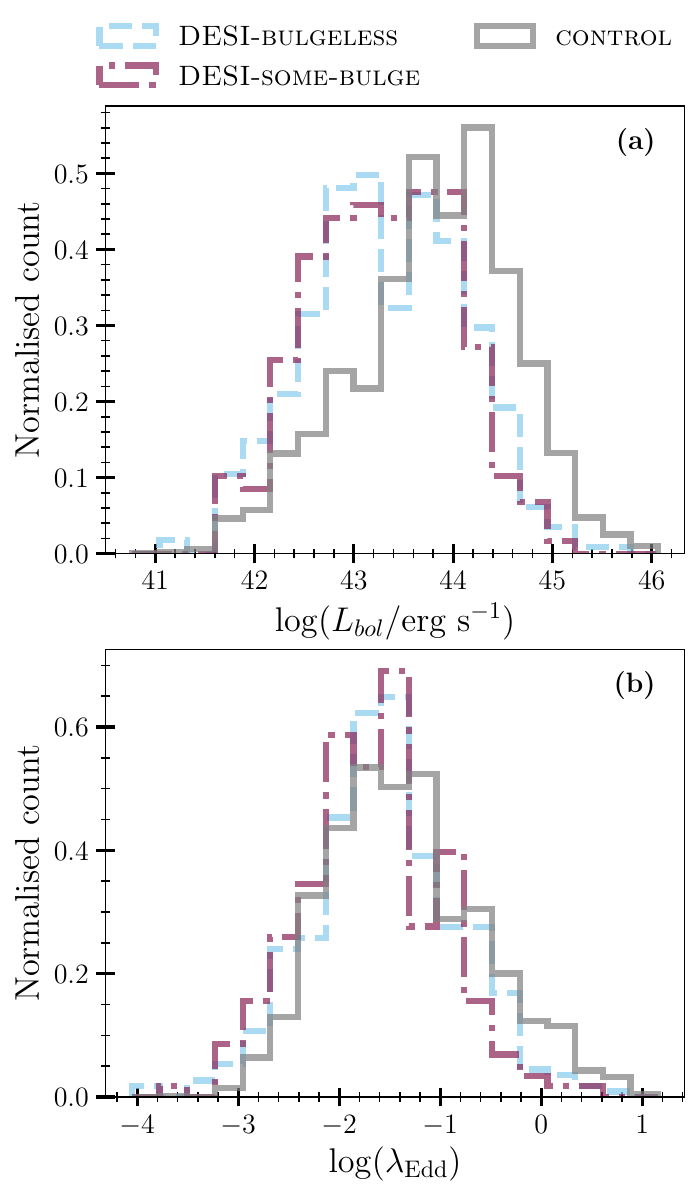}
    \caption{Distributions of AGN bolometric luminosity (top) and Eddington ratio (bottom). The \bulgeless sample is plotted with a blue dashed line ($--$). The \somebulge sample is plotted with a burgundy dash-dotted line ($\cdot-\cdot$) line. The weighted \control sample is plotted as a grey solid line (--). }
    \label{fig:Lbol_edd}
\end{figure}

In Section \ref{subsec:b_v_sb} we noted that the \mbh--\mstar relation for the \bulgeless and \somebulge populations are statistically consistent within $\sim3\sigma$ (see \autoref{fig:mbh_mstar_mbulge_fits}), while their \mbh--\mbulge relations are statistically distinct at $\gtrsim 7\sigma$. This is contrary to what we expect in a merger-driven co-evolution scenario, where the \mbh--\mbulge relation should reflect the co-evolutionary relationship more faithfully than the \mbh--\mstar relation. While stellar bulges can grow through both merger and non-merger processes \citep[e.g.][]{Athanassoula2005, Elmegreen2008, Hopkins2010}, the stellar disk is formed predominantly through secular processes \citep[e.g.][]{White1978, Kormendy2004, vanDokkum2013}. If the BH can only grow through merger-driven processes, we would expect agreement between \bulgeless and \somebulge in their \mbh--\mbulge correlations. Instead we find the \mbh-\mbulge correlations of the \bulgeless and \somebulge samples diverge. This suggests BHs can grow through galaxy-merger-free processes. This is reflected in the agreement between the \mbh-\mstar correlations of the two samples, which we hypothesise is due to including mass from the predominantly secularly grown disk which weakens the contribution of merger-driven effects. As consideration of this secularly grown mass is required to find agreement between the two populations, this suggests that secular processes play a significant role in establishing galaxy-SMBH co-evolution, not only in the \bulgeless population, but in the overall \disk sample.

This result is reminiscent of \citetalias{SSL17} who found a well-defined \mbh--\mstar relation for their sample of SDSS bulgeless galaxies, while upper limits on bulge mass were consistent with there being no correlation between BH mass and bulge mass for the sample. \citetalias{SSL17} interpret this as indicating the significance of secular growth in galaxy-SMBH co-evolution. Our results support this hypothesis, and the factor of $\gtrsim5$ increase in sample size (546 \bulgeless, compared to 101 in \citetalias{SSL17}) strengthens this interpretation.

Bolometric AGN luminosity ($L_{\rm bol}$) provides an insight into the current BH accretion rate. The Eddington ratio, $\lambda_{\rm Edd} \equiv L_{\rm bol}/L_{\rm Edd}$, is then a useful indicator of the observed BH accretion rate relative to the theoretical maximum for a BH of this mass, $L_{\rm Edd}$. The \citet{Siudek2024} catalogue of SED fitting includes the bolometric AGN luminosity outputted by \textsc{CIGALE}. We select only those objects of our sample for which $\rm SNR\geq3$ in all four WISE bands ($W1$, $W2$, $W3$ and $W4$), to ensure sufficient IR information is available for reliable $L_{\rm bol}$ to be determined in the SED fitting process. This gives 1,674 \disk galaxies with robust $L_{\rm bol}$ measurements. We then calculate $L_{\rm Edd}$ from \mbh, and derive $\lambda_{\rm Edd}$. The distributions of $L_{\rm bol}$ and $\lambda_{\rm Edd}$ for the \bulgeless galaxies and \somebulge galaxies, are compared to the \control in panels \textbf{(a)} and \textbf{(b)} of \autoref{fig:Lbol_edd}, respectively. We find a median average $\lambda_{\rm Edd}$ for the \disk (\bulgeless; \somebulge) sample of $0.034$ ($0.027$; $0.021$). Similarly, we find 1,892 galaxies from the \control sample have reliable $L_{\rm bol}$, with a weighted median $\lambda_{\rm Edd}$ of $0.039$. 

\citetalias{SSL17} estimate $L_{\rm bol}$ for their sample of AGN in disk-dominated galaxies by applying a bolometric correction to the WISE $W3$-band luminosity, and thus $\lambda_{\rm Edd}$. They find a large range of $\lambda_{\rm Edd}$ values represented in their sample, with an average $\langle \lambda_{\rm Edd} \rangle=0.15$, significantly larger than the average of our \bulgeless sample. Furthermore, $\sim68\%$ of the \bulgeless sample are accreting at $\lambda_{\rm Edd}<0.05$, whereas only $\sim5\%$ of the \citetalias{SSL17} sample had such low Eddington ratios. This is likely due to the different AGN selection methods used here and in \citetalias{SSL17}; AGN were identified using X-ray detections in \citetalias{SSL17} in order to probe the maximum accretion rates possible through secular processes. This naturally resulted in higher average $L_{\rm bol}$ values across their sample.

The \control sample shows a trend towards high $L_{\rm bol}$, with a peak around $43.97 \leq \log(L_{\rm bol}/\text{erg s}^{-1}) \leq 44.53$. The weighted median $\log(L_{\rm bol})$ of the \control is $\sim0.5$ dex, which is $\sim0.15$ dex higher than the median of the \somebulge and \bulgeless, respectively. We apply a KS test \citep{Massey1951} adapted for weighted samples,\footnote{The KS test is applied to weighted samples by using weights rather counts to calculate cumulative fraction. Weighted sample sizes, rounded to an integer, are used to estimate the $p$ value. For this, we normalise the weights such that the largest weight is 1.0. To compare a weighted and an unweighted sample, we use equal weights of 1.0 for the unweighted sample.} to compare the weighted \control to \bulgeless, finding a KS statistic $0.20$ and $p=0.0003$. This corresponds to $\sim3.6\sigma$ significance, providing strong evidence of a difference in the distributions. Furthermore, the significance is likely underestimated by this weighted KS test, as noted by \citet{Hutchinson-Smith2026}. A higher average $L_{\rm bol}$ suggests that, in general, the AGN in the \control have higher \textit{current} accretion rates and/or are more radiatively efficient than the AGN in the \bulgeless or \somebulge sample. It has been observed that galaxy mergers lead to periods of increased BH growth \citep[e.g.][]{Ellison2011, Ellison2019, Satyapal2014}. Given the early-type morphology of the \control likely corresponds to an active merger history, we expect a tendency towards high $L_{\rm bol}$.

The \logmbh distributions of the \bulgeless and \control samples are compared in panel (a) of \autoref{fig:mbh_mstar_mbulge_bt_hists}. A weighted KS test between them gives statistic $0.14$ and $p=0.006$. This corresponds to $\sim2.7\sigma$ significance, suggesting moderate evidence to reject the null hypothesis that the \control and \bulgeless samples share an underlying \logmbh distribution, but not enough to reject this hypothesis at $3\sigma$. The range of BH masses is also between the two, with $5.13 \leq \log(M_{\rm BH}) \leq 8.58$ for \bulgeless and $5.49 \leq \log(M_{\rm BH} \leq 8.84)$ for \control. The higher average $L_{\rm bol}$ combined with the weaker statistical evidence for a difference in \mbh in \control compared to the \bulgeless, may suggest that, while the AGN in the \control are currently more luminous, this does not translate to significantly higher \mbh. This would suggest that, at a given redshift and for a given \mstar, early-type galaxies generally host BL-AGN that are more radiatively efficient and/or accreting at higher rates than those hosted in disk galaxy counterparts, but these AGN have not necessarily experienced more significant long-term BH growth. This result is tentative, however it is in agreement with \citet{McAlpine2020}, who found, in the EAGLE simulation, galaxy mergers lead to more luminous AGN but not significantly more BH growth.

We compare the $L_{\rm bol}$, \logmbh and $\lambda_{\rm Edd}$ distributions of the \bulgeless and \somebulge using AD tests. In $L_{\rm bol}$, we cannot reject the null hypothesis that the two populations share the same underlying distribution (statistic $0.52$ at $0.45\sigma$ significance). For $\lambda_{\rm Edd}$, there is modest evidence against the null hypothesis (statistic $1.90$ at $1.9\sigma$ significance) but not enough to reject the null at $3\sigma$. Similarly, for \logmbh there is modest evidence against the null (statistic $1.68$ at $1.8\sigma$ significance) but not enough to reject it.

\begin{figure}
    \centering
    \includegraphics[width=\linewidth]{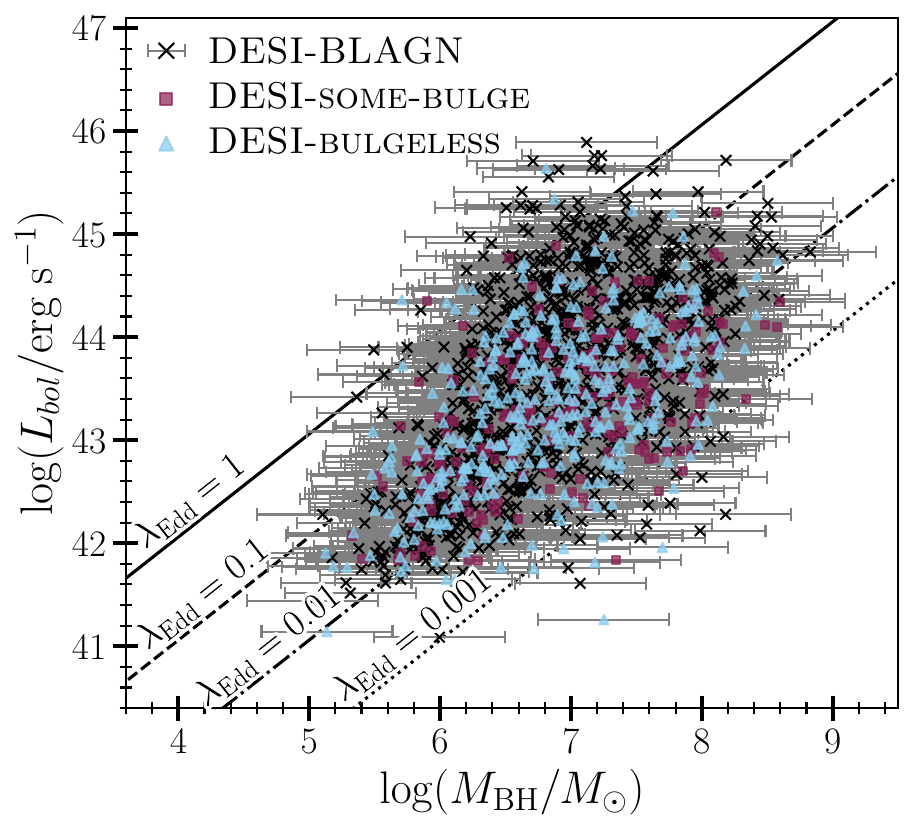}
    \caption{AGN bolometric luminosity against black hole mass for \bulgeless (blue triangles) and \somebulge (burgundy squares) galaxies. The \disk without bulge classifications are also plotted as black 'x's. The error bars are shown in grey. Lines of constant Eddington ratio are also shown: $\lambda_{\rm Edd}=1,\, 0.1,\, 0.01$ and $0.001$ as solid, dashed, dot-dashed and dotted lines, respectively.}
    \label{fig:Lbol_MBH}
\end{figure}

\autoref{fig:Lbol_MBH} shows $\log(L_{\rm bol})$ against \logmbh, with lines of constant $\lambda_{\rm Edd}$ for reference. Of the \disk sample, $51\%$ have $0.01<\lambda_{\rm Edd}<0.1$ and $24\%$ have $0.1 < \lambda_{\rm Edd} < 1$, meaning the vast majority of the sample have $0.01<\lambda_{\rm Edd}<1$. A significant fraction ($22\%$) have very low Eddington fractions at $\lambda_{\rm Edd}<0.01$. Only $3\%$ are super-Eddington. For \bulgeless and \control, these (weighted) fractions vary by $\lesssim2\%$. The \bulgeless and \control galaxies therefore behave similarly in terms of $\lambda_{\rm Edd}$, indicating that AGN in bulgeless galaxies have similar accretion physics to those in early-type galaxies.

AGN luminosity is related to mass accretion rate, $\dot M$, by an accretion efficiency parameter, $\epsilon$, such that $L_{\rm bol} = \epsilon \dot M c^2$. \citet{Elvis2002} find BH mass accretion must be highly efficient, with $\eta \geq 0.15$. Taking $\eta=0.15$, we find the $L_{\rm bol}$ of \disk correspond to $1.44\times10^{-5} \leq \dot M \leq 0.91 \text{ M}_\odot\text{ yr}^{-1}$. We can adopt a simple accretion history model to obtain some measure of the growth time of the BH using these growth rates. We assume the BH formed with seed mass of $10^3 M_\odot$ \citep[considered an intermediate seed mass; see review by][]{Regan2024}, experienced Eddington-limited accretion until it reached the observed $L_{\rm bol}$, then grew with a constant accretion rate (decreasing $\lambda_{\rm Edd}$ until the time of ). Under this simplistic model, we find $\sim85\%$ of the \disk AGN are able to reach their observed \mbh within the age of the Universe at the galaxy's measured redshift. This only decreases to 81\% for \bulgeless galaxies alone. Therefore, if we assume bulgeless galaxies have experienced very little-to-no major merger activity, then major mergers are not required to grow the vast majority of the BHs to their observed masses within a reasonable time, even in this simplified model of Eddington-limited growth.\footnote{This model is fully Eddington-limited, except for those AGN with observed super-Eddington luminosities.}.

This raises the question of what non-merger mechanisms could be driving co-evolution in these systems. We consider some possibilities in Section \ref{subsec:disc:secular-mechanisms} below.

\subsection{Secular mechanisms in galaxy-SMBH co-evolution} \label{subsec:disc:secular-mechanisms}

\subsubsection{Bars and spiral arms}  \label{subsubsec:disc:bars_and_spirals}

Stellar bars play a significant role in the secular evolution of galaxies (observations: \citealp{Cheung2013, Kruk2018, Gadotti2020, Geron2021, deSa-Freitas2025}; simulations: \citealp{Kraljic2012, Kubryk2013, Athanassoula2013, Rosas-Guevara2022}). Simulations find bars are capable of driving gas down to the central $\sim100$ pc region of the galaxy \citep[e.g.][]{Athanassoula1992, Athanassoula2000, Kormendy2004}, and of sustaining inflow rates of order $\sim0.1-1 \text{ M}_\odot\text{ yr}^{-1}$ \citep[e.g.][]{Athanassoula1992, Maciejewski2002, Regan2004, Lin2013}. Bars are, therefore, a merger-free means of transporting gas to the centre of disk galaxies. While highly debated \citep[e.g.][]{Cisternas2013, Cisternas2015, Goulding2017}, there is some evidence of a connection between the presence of a bar and the presence of an AGN in disk galaxies \citep[e.g.][]{Oh2012, Galloway2015, Alonso2018, Silva-Lima2022, Garland2023, Garland2024, LaMarca2025}. 

Using the GZ DESI classifications, we can find the fraction of the sample which are barred using the response fractions to the question "Is there a sign of a bar feature through the centre of the galaxy?". Following previous papers using Galaxy Zoo classifications \citep{Geron2021, Geron2023, Garland2024}, we classify a galaxy as barred if $f_\text{weak bar}+f_\text{strong bar} \geq 0.5$. Under this classification $66\%$ of the \disk sample with reliable GZ bar classifications are barred, and $65\%$ and $68\%$ of the \bulgeless and \somebulge samples respectively. This is somewhat higher than other reported fractions based on optical observations, which find bar fractions of $\sim43-62$ \citep[e.g.][]{Erwin2018, Buta2019, Geron2021}. It is closer to values reported from NIR observations, where bar fractions around $\sim60\%$ are reported \citep[e.g.][]{Eskridge2000, Menendez-Delmestre2007, Marinova2007, Diaz-Garcia2016}. Considering only strong bars (where $f_\text{strong bar}>f_\text{weak bar}$), $27\%$ of the \disk have strong bars ($28\%$ for \bulgeless and $32\%$ for \somebulge). This is in broad agreement with literature values of the strong bar fraction, at $\sim15-29\%$ \citep[e.g.][]{Masters2011, Buta2019, Geron2021}. We note that the prevalence of bars in our sample could bias the surface brightness models, with the \textsc{galfitm} fitting being more likely to fail for barred galaxies.

Comparisons of bar fractions between samples requires careful consideration of galaxy properties such as stellar mass and redshift \citep[see e.g.][]{Erwin2018}. A thorough investigation of the bar fraction is beyond the scope of this paper. However, as stellar bars are prevalent in our sample, if they do indeed contribute to AGN fuelling, we may expect this to be a common fuelling mechanism in our sample. We also note the similar prevalence of bars in \bulgeless and \somebulge ($65\%$ and $68\%$ respectively). This contradicts previous studies such as \citet{Barazza2008} who observe an increase in bar fraction in disk-dominated systems, but also studies such as \citet{Masters2011}, who find the bar fraction increases with the prominence of the stellar bulge. These studies did not take into consideration the presence of AGN, suggesting the presence of an optical BL-AGN may affect the reported trend between bar fraction and bulge prominence, or this trend is different for AGN host galaxies.

There is also evidence that spiral arms can drive mass toward the centre of galaxies by driving angular momentum outwards \citep{Hopkins2010a, Kim2014}. In hydrodynamical simulations, gas inflows along spiral arms at rates of $\sim 0.03-3 \text{ M}_\odot \text{ yr}^{-1}$ \citep[e.g.][]{Maciejewski2004, Kim2014}. In integral field unit (IFU) observations of local spiral galaxies, such gas inflows are detected along spiral arms with rates of $\sim 0.06-1.2 \text{ M}_\odot \text{ yr}^{-1}$ \citep[e.g.][]{Davies2009, Schnorr-Muller2014, Slater2019}. \citet{Davies2009} note that inter-arm outflow, arising from the spiral shock itself, can effect the \textit{net} inflow rate to the very central region of the galaxy. Using hydrodynamical simulations in conjunction with their observations, they estimate the observed $1.2 \text{ M}_\odot \text{ yr}^{-1}$ inflow translates to a net inflow rate of $0.06 \text{ M}_\odot \text{ yr}^{-1}$ to the inner 10 pc. These reported inflow rates are sufficient to sustain the range of mass accretion rates in the \disk AGN, estimated in Section \ref{subsec:disc:secular-growth}, and even the lower limit of $0.03 \text{ M}_\odot\text{ yr}^{-1}$ from simulations is sufficient for 83\% of the \disk sample.   

Again, using GZ DESI classifications, we can find the fraction of our sample with identifiable spiral arms using response fractions to the question "Is there any sign of a spiral arm pattern?". We define a galaxy as having spiral arms if $f_\text{has-spiral-arms-yes} \geq 0.6$, using the threshold suggested from GZ DECaLS \citep{Walmsley2022}. We consider only those galaxies with reliable vote fractions, as described in Section \ref{subsec:GZ}. Under this classification $87\%$ of the \disk sample have spiral arms, and $85\%$ and $76\%$ of the \bulgeless and \somebulge samples respectively. Therefore, gas inflows driven by spiral arms could be a means of transporting gas to the central region in our sample, given their ubiquity across the sample. 

The commonality of stellar bars across the \bulgeless and \somebulge populations may explain the similarity between the two populations in both the \mbh--\mstar relation (see \autoref{fig:mbh_mstar_mbulge_fits}) and in $L_{\rm bol}$ and $\lambda_{\rm Edd}$ (see \autoref{fig:Lbol_edd} and \ref{fig:Lbol_MBH}). One would expect this to be the case where mechanisms of BH growth, which relate wider galaxy properties to the central BH, are shared between the samples. 

\subsubsection{AGN Feedback} \label{subsubsec:disc:AGN_feedback}

AGN feedback is a strong candidate for regulating co-evolution, as it can link properties of a galaxy to the accretion physics of its central SMBH. This is demonstrated in cosmological simulations, as simulations which exclude AGN feedback over-produce massive, star forming galaxies compared to observations, and fail to sufficiently suppress the cooling of gas in massive halos \citep[e.g.][]{McCarthy2010, Dubois2013, Dubois2016, Kaviraj2017, Su2019}.

AGN showing kinetic feedback are observed to typically accrete at $\lambda_{\rm Edd}$ of order 1\% or less, while AGN with radiative outflows typically accrete at higher $\lambda_{\rm Edd}$ \citep{Best2012, Heckman2014}. This observation is consistent with with the theory of advection-dominated accretion flows \citep{Yuan2014}. This is a generalised order-of-magnitude estimate of the transition in energy release mechanism, rather than a precise threshold. As noted in Section \ref{subsec:disc:secular-growth}, $22\%$ of the \disk AGN have $\lambda_{\rm Edd}<0.01$, meaning the remaining $78\%$ have $\lambda_{\rm Edd}>0.01$, and these fractions vary by $\lesssim3\%$ for \bulgeless and \control AGN. Therefore, the majority of our sample is energetically consistent with radiative energy injection, but there is also a considerable fraction consistent with kinetic energy injection. We therefore expect both forms of energy release from the AGN to be operating across the \disk sample, in \bulgeless and \control alike. The \textsc{IllustrisTNG} simulation, \citet{Weinberger2018} find kinetic AGN feedback dominates at late times in galaxies with \mstar$\gtrsim 10^{10.5}$ at $z\sim0$, maintaining a state of inefficient star formation. We suggest that the emergence of the trail of points in the \disk sample discussed in Section \ref{sec:results} (see \autoref{fig:mbh_mstar}) could be caused by kinetic AGN feedback leading to inefficient star formation in these systems, allowing the BH to out-grow the galaxy relative to the co-evolution of the general population. 

Using observations from the Keck Cosmic Web Imager IFU, \citet{Smethurst2021} find [\emissionline{O}{III}] outflows in four bulgeless galaxies from the \citetalias{SSL17} sample at $0.043 < z < 0.073$. In all four, the outflow velocities exceed the host galaxy escape velocities by a factor of $\sim30$, demonstrating that AGN in bulgeless galaxies are capable of large-scale ejective feedback. These galaxies have $44.2 \lesssim \log(L_{\rm bol}/{\text{erg s}^{-1}}) \lesssim 44.8$, whereas the \bulgeless sample has $41.1 \leq \log(L_{\rm bol}/{\text{erg s}^{-1}}) \leq 45.9$, suggesting the \citet{Smethurst2021} AGN are in the general population of bulgeless galaxies. The \bulgeless sample presented in this paper provides a sample from which candidates for further studies of spatially-revolved gas kinematics in merger-free AGN-host galaxies can be identified, contributing towards a population-level study of AGN feedback in bulgeless galaxies.

\subsection{Comparison with simulations} \label{subsec:sims}

\begin{figure*}
    \centering
    \includegraphics[width=\linewidth]{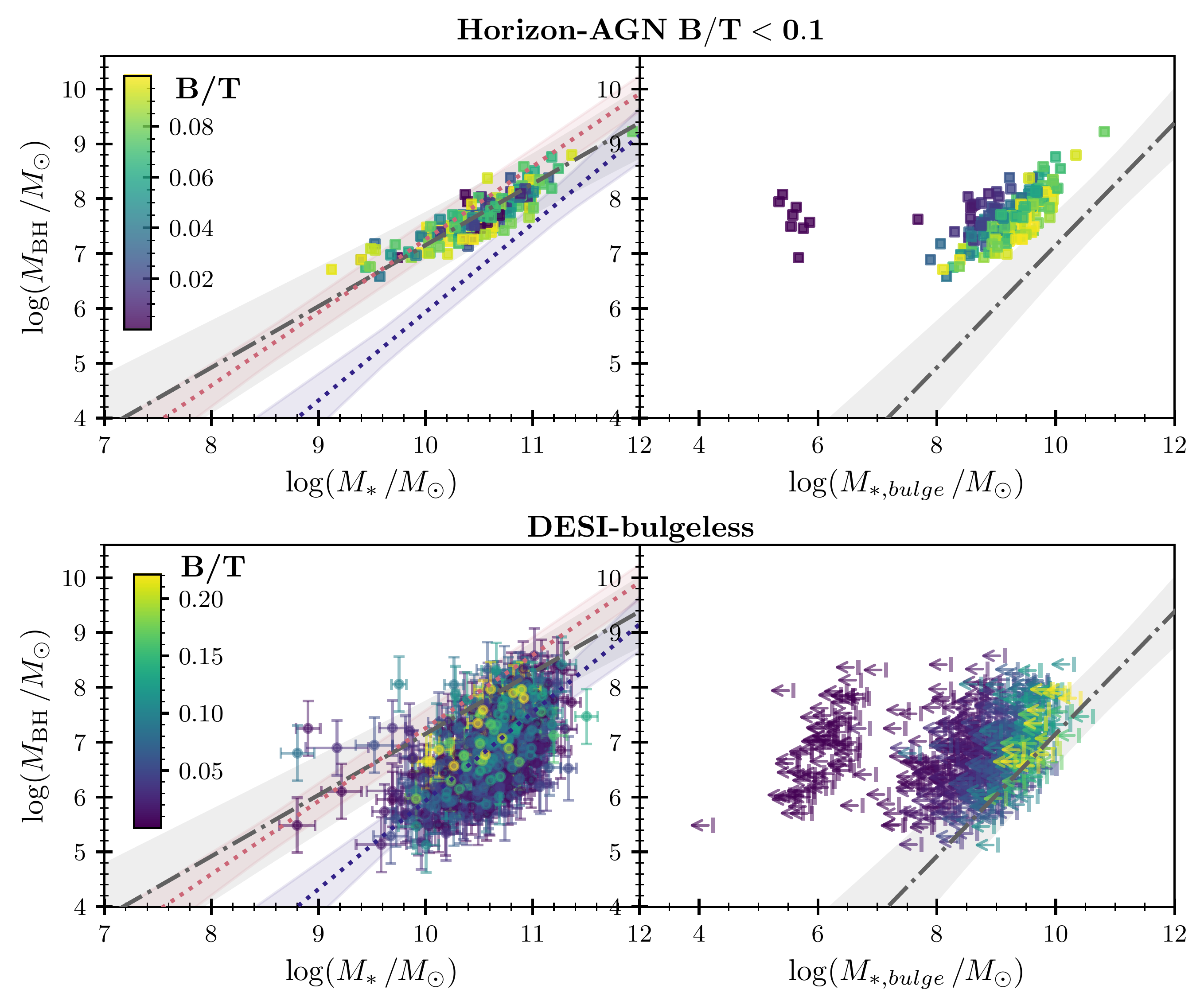}
    \caption{Comparison of the \bulgeless sample with \citep{Smethurst2024} Horizon-AGN simulation results (\textbf{top row}) with \bulgeless galaxies (\textbf{bottom row}), on \mbh--\mstar (left) and \mbh--\mbulge (right). Both samples are coloured by bulge-to-total mass ratio ($\rm B/T$), with a colour bar shown for each row. The $B/T$ scales are different for each sample, as apparent from the colour bars.}
    \label{fig:horizon}
\end{figure*}

We have already mentioned the importance of simulations in understanding how galaxy-SMBH co-evolution occurs, and the influence of mergers on this evolution. In this section we compare our results to those from simulation studies, and discuss the potential implications.

\citet{Smethurst2024} explored merger-free SMBH growth and its role in galaxy-SMBH co-evolution in the \textsc{Horizon-AGN} hydrodynamical simulation \citep{Dubois2014}. They identify a simulated sample of 6,892 galaxies at $z=0.0556$, from which they select two sub-samples: merger-free (having experienced no major of minor merger since $z=2$) and merger-dominated (having experienced $>3$ mergers since $z=2$). Furthermore, they use measurements of bulge mass for the \textsc{Horizon-AGN} galaxies from \citet{Volonteri2016} to calculate $B/T$ to select a simulated bulgeless galaxy sample. The \citet{Volonteri2016} bulge mass values were measured from fitted double-S\'ersic surface brightness profile models, with one S\'ersic having $n=1$ for the disk and the other having $n=1,2,3\text{ or }4$ for the bulge. 

In \autoref{fig:horizon}, we compare the \bulgeless sample to the \citet{Smethurst2024} results on \mbh--\mstar and \mbh--\mbulge. The top row shows the \citet{Smethurst2024} \textsc{Horizon-AGN} galaxies selected as having $B/T<0.1$, while the bottom row shows the \bulgeless sample. Both samples are coloured by $B/T$. In both \mbh--\mstar and \mbh--\mbulge plots, there is a notable lack of agreement between the \citet{Smethurst2024} \textsc{Horizon-AGN} galaxies and the \bulgeless sample. It is clear that the \citet{Smethurst2024} \textsc{Horizon-AGN} galaxies show a lack of low-mass BHs compared to the \bulgeless sample.

There are differences in the composition and properties of the two samples which should be considered when comparing them. The \citet{Smethurst2024} galaxies are taken from a snapshot of \textsc{Horizon-AGN} at $z=0.0556$. Our \bulgeless galaxies have a range of redshifts ($0.0172 \leq z \leq 0.2504$) with median $z=0.0988$, $\sim 558\text{ Myr}$ before the time of the \citet{Smethurst2024} galaxies. However, we would not expect the BH growth which could occur in this time period to account for the lack of low-mass BHs in the sample. The parent \textsc{Horizon-AGN} galaxy sample has a minimum stellar mass of $10^{8.5}\,{\rm M_\odot}$ and should be complete above this mass, in that no additional mass or luminosity cuts are made. Furthermore, reported stellar masses are intrinsic. By comparison, the \disk sample has a minimum stellar mass of $\sim10^{8.2}\,{\rm M_\odot}$, but is not expected to be complete above this. Luminosity cuts were made in the production of the \textsc{GZ DESI} catalogue \citep{Walmsley2023}, which will effect this completeness, along with observational biases such as a bias against low-surface-brightness galaxies \citep[e.g.][]{Disney1983}.

In comparing to the bulgeless \textsc{Horizon-AGN} sample, there is a difference in the ``bulgeless'' criteria. As stated above, these were selected by \citet{Smethurst2024} as having $B/T<0.1$, measured from double-S\'ersic profile fitting on the simulated galaxy images. This selection allows for some small bulge component to be present in the model, whereas our selection excludes models containing any bulge component (see Section \ref{subsec:galfit:bls_sb}). The resulting difference in the scale of $B/T$ represented is apparent in the difference in colour scales in \autoref{fig:horizon}. For the selection based on $B/T$, more massive bulges in more massive galaxies will meet this criteria. Therefore we may expect the difference in selection to have a more significant effect at higher stellar masses. The \citet{Volonteri2016} bulge-disk decompositions are limited by the spatial resolution of \textsc{Horizon-AGN}, $ 1\text{ kpc}$, meaning bulges in physically small and/or low-mass galaxies are likely missed. In  qualitative comparison with observed bulge-disk decompositions, \citet{Volonteri2016} find the simulation bulge-disk decompositions tend to overestimate $B/T$ for $10<\text{\mstar}<11$, meaning some bulgeless galaxies in this mass range have likely been missed in the $B/T<0.1$ selection. Furthermore, the \citet{Volonteri2016} bulge components were fitted using 3D averaged density profiles, compared to the 2D density profiles used in our bulge-disk decomposition on observational images (Section \ref{subsec:galfitm}). As such the do not suffer from observational complications such as inclination effects and signal-to-noise ratio, which will affect our science sample.

Nevertheless, the comparison of these samples is intriguing. The lack of agreement between the sample is perhaps unsurprising, as the sub-grid physics in \textsc{Horizon-AGN} is calibrated to reproduce the local \citet{Haring2004} \mbh--\mstar and \citet{Tremaine2002} \mbh-$\sigma$ relations \citep[see][]{Dubois2012, Volonteri2016}. A number of parameters pertaining to the AGN feedback are chosen to ensure reasonable agreement with these local relations. As both relations are derived from samples of massive elliptical or bulge-dominated galaxies, we expect to find more massive BHs in such galaxies. This may contribute, therefore, to the lack of low black hole masses seen in comparison to the observed \bulgeless sample. This has been noted previously by \citet{Habouzit2021}. We note that \citet{Smethurst2024} find that the majority of SMBHs in their \textsc{Horizon-AGN} simulated sample grow in the absence of galaxy merger activity. However, the bias towards high \mbh, potentially introduced from the calibration of the sub-grid physics, remains despite this.

In \textsc{Horizon-AGN}, BHs are seeded with mass $10^5\text{ M}_\odot$ therefore the simulation cannot produce BHs of masses lower than this by design \citep{Dubois2014}. This may also explain the lack of scatter compared to observations: the first $10^5\text{ M}_\odot$ of BH mass is not accreted but placed, meaning this early BH evolution is not accounted for and any scatter arising from this early growth is not captured. Furthermore, the BH mass accretion rate in \textsc{Horizon-AGN} is mass-dependent. A lack of initial scatter therefore leads to populations remaining more homogenous over time.

It is also important to consider that \textsc{Horizon-AGN} has a resolution of $1\text{ kpc}$, and all gas properties in the simulation are smoothed over this scale. This effect may allow BHs in the simulation to accrete gas from unphysically large distances, especially low-mass BHs as, unlike in the physical universe, the gas accretion radius is independent of BH mass. Furthermore, the properties of the interstellar medium (ISM) are also smoothed across this resolution scale, meaning any stochasticity in the feeding of the BH cannot be captured in such low-resolution simulations. This issue remains irrespective of sub-grid physics calibrations such as the choice of feedback parameter. This is, therefore, a more general issue across many cosmological simulations, which may explain the lack of scatter compared to observations.

This is not an issue unique to \textsc{Horizon-AGN}. A lack of low-to-intermediate BHs is a common trait across multiple cosmological hydrodynamical simulations. \cite{Habouzit2021} investigate the impact of sub-grid physics on the \mbh-\mstar relation in the Illustris, TNG100, TNG300, \textsc{Horizon-AGN}, EAGLE, and SIMBA simulations, and find all simulations struggle to form BHs with \logmbh$\lesssim7.5$ in galaxies with $10.5 \lesssim$\logmstar$\lesssim 11.5$, and show an excess of BHs with \logmbh$\gtrsim9$, compared to observations of the local Universe. 

Difficulty with producing physically representative populations of disk galaxies is also a common issue for many simulations \citep[e.g.][]{Rodriguez-Gomez2019, Haslbauer2022}. The $1\text{ kpc}$ resolution of the \textsc{Horizon-AGN} simulation also has implications for the measured galaxy properties. The ability to model sub-structures such as stellar bulges, separately from broader stellar components such as the disk, is limited by the resolution. This is also dependent on the size of the galaxy, as smaller galaxies contain fewer resolution elements meaning structures such as bulges are difficult to resolve in these galaxies.

The discrepancy between the \bulgeless sample and the \textsc{Horizon-AGN} galaxies seen in \autoref{fig:horizon} highlights the significance of this issue, indicating a large population of SMBHs are severely under-represented in the simulation. This may reflect the inability of low-resolution simulations to capture stochastic BH feeding. It may also suggest that the feedback parameters, such as feedback efficiency, or the size of the radio jet / energy bubble in radio/quasar feedback modes, chosen to calibrate the sub-grid physics of the simulation against local relations, may not be reflective of the underlying physics operating within these systems. Future simulations calibrated against different observed correlations, explicitly derived for disk galaxies or including mixed morphology hosts, may help to shed light on this issue. This further highlights the importance of obtaining robust BH masses for local low-mass black holes and for a wide variety of host galaxy morphologies, that can be included in local scaling relations, potentially improving the accuracy of simulations in reproducing observations.


\section{\label{sec:conclusions} Conclusions}

We have identified a sample of 2,435 disk galaxies hosting optical BL-AGN in DESI using GZ DESI morphology classifications \citep{Walmsley2023} and the \textsc{EmFit} emission line catalogue \citep{Pucha2025, Pucha2026}. We performed bulge-disk decompositions of the surface brightness profile on the $r-$band imaging of this sample using \textsc{galfitm}. From this, we identified a sample of 546 bulgeless hosts (\bulgeless) and of a 240 hosts with some bulge component in the model (\somebulge). We also composed a \control sample of early-type BL-AGN host galaxies, weighted to match \disk is redshift and stellar mass.

We considered the distribution of the full \disk sample, and \bulgeless, and \somebulge sub-populations on the \mbh--\mstar and \mbh--\mbulge planes, fitting straight line models to the distributions. The key results and discussion points are as follows:

\begin{itemize}
    \item There is a correlation between BH mass and total stellar mass in the disk-galaxy BL-AGN sample ($r=0.536$), with the fitted form $\log(M_{\rm BH}/{\rm M_\odot})=-3.7^{+0.3}_{-0.3}+1.00^{+0.03}_{-0.03}\log(M_*/{\rm M_\odot})$ (\autoref{fig:mbh_mstar}). The slope of this relation is in agreement with previous literature scaling relations within the reported errors (Section \ref{sec:disc:mbh_mstar}).
    \item The \mbh--\mstar correlation persists for the \bulgeless sample (\autoref{fig:mbh_mstar_mbulge_fits}). Under the interpretation of bulgeless disks as merger-free systems, this demonstrates that co-evolution is occurring in secularly grown systems (Section \ref{subsec:disc:secular-growth}).
    \item The \mbh--\mstar relations for \bulgeless and \somebulge galaxies are statistically consistent, whereas their \mbh--\mbulge relations are statistically distinct. This is contrary to what we expect in a highly merger-driven co-evolution scenario, advocating for significant contribution from secular mechanisms to the observed galaxy-SMBH co-evolution (Section \ref{subsec:disc:secular-growth}).
    \item A population of BHs that are over-massive with respect to the general population emerge at $\log(M_*)\lesssim10.5$ (\autoref{fig:mbh_mstar}). We speculate this may be related to kinetic AGN feedback activity (Section \ref{sec:disc:mbh_mstar}).
    \item The \control sample shows a significant tendency towards higher $L_{\rm bol}$ than the \bulgeless sample, but only moderate statistical evidence for a difference in \mbh distribution (\autoref{fig:Lbol_edd}). This suggests that the higher prevalence of galaxy mergers in the evolution of these early-type galaxies may not lead to significantly increased long-term BH growth compared to disk galaxies, despite a notable increase in observed $L_{\rm bol}$ (Section \ref{subsec:disc:secular-growth}).
    \item A variety of secular mechanisms including AGN feedback, stellar bars and spiral arms, may facilitate galaxy-SMBH co-evolution in secularly evolving systems (Section \ref{subsec:disc:secular-mechanisms}).
    \item Comparing with simulated bulgeless galaxies from \textsc{Horizon-AGN} \citep{Smethurst2024}, we find the simulation is missing a significant population of low-mass BHs (\autoref{fig:horizon}). As highlighted by e.g. \citet{Habouzit2021}, this is a consistent problem across multiple cosmological simulations. We suggest this is due to bias imparted by calibrating sub-grid physics to local early-type correlations, which exclude contributions from secular SMBH growth at low-redshift, and/or the effects of the ISM being smoothed across the spatial resolution scale. BH seed mass and difficulty in reproducing physically representative disk galaxy population may also contribute (Section \ref{subsec:sims}).
\end{itemize}

Telescopes such as Euclid and Roman will greatly aid the study of bulgeless galaxies. Their high-resolution, space-based photometry will better resolve stellar bulge components, and allow less massive bulges to be resolved. This will also allow bulgeless galaxies to be studied out to higher redshifts, allowing us to test the significance of secular BH growth over a longer time frame. The \disk and \bulgeless samples presented here can also be used in future work, to further investigate the properties of BL-AGN in disk-dominated galaxies.


\section*{Acknowledgements}

SMJ thanks the anonymous reviewer for their comments and suggestions. SMJ gratefully acknowledges support through the Oxford-Wolfson Marriott Scholarship, made possible by a donation to Wolfson College from Dr Francis Marriott, and the STFC [grant number ST/Y509474/1]. RJS gratefully acknowledges support through the Royal Astronomical Society Research Fellowship. BDS acknowledges support through a UK Research and Innovation Future Leaders Fellowship [grant number MR/T044136/1] and its renewal [grant number MR/Z000076/1]. ILG has received the support from the Czech Science Foundation Junior Star grant no. GM24-10599M. RSB acknowledges support from a UKRI Future Leaders Fellowship (grant code: MR/Y015517/1). R.P. is currently supported by the University of Utah, and was also previously supported by the University of Arizona, and in part by NSF NOIRLab. The Dunlap Institute is funded through an endowment established by the David Dunlap family and the University of Toronto. 

This research used data obtained with the Dark Energy Spectroscopic Instrument (DESI). DESI construction and operations is managed by the Lawrence Berkeley National Laboratory. This material is based upon work supported by the U.S. Department of Energy, Office of Science, Office of High-Energy Physics, under Contract No. DE–AC02–05CH11231, and by the National Energy Research Scientific Computing Center, a DOE Office of Science User Facility under the same contract. Additional support for DESI was provided by the U.S. National Science Foundation (NSF), Division of Astronomical Sciences under Contract No. AST-0950945 to the NSF’s National Optical-Infrared Astronomy Research Laboratory; the Science and Technology Facilities Council of the United Kingdom; the Gordon and Betty Moore Foundation; the Heising-Simons Foundation; the French Alternative Energies and Atomic Energy Commission (CEA); the National Council of Humanities, Science and Technology of Mexico (CONAHCYT); the Ministry of Science and Innovation of Spain (MICINN), and by the DESI Member Institutions: www.desi.lbl.gov/collaborating-institutions. The DESI collaboration is honoured to be permitted to conduct scientific research on I’oligam Du’ag (Kitt Peak), a mountain with particular significance to the Tohono O’odham Nation. Any opinions, findings, and conclusions or recommendations expressed in this material are those of the author(s) and do not necessarily reflect the views of the U.S. National Science Foundation, the U.S. Department of Energy, or any of the listed funding agencies.

The Legacy Surveys consist of three individual and complementary projects: the Dark Energy Camera Legacy Survey (DECaLS; Proposal ID \#2014B-0404; PIs: David Schlegel and Arjun Dey), the Beijing-Arizona Sky Survey (BASS; NOAO Prop. ID \#2015A-0801; PIs: Zhou Xu and Xiaohui Fan), and the Mayall z-band Legacy Survey (MzLS; Prop. ID \#2016A-0453; PI: Arjun Dey). DECaLS, BASS and MzLS together include data obtained, respectively, at the Blanco telescope, Cerro Tololo Inter-American Observatory, NSF’s NOIRLab; the Bok telescope, Steward Observatory, University of Arizona; and the Mayall telescope, Kitt Peak National Observatory, NOIRLab. Pipeline processing and analyses of the data were supported by NOIRLab and the Lawrence Berkeley National Laboratory (LBNL). The Legacy Surveys project is honoured to be permitted to conduct astronomical research on Iolkam Du’ag (Kitt Peak), a mountain with particular significance to the Tohono O’odham Nation. The complete acknowledgments can be found at \href{https://www.legacysurvey.org/acknowledgment/}{www.legacysurvey.org/acknowledgment/}

This research made use of \textsc{astropy}, a community-developed core \textsc{python} package for Astronomy \citep{astropy2013, astropy2018, astropy2022}, as well as other \textsc{python} packages, such as \textsc{matplotlib} \citep{Hunter2007}, \textsc{numpy} \citep{Harris2020}, \textsc{linmix} \citep{Kelly2007}, \textsc{pandas} \citep{Mckinney2010, reback2020pandas}, and \textsc{scipy} \citep{SciPy2020}.

\section{Data availability}

The data used in this paper is publicly accessible. The DESI data is available through the DESI Public Data Release portal (\href{https://data.desi.lbl.gov/doc/}{data.desi.lbl.gov/doc/}). The LS is available through the DESI Legacy Imaging Surveys portal (\href{https://www.legacysurvey.org}{www.legacysurvey.org}). The GZ DESI catalogue is available with Zonodo: \href{https://zenodo.org/record/7786416}{zenodo.org/record/7786416} \citep{Walmsley2023}, and the code and trained model used to produce this catalogue is available at \href{https://github.com/mwalmsley/zoobot}{github.com/mwalmsley/zoobot}.


\bibliographystyle{mnras}
\bibliography{refs}

@ARTICLE{Alonso2018,
       author = {{Alonso}, Sol and {Coldwell}, Georgina and {Duplancic}, Fernanda and {Mesa}, Valeria and {Lambas}, Diego G.},
        title = "{The impact of bars and interactions on optically selected AGNs in spiral galaxies}",
      journal = {\aap},
         year = 2018,
        month = oct,
       volume = {618},
          eid = {A149},
        pages = {A149},
          doi = {10.1051/0004-6361/201832796},
archivePrefix = {arXiv},
       eprint = {1808.05536},
 primaryClass = {astro-ph.GA},
       adsurl = {https://ui.adsabs.harvard.edu/abs/2018A&A...618A.149A}
}

@ARTICLE{Anand2024,
       author = {{Anand}, Abhijeet and {Guy}, Julien and {Bailey}, Stephen and {Moustakas}, John and {Aguilar}, J. and {Ahlen}, S. and {Bolton}, A.~S. and {Brodzeller}, A. and {Brooks}, D. and {Claybaugh}, T. and {Cole}, S. and {de la Macorra}, A. and {Dey}, Biprateep and {Fanning}, K. and {Forero-Romero}, J.~E. and {Gazta{\~n}aga}, E. and {Gontcho A Gontcho}, S. and {Gutierrez}, G. and {Honscheid}, K. and {Howlett}, C. and {Juneau}, S. and {Kirkby}, D. and {Kisner}, T. and {Kremin}, A. and {Lambert}, A. and {Landriau}, M. and {Le Guillou}, L. and {Manera}, M. and {Meisner}, A. and {Miquel}, R. and {Mueller}, E. and {Niz}, G. and {Palanque-Delabrouille}, N. and {Percival}, W.~J. and {Poppett}, C. and {Prada}, F. and {Raichoor}, A. and {Rezaie}, M. and {Rossi}, G. and {Sanchez}, E. and {Schlafly}, E.~F. and {Schlegel}, D. and {Schubnell}, M. and {Sprayberry}, D. and {Tarl{\'e}}, G. and {Warner}, C. and {Weaver}, B.~A. and {Zhou}, R. and {Zou}, H.},
        title = "{Archetype-based Redshift Estimation for the Dark Energy Spectroscopic Instrument Survey}",
      journal = {\aj},
         year = 2024,
        month = sep,
       volume = {168},
       number = {3},
          eid = {124},
        pages = {124},
          doi = {10.3847/1538-3881/ad60c2},
archivePrefix = {arXiv},
       eprint = {2405.19288},
 primaryClass = {astro-ph.CO},
       adsurl = {https://ui.adsabs.harvard.edu/abs/2024AJ....168..124A}
}

@article{Anderson1954,
author = {T. W. Anderson and D. A. Darling},
title = {A Test of Goodness of Fit},
journal = {Journal of the American Statistical Association},
volume = {49},
number = {268},
pages = {765--769},
year = {1954},
publisher = {Taylor \& Francis},
doi = {10.1080/01621459.1954.10501232},
URL = {https://www.tandfonline.com/doi/abs/10.1080/01621459.1954.10501232},
eprint = {https://www.tandfonline.com/doi/pdf/10.1080/01621459.1954.10501232}
}

@ARTICLE{Athanassoula1992,
       author = {{Athanassoula}, E.},
        title = "{The existence and shapes of dust lanes in galactic bars.}",
      journal = {\mnras},
         year = 1992,
        month = nov,
       volume = {259},
        pages = {345-364},
          doi = {10.1093/mnras/259.2.345},
       adsurl = {https://ui.adsabs.harvard.edu/abs/1992MNRAS.259..345A}
}

@INPROCEEDINGS{Athanassoula2000,
       author = {{Athanassoula}, E.},
        title = "{Gas Flow in Barred Galaxies}",
    booktitle = {Stars, Gas and Dust in Galaxies: Exploring the Links},
         year = 2000,
       editor = {{Alloin}, Danielle and {Olsen}, Knut and {Galaz}, Gaspar},
       series = {Astronomical Society of the Pacific Conference Series},
       volume = {221},
        month = jan,
        pages = {243},
          doi = {10.48550/arXiv.astro-ph/0006403},
archivePrefix = {arXiv},
       eprint = {astro-ph/0006403},
 primaryClass = {astro-ph},
       adsurl = {https://ui.adsabs.harvard.edu/abs/2000ASPC..221..243A}
}

@ARTICLE{Athanassoula2005,
       author = {{Athanassoula}, E.},
        title = "{On the nature of bulges in general and of box/peanut bulges in particular: input from N-body simulations}",
      journal = {\mnras},
         year = 2005,
        month = apr,
       volume = {358},
       number = {4},
        pages = {1477-1488},
          doi = {10.1111/j.1365-2966.2005.08872.x},
archivePrefix = {arXiv},
       eprint = {astro-ph/0502316},
 primaryClass = {astro-ph},
       adsurl = {https://ui.adsabs.harvard.edu/abs/2005MNRAS.358.1477A}
}

@ARTICLE{Athanassoula2013,
       author = {{Athanassoula}, E. and {Machado}, Rubens E.~G. and {Rodionov}, S.~A.},
        title = "{Bar formation and evolution in disc galaxies with gas and a triaxial halo: morphology, bar strength and halo properties}",
      journal = {\mnras},
         year = 2013,
        month = mar,
       volume = {429},
       number = {3},
        pages = {1949-1969},
          doi = {10.1093/mnras/sts452},
archivePrefix = {arXiv},
       eprint = {1211.6754},
 primaryClass = {astro-ph.CO},
       adsurl = {https://ui.adsabs.harvard.edu/abs/2013MNRAS.429.1949A}
}

@ARTICLE{Baldassare2020,
       author = {{Baldassare}, Vivienne F. and {Dickey}, Claire and {Geha}, Marla and {Reines}, Amy E.},
        title = "{Populating the Low-mass End of the M$_{BH}$- \{\textbackslashsigma \}\_\{* \} Relation}",
      journal = {\apjl},
         year = 2020,
        month = jul,
       volume = {898},
       number = {1},
          eid = {L3},
        pages = {L3},
          doi = {10.3847/2041-8213/aba0c1},
archivePrefix = {arXiv},
       eprint = {2006.15150},
 primaryClass = {astro-ph.GA},
       adsurl = {https://ui.adsabs.harvard.edu/abs/2020ApJ...898L...3B}
}

@ARTICLE{Baldry2004,
       author = {{Baldry}, Ivan K. and {Glazebrook}, Karl and {Brinkmann}, Jon and {Ivezi{\'c}}, {\v{Z}}eljko and {Lupton}, Robert H. and {Nichol}, Robert C. and {Szalay}, Alexander S.},
        title = "{Quantifying the Bimodal Color-Magnitude Distribution of Galaxies}",
      journal = {\apj},
         year = 2004,
        month = jan,
       volume = {600},
       number = {2},
        pages = {681-694},
          doi = {10.1086/380092},
archivePrefix = {arXiv},
       eprint = {astro-ph/0309710},
 primaryClass = {astro-ph},
       adsurl = {https://ui.adsabs.harvard.edu/abs/2004ApJ...600..681B}
}

@ARTICLE{Baldwin1981,
       author = {{Baldwin}, J.~A. and {Phillips}, M.~M. and {Terlevich}, R.},
        title = "{Classification parameters for the emission-line spectra of extragalactic objects.}",
      journal = {\pasp},
         year = 1981,
        month = feb,
       volume = {93},
        pages = {5-19},
          doi = {10.1086/130766},
       adsurl = {https://ui.adsabs.harvard.edu/abs/1981PASP...93....5B}
}

@INPROCEEDINGS{Bamford2011,
       author = {{Bamford}, S.~P. and {H{\"a}u{\ss}ler}, B. and {Rojas}, A. and {Borch}, A.},
        title = "{Measuring the Physical Properties of Galaxy Components Using Modern Surveys}",
    booktitle = {Astronomical Data Analysis Software and Systems XX},
         year = 2011,
       editor = {{Evans}, I.~N. and {Accomazzi}, A. and {Mink}, D.~J. and {Rots}, A.~H.},
       series = {Astronomical Society of the Pacific Conference Series},
       volume = {442},
        month = jul,
        pages = {479},
       adsurl = {https://ui.adsabs.harvard.edu/abs/2011ASPC..442..479B}
}

@ARTICLE{Barazza2008,
       author = {{Barazza}, Fabio D. and {Jogee}, Shardha and {Marinova}, Irina},
        title = "{Bars in Disk-dominated and Bulge-dominated Galaxies at z \raisebox{-0.5ex}\textasciitilde 0: New Insights from \raisebox{-0.5ex}\textasciitilde3600 SDSS Galaxies}",
      journal = {\apj},
         year = 2008,
        month = mar,
       volume = {675},
       number = {2},
        pages = {1194-1212},
          doi = {10.1086/526510},
archivePrefix = {arXiv},
       eprint = {0710.4674},
 primaryClass = {astro-ph},
       adsurl = {https://ui.adsabs.harvard.edu/abs/2008ApJ...675.1194B}
}

@ARTICLE{Bentz2009,
       author = {{Bentz}, Misty C. and {Peterson}, Bradley M. and {Netzer}, Hagai and {Pogge}, Richard W. and {Vestergaard}, Marianne},
        title = "{The Radius-Luminosity Relationship for Active Galactic Nuclei: The Effect of Host-Galaxy Starlight on Luminosity Measurements. II. The Full Sample of Reverberation-Mapped AGNs}",
      journal = {\apj},
         year = 2009,
        month = may,
       volume = {697},
       number = {1},
        pages = {160-181},
          doi = {10.1088/0004-637X/697/1/160},
archivePrefix = {arXiv},
       eprint = {0812.2283},
 primaryClass = {astro-ph},
       adsurl = {https://ui.adsabs.harvard.edu/abs/2009ApJ...697..160B}
}

@ARTICLE{Bentz2009b,
       author = {{Bentz}, Misty C. and {Walsh}, Jonelle L. and {Barth}, Aaron J. and {Baliber}, Nairn and {Bennert}, Vardha Nicola and {Canalizo}, Gabriela and {Filippenko}, Alexei V. and {Ganeshalingam}, Mohan and {Gates}, Elinor L. and {Greene}, Jenny E. and {Hidas}, Marton G. and {Hiner}, Kyle D. and {Lee}, Nicholas and {Li}, Weidong and {Malkan}, Matthew A. and {Minezaki}, Takeo and {Sakata}, Yu and {Serduke}, Frank J.~D. and {Silverman}, Jeffrey M. and {Steele}, Thea N. and {Stern}, Daniel and {Street}, Rachel A. and {Thornton}, Carol E. and {Treu}, Tommaso and {Wang}, Xiaofeng and {Woo}, Jong-Hak and {Yoshii}, Yuzuru},
        title = "{The Lick AGN Monitoring Project: Broad-line Region Radii and Black Hole Masses from Reverberation Mapping of H{\ensuremath{\beta}}}",
      journal = {\apj},
         year = 2009,
        month = nov,
       volume = {705},
       number = {1},
        pages = {199-217},
          doi = {10.1088/0004-637X/705/1/199},
archivePrefix = {arXiv},
       eprint = {0908.0003},
 primaryClass = {astro-ph.CO},
       adsurl = {https://ui.adsabs.harvard.edu/abs/2009ApJ...705..199B}
}

@ARTICLE{Bentz2013,
       author = {{Bentz}, Misty C. and {Denney}, Kelly D. and {Grier}, Catherine J. and {Barth}, Aaron J. and {Peterson}, Bradley M. and {Vestergaard}, Marianne and {Bennert}, Vardha N. and {Canalizo}, Gabriela and {De Rosa}, Gisella and {Filippenko}, Alexei V. and {Gates}, Elinor L. and {Greene}, Jenny E. and {Li}, Weidong and {Malkan}, Matthew A. and {Pogge}, Richard W. and {Stern}, Daniel and {Treu}, Tommaso and {Woo}, Jong-Hak},
        title = "{The Low-luminosity End of the Radius-Luminosity Relationship for Active Galactic Nuclei}",
      journal = {\apj},
         year = 2013,
        month = apr,
       volume = {767},
       number = {2},
          eid = {149},
        pages = {149},
          doi = {10.1088/0004-637X/767/2/149},
archivePrefix = {arXiv},
       eprint = {1303.1742},
 primaryClass = {astro-ph.CO},
       adsurl = {https://ui.adsabs.harvard.edu/abs/2013ApJ...767..149B}
}

@INPROCEEDINGS{Bertin2011,
       author = {{Bertin}, E.},
        title = "{Automated Morphometry with SExtractor and PSFEx}",
    booktitle = {Astronomical Data Analysis Software and Systems XX},
         year = 2011,
       editor = {{Evans}, I.~N. and {Accomazzi}, A. and {Mink}, D.~J. and {Rots}, A.~H.},
       series = {Astronomical Society of the Pacific Conference Series},
       volume = {442},
        month = jul,
        pages = {435},
       adsurl = {https://ui.adsabs.harvard.edu/abs/2011ASPC..442..435B}
}

@ARTICLE{Best2012,
       author = {{Best}, P.~N. and {Heckman}, T.~M.},
        title = "{On the fundamental dichotomy in the local radio-AGN population: accretion, evolution and host galaxy properties}",
      journal = {\mnras},
         year = 2012,
        month = apr,
       volume = {421},
       number = {2},
        pages = {1569-1582},
          doi = {10.1111/j.1365-2966.2012.20414.x},
archivePrefix = {arXiv},
       eprint = {1201.2397},
 primaryClass = {astro-ph.CO},
       adsurl = {https://ui.adsabs.harvard.edu/abs/2012MNRAS.421.1569B}
}

@ARTICLE{Birchall2020,
       author = {{Birchall}, Keir L. and {Watson}, M.~G. and {Aird}, J.},
        title = "{X-ray detected AGN in SDSS dwarf galaxies}",
      journal = {\mnras},
         year = 2020,
        month = feb,
       volume = {492},
       number = {2},
        pages = {2268-2284},
          doi = {10.1093/mnras/staa040},
archivePrefix = {arXiv},
       eprint = {2001.03135},
 primaryClass = {astro-ph.GA},
       adsurl = {https://ui.adsabs.harvard.edu/abs/2020MNRAS.492.2268B}
}

@book{Bishop2008,
    author = {{Bishop}, Christopher M.},
    booktitle = {Pattern Recognition and Machine Learning :},
    isbn = {9780387563282},
    language = {eng},
    publisher = {Springer},
    title = {Pattern Recognition and Machine Learning : },
    year = {2008},
}

@ARTICLE{Boquien2019,
       author = {{Boquien}, M. and {Burgarella}, D. and {Roehlly}, Y. and {Buat}, V. and {Ciesla}, L. and {Corre}, D. and {Inoue}, A.~K. and {Salas}, H.},
        title = "{CIGALE: a python Code Investigating GALaxy Emission}",
      journal = {\aap},
         year = 2019,
        month = feb,
       volume = {622},
          eid = {A103},
        pages = {A103},
          doi = {10.1051/0004-6361/201834156},
archivePrefix = {arXiv},
       eprint = {1811.03094},
 primaryClass = {astro-ph.GA},
       adsurl = {https://ui.adsabs.harvard.edu/abs/2019A&A...622A.103B}
}

@ARTICLE{Brinchmann2004,
       author = {{Brinchmann}, J. and {Charlot}, S. and {White}, S.~D.~M. and {Tremonti}, C. and {Kauffmann}, G. and {Heckman}, T. and {Brinkmann}, J.},
        title = "{The physical properties of star-forming galaxies in the low-redshift Universe}",
      journal = {\mnras},
         year = 2004,
        month = jul,
       volume = {351},
       number = {4},
        pages = {1151-1179},
          doi = {10.1111/j.1365-2966.2004.07881.x},
archivePrefix = {arXiv},
       eprint = {astro-ph/0311060},
 primaryClass = {astro-ph},
       adsurl = {https://ui.adsabs.harvard.edu/abs/2004MNRAS.351.1151B}
}

@ARTICLE{Brodzeller2023,
       author = {{Brodzeller}, Allyson and {Dawson}, Kyle and {Bailey}, Stephen and {Yu}, Jiaxi and {Ross}, A.~J. and {Bault}, A. and {Filbert}, S. and {Aguilar}, J. and {Ahlen}, S. and {Alexander}, David M. and {Armengaud}, E. and {Berti}, A. and {Brooks}, D. and {Chaussidon}, E. and {de la Macorra}, A. and {Doel}, P. and {Fanning}, K. and {Fawcett}, V.~A. and {Font-Ribera}, A. and {A Gontcho}, S. Gontcho and {Guy}, J. and {Honscheid}, K. and {Juneau}, S. and {Kehoe}, R. and {Kisner}, T. and {Kremin}, Anthony and {Lan}, Ting-Wen and {Landriau}, M. and {Levi}, Michael E. and {Magneville}, C. and {Martini}, Paul and {Meisner}, Aaron M. and {Miquel}, R. and {Moustakas}, J. and {Palanque-Delabrouille}, N. and {Percival}, W.~J. and {Prada}, F. and {Ravoux}, C. and {Rossi}, Graziano and {Saulder}, C. and {Siudek}, M. and {Tarl{\'e}}, Gregory and {Weaver}, B.~A. and {Youles}, S. and {Zheng}, Zheng and {Zhou}, Rongpu and {Zhou}, Zhimin},
        title = "{Performance of the Quasar Spectral Templates for the Dark Energy Spectroscopic Instrument}",
      journal = {\aj},
         year = 2023,
        month = aug,
       volume = {166},
       number = {2},
          eid = {66},
        pages = {66},
          doi = {10.3847/1538-3881/ace35d},
archivePrefix = {arXiv},
       eprint = {2305.10426},
 primaryClass = {astro-ph.IM},
       adsurl = {https://ui.adsabs.harvard.edu/abs/2023AJ....166...66B}
}

@ARTICLE{Bruce2016,
       author = {{Bruce}, V.~A. and {Dunlop}, J.~S. and {Mortlock}, A. and {Kocevski}, D.~D. and {McGrath}, E.~J. and {Rosario}, D.~J.},
        title = "{The bulge-disc decomposition of AGN host galaxies}",
      journal = {\mnras},
         year = 2016,
        month = may,
       volume = {458},
       number = {3},
        pages = {2391-2404},
          doi = {10.1093/mnras/stw467},
archivePrefix = {arXiv},
       eprint = {1510.03870},
 primaryClass = {astro-ph.GA},
       adsurl = {https://ui.adsabs.harvard.edu/abs/2016MNRAS.458.2391B}
}

@ARTICLE{Bruzual2003,
       author = {{Bruzual}, G. and {Charlot}, S.},
        title = "{Stellar population synthesis at the resolution of 2003}",
      journal = {\mnras},
         year = 2003,
        month = oct,
       volume = {344},
       number = {4},
        pages = {1000-1028},
          doi = {10.1046/j.1365-8711.2003.06897.x},
archivePrefix = {arXiv},
       eprint = {astro-ph/0309134},
 primaryClass = {astro-ph},
       adsurl = {https://ui.adsabs.harvard.edu/abs/2003MNRAS.344.1000B}
}

@ARTICLE{Buta2019,
       author = {{Buta}, Ronald J. and {Verdes-Montenegro}, Lourdes and {Damas-Segovia}, Ancor and {Jones}, Michael and {Blasco}, Javier and {Fern{\'a}ndez-Lorenzo}, Mirian and {Sanchez}, Susana and {Garrido}, Julian and {Ramirez-Moreta}, Pablo and {Sulentic}, Jack W.},
        title = "{A comprehensive examination of the optical morphologies of 719 isolated galaxies in the AMIGA sample}",
      journal = {\mnras},
         year = 2019,
        month = sep,
       volume = {488},
       number = {2},
        pages = {2175-2189},
          doi = {10.1093/mnras/stz1780},
archivePrefix = {arXiv},
       eprint = {1906.11677},
 primaryClass = {astro-ph.GA},
       adsurl = {https://ui.adsabs.harvard.edu/abs/2019MNRAS.488.2175B}
}

@ARTICLE{Calzetti2000,
       author = {{Calzetti}, Daniela and {Armus}, Lee and {Bohlin}, Ralph C. and {Kinney}, Anne L. and {Koornneef}, Jan and {Storchi-Bergmann}, Thaisa},
        title = "{The Dust Content and Opacity of Actively Star-forming Galaxies}",
      journal = {\apj},
         year = 2000,
        month = apr,
       volume = {533},
       number = {2},
        pages = {682-695},
          doi = {10.1086/308692},
archivePrefix = {arXiv},
       eprint = {astro-ph/9911459},
 primaryClass = {astro-ph},
       adsurl = {https://ui.adsabs.harvard.edu/abs/2000ApJ...533..682C}
}

@ARTICLE{Chabrier2003,
       author = {{Chabrier}, Gilles},
        title = "{Galactic Stellar and Substellar Initial Mass Function}",
      journal = {\pasp},
         year = 2003,
        month = jul,
       volume = {115},
       number = {809},
        pages = {763-795},
          doi = {10.1086/376392},
archivePrefix = {arXiv},
       eprint = {astro-ph/0304382},
 primaryClass = {astro-ph},
       adsurl = {https://ui.adsabs.harvard.edu/abs/2003PASP..115..763C}
}

@ARTICLE{Charlot2000,
       author = {{Charlot}, St{\'e}phane and {Fall}, S. Michael},
        title = "{A Simple Model for the Absorption of Starlight by Dust in Galaxies}",
      journal = {\apj},
         year = 2000,
        month = aug,
       volume = {539},
       number = {2},
        pages = {718-731},
          doi = {10.1086/309250},
archivePrefix = {arXiv},
       eprint = {astro-ph/0003128},
 primaryClass = {astro-ph},
       adsurl = {https://ui.adsabs.harvard.edu/abs/2000ApJ...539..718C}
}

@ARTICLE{Cheung2013,
       author = {{Cheung}, Edmond and {Athanassoula}, E. and {Masters}, Karen L. and {Nichol}, Robert C. and {Bosma}, A. and {Bell}, Eric F. and {Faber}, S.~M. and {Koo}, David C. and {Lintott}, Chris and {Melvin}, Thomas and {Schawinski}, Kevin and {Skibba}, Ramin A. and {Willett}, Kyle W.},
        title = "{Galaxy Zoo: Observing Secular Evolution through Bars}",
      journal = {\apj},
         year = 2013,
        month = dec,
       volume = {779},
       number = {2},
          eid = {162},
        pages = {162},
          doi = {10.1088/0004-637X/779/2/162},
archivePrefix = {arXiv},
       eprint = {1310.2941},
 primaryClass = {astro-ph.CO},
       adsurl = {https://ui.adsabs.harvard.edu/abs/2013ApJ...779..162C}
}

@ARTICLE{Ciesla2015,
       author = {{Ciesla}, L. and {Charmandaris}, V. and {Georgakakis}, A. and {Bernhard}, E. and {Mitchell}, P.~D. and {Buat}, V. and {Elbaz}, D. and {LeFloc'h}, E. and {Lacey}, C.~G. and {Magdis}, G.~E. and {Xilouris}, M.},
        title = "{Constraining the properties of AGN host galaxies with spectral energy distribution modelling}",
      journal = {\aap},
         year = 2015,
        month = apr,
       volume = {576},
          eid = {A10},
        pages = {A10},
          doi = {10.1051/0004-6361/201425252},
archivePrefix = {arXiv},
       eprint = {1501.03672},
 primaryClass = {astro-ph.GA},
       adsurl = {https://ui.adsabs.harvard.edu/abs/2015A&A...576A..10C}
}

@ARTICLE{Cisternas2011,
       author = {{Cisternas}, Mauricio and {Jahnke}, Knud and {Bongiorno}, Angela and {Inskip}, Katherine J. and {Impey}, Chris D. and {Koekemoer}, Anton M. and {Merloni}, Andrea and {Salvato}, Mara and {Trump}, Jonathan R.},
        title = "{Secular Evolution and a Non-evolving Black-hole-to-galaxy Mass Ratio in the Last 7 Gyr}",
      journal = {\apjl},
         year = 2011,
        month = nov,
       volume = {741},
       number = {1},
          eid = {L11},
        pages = {L11},
          doi = {10.1088/2041-8205/741/1/L11},
archivePrefix = {arXiv},
       eprint = {1109.4633},
 primaryClass = {astro-ph.CO},
       adsurl = {https://ui.adsabs.harvard.edu/abs/2011ApJ...741L..11C}
}

@ARTICLE{Cisternas2013,
       author = {{Cisternas}, Mauricio and {Gadotti}, Dimitri A. and {Knapen}, Johan H. and {Kim}, Taehyun and {D{\'\i}az-Garc{\'\i}a}, Sim{\'o}n and {Laurikainen}, Eija and {Salo}, Heikki and {Gonz{\'a}lez-Mart{\'\i}n}, Omaira and {Ho}, Luis C. and {Elmegreen}, Bruce G. and {Zaritsky}, Dennis and {Sheth}, Kartik and {Athanassoula}, E. and {Bosma}, Albert and {Comer{\'o}n}, S{\'e}bastien and {Erroz-Ferrer}, Santiago and {Gil de Paz}, Armando and {Hinz}, Joannah L. and {Holwerda}, Benne W. and {Laine}, Jarkko and {Meidt}, Sharon and {Men{\'e}ndez-Delmestre}, Kar{\'\i}n and {Mizusawa}, Trisha and {Mu{\~n}oz-Mateos}, Juan Carlos and {Regan}, Michael W. and {Seibert}, Mark},
        title = "{X-Ray Nuclear Activity in S$^{4}$G Barred Galaxies: No Link between Bar Strength and Co-occurrent Supermassive Black Hole Fueling}",
      journal = {\apj},
         year = 2013,
        month = oct,
       volume = {776},
       number = {1},
          eid = {50},
        pages = {50},
          doi = {10.1088/0004-637X/776/1/50},
archivePrefix = {arXiv},
       eprint = {1307.7709},
 primaryClass = {astro-ph.CO},
       adsurl = {https://ui.adsabs.harvard.edu/abs/2013ApJ...776...50C}
}

@ARTICLE{Cisternas2015,
       author = {{Cisternas}, Mauricio and {Sheth}, Kartik and {Salvato}, Mara and {Knapen}, Johan H. and {Civano}, Francesca and {Santini}, Paola},
        title = "{The Role of Bars in AGN Fueling in Disk Galaxies Over the Last Seven Billion Years}",
      journal = {\apj},
         year = 2015,
        month = apr,
       volume = {802},
       number = {2},
          eid = {137},
        pages = {137},
          doi = {10.1088/0004-637X/802/2/137},
archivePrefix = {arXiv},
       eprint = {1409.2871},
 primaryClass = {astro-ph.GA},
       adsurl = {https://ui.adsabs.harvard.edu/abs/2015ApJ...802..137C}
}

@ARTICLE{Crain2015,
       author = {{Crain}, Robert A. and {Schaye}, Joop and {Bower}, Richard G. and {Furlong}, Michelle and {Schaller}, Matthieu and {Theuns}, Tom and {Dalla Vecchia}, Claudio and {Frenk}, Carlos S. and {McCarthy}, Ian G. and {Helly}, John C. and {Jenkins}, Adrian and {Rosas-Guevara}, Yetli M. and {White}, Simon D.~M. and {Trayford}, James W.},
        title = "{The EAGLE simulations of galaxy formation: calibration of subgrid physics and model variations}",
      journal = {\mnras},
         year = 2015,
        month = jun,
       volume = {450},
       number = {2},
        pages = {1937-1961},
          doi = {10.1093/mnras/stv725},
archivePrefix = {arXiv},
       eprint = {1501.01311},
 primaryClass = {astro-ph.GA},
       adsurl = {https://ui.adsabs.harvard.edu/abs/2015MNRAS.450.1937C}
}

@ARTICLE{Dale2014,
       author = {{Dale}, Daniel A. and {Helou}, George and {Magdis}, Georgios E. and {Armus}, Lee and {D{\'\i}az-Santos}, Tanio and {Shi}, Yong},
        title = "{A Two-parameter Model for the Infrared/Submillimeter/Radio Spectral Energy Distributions of Galaxies and Active Galactic Nuclei}",
      journal = {\apj},
         year = 2014,
        month = mar,
       volume = {784},
       number = {1},
          eid = {83},
        pages = {83},
          doi = {10.1088/0004-637X/784/1/83},
archivePrefix = {arXiv},
       eprint = {1402.1495},
 primaryClass = {astro-ph.GA},
       adsurl = {https://ui.adsabs.harvard.edu/abs/2014ApJ...784...83D}
}

@ARTICLE{Davies2009,
       author = {{Davies}, R.~I. and {Maciejewski}, W. and {Hicks}, E.~K.~S. and {Tacconi}, L.~J. and {Genzel}, R. and {Engel}, H.},
        title = "{Stellar and Molecular Gas Kinematics Of NGC 1097: Inflow Driven by a Nuclear Spiral}",
      journal = {\apj},
         year = 2009,
        month = sep,
       volume = {702},
       number = {1},
        pages = {114-128},
          doi = {10.1088/0004-637X/702/1/114},
archivePrefix = {arXiv},
       eprint = {0903.0313},
 primaryClass = {astro-ph.CO},
       adsurl = {https://ui.adsabs.harvard.edu/abs/2009ApJ...702..114D}
}

@ARTICLE{Davis2018,
       author = {{Davis}, Benjamin L. and {Graham}, Alister W. and {Cameron}, Ewan},
        title = "{Black Hole Mass Scaling Relations for Spiral Galaxies. II. M $_{BH}$-M $_{*,tot}$ and M $_{BH}$-M $_{*,disk}$}",
      journal = {\apj},
         year = 2018,
        month = dec,
       volume = {869},
       number = {2},
          eid = {113},
        pages = {113},
          doi = {10.3847/1538-4357/aae820},
archivePrefix = {arXiv},
       eprint = {1810.04888},
 primaryClass = {astro-ph.GA},
       adsurl = {https://ui.adsabs.harvard.edu/abs/2018ApJ...869..113D}
}

@ARTICLE{Debattista2006,
       author = {{Debattista}, Victor P. and {Mayer}, Lucio and {Carollo}, C. Marcella and {Moore}, Ben and {Wadsley}, James and {Quinn}, Thomas},
        title = "{The Secular Evolution of Disk Structural Parameters}",
      journal = {\apj},
         year = 2006,
        month = jul,
       volume = {645},
       number = {1},
        pages = {209-227},
          doi = {10.1086/504147},
archivePrefix = {arXiv},
       eprint = {astro-ph/0509310},
 primaryClass = {astro-ph},
       adsurl = {https://ui.adsabs.harvard.edu/abs/2006ApJ...645..209D}
}

@ARTICLE{Deeley2017,
       author = {{Deeley}, Simon and {Drinkwater}, Michael J. and {Cunnama}, Daniel and {Bland-Hawthorn}, Joss and {Brough}, Sarah and {Cluver}, Michelle and {Colless}, Matthew and {Davies}, Luke J.~M. and {Driver}, Simon P. and {Foster}, Caroline and {Grootes}, Meiert W. and {Hopkins}, Andrew M. and {Kafle}, Prajwal R. and {Lara-Lopez}, Maritza A. and {Liske}, Jochen and {Mahajan}, Smriti and {Phillipps}, Steven and {Power}, Chris and {Robotham}, Aaron},
        title = "{Galaxy and Mass Assembly (GAMA): formation and growth of elliptical galaxies in the group environment}",
      journal = {\mnras},
         year = 2017,
        month = jun,
       volume = {467},
       number = {4},
        pages = {3934-3943},
          doi = {10.1093/mnras/stx441},
archivePrefix = {arXiv},
       eprint = {1702.07641},
 primaryClass = {astro-ph.GA},
       adsurl = {https://ui.adsabs.harvard.edu/abs/2017MNRAS.467.3934D}
}

@ARTICLE{Dekel2006,
       author = {{Dekel}, Avishai and {Birnboim}, Yuval},
        title = "{Galaxy bimodality due to cold flows and shock heating}",
      journal = {\mnras},
         year = 2006,
        month = may,
       volume = {368},
       number = {1},
        pages = {2-20},
          doi = {10.1111/j.1365-2966.2006.10145.x},
archivePrefix = {arXiv},
       eprint = {astro-ph/0412300},
 primaryClass = {astro-ph},
       adsurl = {https://ui.adsabs.harvard.edu/abs/2006MNRAS.368....2D}
}

@ARTICLE{DESI2016a,
       author = {{DESI Collaboration} and {Aghamousa}, Amir and {Aguilar}, Jessica and {Ahlen}, Steve and {Alam}, Shadab and {Allen}, Lori E. and {Allende Prieto}, Carlos and {Annis}, James and {Bailey}, Stephen and {Balland}, Christophe and {Ballester}, Otger and {Baltay}, Charles and {Beaufore}, Lucas and {Bebek}, Chris and {Beers}, Timothy C. and {Bell}, Eric F. and {Bernal}, Jos{\'e} Luis and {Besuner}, Robert and {Beutler}, Florian and {Blake}, Chris and {Bleuler}, Hannes and {Blomqvist}, Michael and {Blum}, Robert and {Bolton}, Adam S. and {Briceno}, Cesar and {Brooks}, David and {Brownstein}, Joel R. and {Buckley-Geer}, Elizabeth and {Burden}, Angela and {Burtin}, Etienne and {Busca}, Nicolas G. and {Cahn}, Robert N. and {Cai}, Yan-Chuan and {Cardiel-Sas}, Laia and {Carlberg}, Raymond G. and {Carton}, Pierre-Henri and {Casas}, Ricard and {Castander}, Francisco J. and {Cervantes-Cota}, Jorge L. and {Claybaugh}, Todd M. and {Close}, Madeline and {Coker}, Carl T. and {Cole}, Shaun and {Comparat}, Johan and {Cooper}, Andrew P. and {Cousinou}, M. -C. and {Crocce}, Martin and {Cuby}, Jean-Gabriel and {Cunningham}, Daniel P. and {Davis}, Tamara M. and {Dawson}, Kyle S. and {de la Macorra}, Axel and {De Vicente}, Juan and {Delubac}, Timoth{\'e}e and {Derwent}, Mark and {Dey}, Arjun and {Dhungana}, Govinda and {Ding}, Zhejie and {Doel}, Peter and {Duan}, Yutong T. and {Ealet}, Anne and {Edelstein}, Jerry and {Eftekharzadeh}, Sarah and {Eisenstein}, Daniel J. and {Elliott}, Ann and {Escoffier}, St{\'e}phanie and {Evatt}, Matthew and {Fagrelius}, Parker and {Fan}, Xiaohui and {Fanning}, Kevin and {Farahi}, Arya and {Farihi}, Jay and {Favole}, Ginevra and {Feng}, Yu and {Fernandez}, Enrique and {Findlay}, Joseph R. and {Finkbeiner}, Douglas P. and {Fitzpatrick}, Michael J. and {Flaugher}, Brenna and {Flender}, Samuel and {Font-Ribera}, Andreu and {Forero-Romero}, Jaime E. and {Fosalba}, Pablo and {Frenk}, Carlos S. and {Fumagalli}, Michele and {Gaensicke}, Boris T. and {Gallo}, Giuseppe and {Garcia-Bellido}, Juan and {Gaztanaga}, Enrique and {Pietro Gentile Fusillo}, Nicola and {Gerard}, Terry and {Gershkovich}, Irena and {Giannantonio}, Tommaso and {Gillet}, Denis and {Gonzalez-de-Rivera}, Guillermo and {Gonzalez-Perez}, Violeta and {Gott}, Shelby and {Graur}, Or and {Gutierrez}, Gaston and {Guy}, Julien and {Habib}, Salman and {Heetderks}, Henry and {Heetderks}, Ian and {Heitmann}, Katrin and {Hellwing}, Wojciech A. and {Herrera}, David A. and {Ho}, Shirley and {Holland}, Stephen and {Honscheid}, Klaus and {Huff}, Eric and {Hutchinson}, Timothy A. and {Huterer}, Dragan and {Hwang}, Ho Seong and {Illa Laguna}, Joseph Maria and {Ishikawa}, Yuzo and {Jacobs}, Dianna and {Jeffrey}, Niall and {Jelinsky}, Patrick and {Jennings}, Elise and {Jiang}, Linhua and {Jimenez}, Jorge and {Johnson}, Jennifer and {Joyce}, Richard and {Jullo}, Eric and {Juneau}, St{\'e}phanie and {Kama}, Sami and {Karcher}, Armin and {Karkar}, Sonia and {Kehoe}, Robert and {Kennamer}, Noble and {Kent}, Stephen and {Kilbinger}, Martin and {Kim}, Alex G. and {Kirkby}, David and {Kisner}, Theodore and {Kitanidis}, Ellie and {Kneib}, Jean-Paul and {Koposov}, Sergey and {Kovacs}, Eve and {Koyama}, Kazuya and {Kremin}, Anthony and {Kron}, Richard and {Kronig}, Luzius and {Kueter-Young}, Andrea and {Lacey}, Cedric G. and {Lafever}, Robin and {Lahav}, Ofer and {Lambert}, Andrew and {Lampton}, Michael and {Landriau}, Martin and {Lang}, Dustin and {Lauer}, Tod R. and {Le Goff}, Jean-Marc and {Le Guillou}, Laurent and {Le Van Suu}, Auguste and {Lee}, Jae Hyeon and {Lee}, Su-Jeong and {Leitner}, Daniela and {Lesser}, Michael and {Levi}, Michael E. and {L'Huillier}, Benjamin and {Li}, Baojiu and {Liang}, Ming and {Lin}, Huan and {Linder}, Eric and {Loebman}, Sarah R. and {Luki{\'c}}, Zarija and {Ma}, Jun and {MacCrann}, Niall and {Magneville}, Christophe and {Makarem}, Laleh and {Manera}, Marc and {Manser}, Christopher J. and {Marshall}, Robert and {Martini}, Paul and {Massey}, Richard and {Matheson}, Thomas and {McCauley}, Jeremy and {McDonald}, Patrick and {McGreer}, Ian D. and {Meisner}, Aaron and {Metcalfe}, Nigel and {Miller}, Timothy N. and {Miquel}, Ramon and {Moustakas}, John and {Myers}, Adam and {Naik}, Milind and {Newman}, Jeffrey A. and {Nichol}, Robert C. and {Nicola}, Andrina and {Nicolati da Costa}, Luiz and {Nie}, Jundan and {Niz}, Gustavo and {Norberg}, Peder and {Nord}, Brian and {Norman}, Dara and {Nugent}, Peter and {O'Brien}, Thomas and {Oh}, Minji and {Olsen}, Knut A.~G.},
        title = "{The DESI Experiment Part I: Science,Targeting, and Survey Design}",
      journal = {arXiv e-prints},
         year = 2016,
        month = oct,
          eid = {arXiv:1611.00036},
        pages = {arXiv:1611.00036},
          doi = {10.48550/arXiv.1611.00036},
archivePrefix = {arXiv},
       eprint = {1611.00036},
 primaryClass = {astro-ph.IM},
       adsurl = {https://ui.adsabs.harvard.edu/abs/2016arXiv161100036D}
}

@ARTICLE{DESI2016b,
       author = {{DESI Collaboration} and {Aghamousa}, Amir and {Aguilar}, Jessica and {Ahlen}, Steve and {Alam}, Shadab and {Allen}, Lori E. and {Allende Prieto}, Carlos and {Annis}, James and {Bailey}, Stephen and {Balland}, Christophe and {Ballester}, Otger and {Baltay}, Charles and {Beaufore}, Lucas and {Bebek}, Chris and {Beers}, Timothy C. and {Bell}, Eric F. and {Bernal}, Jos{\'e} Luis and {Besuner}, Robert and {Beutler}, Florian and {Blake}, Chris and {Bleuler}, Hannes and {Blomqvist}, Michael and {Blum}, Robert and {Bolton}, Adam S. and {Briceno}, Cesar and {Brooks}, David and {Brownstein}, Joel R. and {Buckley-Geer}, Elizabeth and {Burden}, Angela and {Burtin}, Etienne and {Busca}, Nicolas G. and {Cahn}, Robert N. and {Cai}, Yan-Chuan and {Cardiel-Sas}, Laia and {Carlberg}, Raymond G. and {Carton}, Pierre-Henri and {Casas}, Ricard and {Castander}, Francisco J. and {Cervantes-Cota}, Jorge L. and {Claybaugh}, Todd M. and {Close}, Madeline and {Coker}, Carl T. and {Cole}, Shaun and {Comparat}, Johan and {Cooper}, Andrew P. and {Cousinou}, M. -C. and {Crocce}, Martin and {Cuby}, Jean-Gabriel and {Cunningham}, Daniel P. and {Davis}, Tamara M. and {Dawson}, Kyle S. and {de la Macorra}, Axel and {De Vicente}, Juan and {Delubac}, Timoth{\'e}e and {Derwent}, Mark and {Dey}, Arjun and {Dhungana}, Govinda and {Ding}, Zhejie and {Doel}, Peter and {Duan}, Yutong T. and {Ealet}, Anne and {Edelstein}, Jerry and {Eftekharzadeh}, Sarah and {Eisenstein}, Daniel J. and {Elliott}, Ann and {Escoffier}, St{\'e}phanie and {Evatt}, Matthew and {Fagrelius}, Parker and {Fan}, Xiaohui and {Fanning}, Kevin and {Farahi}, Arya and {Farihi}, Jay and {Favole}, Ginevra and {Feng}, Yu and {Fernandez}, Enrique and {Findlay}, Joseph R. and {Finkbeiner}, Douglas P. and {Fitzpatrick}, Michael J. and {Flaugher}, Brenna and {Flender}, Samuel and {Font-Ribera}, Andreu and {Forero-Romero}, Jaime E. and {Fosalba}, Pablo and {Frenk}, Carlos S. and {Fumagalli}, Michele and {Gaensicke}, Boris T. and {Gallo}, Giuseppe and {Garcia-Bellido}, Juan and {Gaztanaga}, Enrique and {Pietro Gentile Fusillo}, Nicola and {Gerard}, Terry and {Gershkovich}, Irena and {Giannantonio}, Tommaso and {Gillet}, Denis and {Gonzalez-de-Rivera}, Guillermo and {Gonzalez-Perez}, Violeta and {Gott}, Shelby and {Graur}, Or and {Gutierrez}, Gaston and {Guy}, Julien and {Habib}, Salman and {Heetderks}, Henry and {Heetderks}, Ian and {Heitmann}, Katrin and {Hellwing}, Wojciech A. and {Herrera}, David A. and {Ho}, Shirley and {Holland}, Stephen and {Honscheid}, Klaus and {Huff}, Eric and {Hutchinson}, Timothy A. and {Huterer}, Dragan and {Hwang}, Ho Seong and {Illa Laguna}, Joseph Maria and {Ishikawa}, Yuzo and {Jacobs}, Dianna and {Jeffrey}, Niall and {Jelinsky}, Patrick and {Jennings}, Elise and {Jiang}, Linhua and {Jimenez}, Jorge and {Johnson}, Jennifer and {Joyce}, Richard and {Jullo}, Eric and {Juneau}, St{\'e}phanie and {Kama}, Sami and {Karcher}, Armin and {Karkar}, Sonia and {Kehoe}, Robert and {Kennamer}, Noble and {Kent}, Stephen and {Kilbinger}, Martin and {Kim}, Alex G. and {Kirkby}, David and {Kisner}, Theodore and {Kitanidis}, Ellie and {Kneib}, Jean-Paul and {Koposov}, Sergey and {Kovacs}, Eve and {Koyama}, Kazuya and {Kremin}, Anthony and {Kron}, Richard and {Kronig}, Luzius and {Kueter-Young}, Andrea and {Lacey}, Cedric G. and {Lafever}, Robin and {Lahav}, Ofer and {Lambert}, Andrew and {Lampton}, Michael and {Landriau}, Martin and {Lang}, Dustin and {Lauer}, Tod R. and {Le Goff}, Jean-Marc and {Le Guillou}, Laurent and {Le Van Suu}, Auguste and {Lee}, Jae Hyeon and {Lee}, Su-Jeong and {Leitner}, Daniela and {Lesser}, Michael and {Levi}, Michael E. and {L'Huillier}, Benjamin and {Li}, Baojiu and {Liang}, Ming and {Lin}, Huan and {Linder}, Eric and {Loebman}, Sarah R. and {Luki{\'c}}, Zarija and {Ma}, Jun and {MacCrann}, Niall and {Magneville}, Christophe and {Makarem}, Laleh and {Manera}, Marc and {Manser}, Christopher J. and {Marshall}, Robert and {Martini}, Paul and {Massey}, Richard and {Matheson}, Thomas and {McCauley}, Jeremy and {McDonald}, Patrick and {McGreer}, Ian D. and {Meisner}, Aaron and {Metcalfe}, Nigel and {Miller}, Timothy N. and {Miquel}, Ramon and {Moustakas}, John and {Myers}, Adam and {Naik}, Milind and {Newman}, Jeffrey A. and {Nichol}, Robert C. and {Nicola}, Andrina and {Nicolati da Costa}, Luiz and {Nie}, Jundan and {Niz}, Gustavo and {Norberg}, Peder and {Nord}, Brian and {Norman}, Dara and {Nugent}, Peter and {O'Brien}, Thomas and {Oh}, Minji and {Olsen}, Knut A.~G.},
        title = "{The DESI Experiment Part II: Instrument Design}",
      journal = {arXiv e-prints},
         year = 2016,
        month = oct,
          eid = {arXiv:1611.00037},
        pages = {arXiv:1611.00037},
          doi = {10.48550/arXiv.1611.00037},
archivePrefix = {arXiv},
       eprint = {1611.00037},
 primaryClass = {astro-ph.IM},
       adsurl = {https://ui.adsabs.harvard.edu/abs/2016arXiv161100037D}
}

@ARTICLE{DESI2024,
       author = {{DESI Collaboration} and {Adame}, A.~G. and {Aguilar}, J. and {Ahlen}, S. and {Alam}, S. and {Aldering}, G. and {Alexander}, D.~M. and {Alfarsy}, R. and {Allende Prieto}, C. and {Alvarez}, M. and {Alves}, O. and {Anand}, A. and {Andrade-Oliveira}, F. and {Armengaud}, E. and {Asorey}, J. and {Avila}, S. and {Aviles}, A. and {Bailey}, S. and {Balaguera-Antol{\'\i}nez}, A. and {Ballester}, O. and {Baltay}, C. and {Bault}, A. and {Bautista}, J. and {Behera}, J. and {Beltran}, S.~F. and {BenZvi}, S. and {Beraldo e Silva}, L. and {Bermejo-Climent}, J.~R. and {Berti}, A. and {Besuner}, R. and {Beutler}, F. and {Bianchi}, D. and {Blake}, C. and {Blum}, R. and {Bolton}, A.~S. and {Brieden}, S. and {Brodzeller}, A. and {Brooks}, D. and {Brown}, Z. and {Buckley-Geer}, E. and {Burtin}, E. and {Cabayol-Garcia}, L. and {Cai}, Z. and {Canning}, R. and {Cardiel-Sas}, L. and {Carnero Rosell}, A. and {Castander}, F.~J. and {Cervantes-Cota}, J.~L. and {Chabanier}, S. and {Chaussidon}, E. and {Chaves-Montero}, J. and {Chen}, S. and {Chen}, X. and {Chuang}, C. and {Claybaugh}, T. and {Cole}, S. and {Cooper}, A.~P. and {Cuceu}, A. and {Davis}, T.~M. and {Dawson}, K. and {de Belsunce}, R. and {de la Cruz}, R. and {de la Macorra}, A. and {Della Costa}, J. and {de Mattia}, A. and {Demina}, R. and {Demirbozan}, U. and {DeRose}, J. and {Dey}, A. and {Dey}, B. and {Dhungana}, G. and {Ding}, J. and {Ding}, Z. and {Doel}, P. and {Doshi}, R. and {Douglass}, K. and {Edge}, A. and {Eftekharzadeh}, S. and {Eisenstein}, D.~J. and {Elliott}, A. and {Ereza}, J. and {Escoffier}, S. and {Fagrelius}, P. and {Fan}, X. and {Fanning}, K. and {Fawcett}, V.~A. and {Ferraro}, S. and {Flaugher}, B. and {Font-Ribera}, A. and {Forero-Romero}, J.~E. and {Forero-S{\'a}nchez}, D. and {Frenk}, C.~S. and {G{\"a}nsicke}, B.~T. and {Garc{\'\i}a}, L. {\'A}. and {Garc{\'\i}a-Bellido}, J. and {Garcia-Quintero}, C. and {Garrison}, L.~H. and {Gil-Mar{\'\i}n}, H. and {Golden-Marx}, J. and {Gontcho A Gontcho}, S. and {Gonzalez-Morales}, A.~X. and {Gonzalez-Perez}, V. and {Gordon}, C. and {Graur}, O. and {Green}, D. and {Gruen}, D. and {Guy}, J. and {Hadzhiyska}, B. and {Hahn}, C. and {Han}, J.~J. and {Hanif}, M.~M.~S. and {Herrera-Alcantar}, H.~K. and {Honscheid}, K. and {Hou}, J. and {Howlett}, C. and {Huterer}, D. and {Ir{\v{s}}i{\v{c}}}, V. and {Ishak}, M. and {Jacques}, A. and {Jana}, A. and {Jiang}, L. and {Jimenez}, J. and {Jing}, Y.~P. and {Joudaki}, S. and {Joyce}, R. and {Jullo}, E. and {Juneau}, S. and {Kara{\c{c}}ayl{\i}}, N.~G. and {Karim}, T. and {Kehoe}, R. and {Kent}, S. and {Khederlarian}, A. and {Kim}, S. and {Kirkby}, D. and {Kisner}, T. and {Kitaura}, F. and {Kizhuprakkat}, N. and {Kneib}, J. and {Koposov}, S.~E. and {Kov{\'a}cs}, A. and {Kremin}, A. and {Krolewski}, A. and {L'Huillier}, B. and {Lahav}, O. and {Lambert}, A. and {Lamman}, C. and {Lan}, T.-W. and {Landriau}, M. and {Lang}, D. and {Lange}, J.~U. and {Lasker}, J. and {Leauthaud}, A. and {Le Guillou}, L. and {Levi}, M.~E. and {Li}, T.~S. and {Linder}, E. and {Lyons}, A. and {Magneville}, C. and {Manera}, M. and {Manser}, C.~J. and {Margala}, D. and {Martini}, P. and {McDonald}, P. and {Medina}, G.~E. and {Medina-Varela}, L. and {Meisner}, A. and {Mena-Fern{\'a}ndez}, J. and {Meneses-Rizo}, J. and {Mezcua}, M. and {Miquel}, R. and {Montero-Camacho}, P. and {Moon}, J. and {Moore}, S. and {Moustakas}, J. and {Mueller}, E. and {Mundet}, J. and {Mu{\~n}oz-Guti{\'e}rrez}, A. and {Myers}, A.~D. and {Nadathur}, S. and {Napolitano}, L. and {Neveux}, R. and {Newman}, J.~A. and {Nie}, J. and {Nikutta}, R. and {Niz}, G. and {Norberg}, P. and {Noriega}, H.~E. and {Paillas}, E. and {Palanque-Delabrouille}, N. and {Palmese}, A. and {Pan}, Z. and {Parkinson}, D. and {Penmetsa}, S. and {Percival}, W.~J. and {P{\'e}rez-Fern{\'a}ndez}, A. and {P{\'e}rez-R{\`a}fols}, I. and {Pieri}, M. and {Poppett}, C. and {Porredon}, A. and {Pothier}, S.},
        title = "{The Early Data Release of the Dark Energy Spectroscopic Instrument}",
      journal = {\aj},
         year = 2024,
        month = aug,
       volume = {168},
       number = {2},
          eid = {58},
        pages = {58},
          doi = {10.3847/1538-3881/ad3217},
archivePrefix = {arXiv},
       eprint = {2306.06308},
 primaryClass = {astro-ph.CO},
       adsurl = {https://ui.adsabs.harvard.edu/abs/2024AJ....168...58D}
}

@ARTICLE{DESI2025,
       author = {{DESI Collaboration} and {Karim}, M. Abdul and {Adame}, A.~G. and {Aguado}, D. and {Aguilar}, J. and {Ahlen}, S. and {Alam}, S. and {Aldering}, G. and {Alexander}, D.~M. and {Alfarsy}, R. and {Allen}, L. and {Allende Prieto}, C. and {Alves}, O. and {Anand}, A. and {Andrade}, U. and {Armengaud}, E. and {Avila}, S. and {Aviles}, A. and {Awan}, H. and {Bailey}, S. and {Baleato Lizancos}, A. and {Ballester}, O. and {Bault}, A. and {Bautista}, J. and {Bean}, R. and {Behera}, J. and {BenZvi}, S. and {Beraldo e Silva}, L. and {Bermejo-Climent}, J.~R. and {Beutler}, F. and {Bianchi}, D. and {Blake}, C. and {Blum}, R. and {Bolton}, A.~S. and {Bonici}, M. and {Brieden}, S. and {Brodzeller}, A. and {Brooks}, D. and {Buckley-Geer}, E. and {Burtin}, E. and {Bystr{\"o}m}, A. and {Canning}, R. and {Carnero Rosell}, A. and {Carr}, A. and {Carrilho}, P. and {Casas}, L. and {Castander}, F.~J. and {Cereskaite}, R. and {Cervantes-Cota}, J.~L. and {Chaussidon}, E. and {Chaves-Montero}, J. and {Chen}, S. and {Chen}, X. and {Circosta}, C. and {Claybaugh}, T. and {Cole}, S. and {Cooper}, A.~P. and {Cousinou}, M.-C. and {Cuceu}, A. and {Davis}, T.~M. and {Dawson}, K.~S. and {de Belsunce}, R. and {de la Cruz}, R. and {de la Macorra}, A. and {de Mattia}, A. and {Deiosso}, N. and {Della Costa}, J. and {Demina}, R. and {Demirbozan}, U. and {DeRose}, J. and {Dey}, A. and {Dey}, B. and {Ding}, J. and {Ding}, Z. and {Doel}, P. and {Douglass}, K. and {Dowicz}, M. and {Ebina}, H. and {Edelstein}, J. and {Eisenstein}, D.~J. and {Elbers}, W. and {Emas}, N. and {Escoffier}, S. and {Fagrelius}, P. and {Fan}, X. and {Fanning}, K. and {Favole}, G. and {Fawcett}, V.~A. and {Fern{\'a}ndez-Garc{\'\i}a}, E. and {Ferraro}, S. and {Findlay}, N. and {Font-Ribera}, A. and {Forero-Romero}, J.~E. and {Forero-S{\'a}nchez}, D. and {Frenk}, C.~S. and {G{\"a}nsicke}, B.~T. and {Galbany}, L. and {Garc{\'\i}a-Bellido}, J. and {Garcia-Quintero}, C. and {Garrison}, L.~H. and {Gazta{\~n}aga}, E. and {Gil-Mar{\'\i}n}, H. and {Gloudemans}, A. and {Gnedin}, O.~Y. and {Gontcho}, S. Gontcho A and {Gonzalez}, D. and {Gonzalez-Morales}, A.~X. and {Gonzalez-Perez}, V. and {Gordon}, C. and {Graur}, O. and {Green}, D. and {Gruen}, D. and {Gsponer}, R. and {Guandalin}, C. and {Gutierrez}, G. and {Guy}, J. and {Hahn}, C. and {Han}, J.~J. and {Han}, J. and {He}, S. and {Herrera-Alcantar}, H.~K. and {Heydenreich}, S. and {Honscheid}, K. and {Hou}, J. and {Howlett}, C. and {Huterer}, D. and {Ir{\v{s}}i{\v{c}}}, V. and {Ishak}, M. and {Jacques}, A. and {Jiang}, L. and {Jimenez}, J. and {Jing}, Y.~P. and {Joachimi}, B. and {Joudaki}, S. and {Joyce}, R. and {Jullo}, E. and {Juneau}, S. and {Kara{\c{c}}ayl{\i}}, N.~G. and {Karim}, T. and {Kehoe}, R. and {Kent}, S. and {Khederlarian}, A. and {Kirkby}, D. and {Kisner}, T. and {Kitaura}, F.-S. and {Kizhuprakkat}, N. and {Kong}, H. and {Koposov}, S.~E. and {Kremin}, A. and {Krolewski}, A. and {Lahav}, O. and {Lai}, Y. and {Lamman}, C. and {Lan}, T.-W. and {Landriau}, M. and {Lang}, D. and {Lange}, J.~U. and {Lasker}, J. and {Le Goff}, J.~M. and {Le Guillou}, L. and {Leauthaud}, A. and {Levi}, M.~E. and {Li}, S. and {Li}, T.~S. and {Liu}, W. and {Lodha}, K. and {Lokken}, M. and {Luo}, Y. and {Luo}, Y. and {Magneville}, C. and {Manera}, M. and {Manser}, C.~J. and {Margala}, D. and {Martini}, P. and {Maus}, M. and {McCullough}, J. and {McDonald}, P. and {Medina}, G.~E. and {Medina-Varela}, L. and {Meisner}, A. and {Mena-Fern{\'a}ndez}, J. and {Menegas}, A. and {Meneses-Rizo}, J. and {Mezcua}, M. and {Miquel}, R. and {Montero-Camacho}, P. and {Moon}, J. and {Moustakas}, J. and {Mu{\~n}oz-Guti{\'e}rrez}, A. and {Mu{\~n}oz-Santos}, D. and {Myers}, A.~D. and {Myles}, J. and {Nadathur}, S. and {Najita}, J. and {Napolitano}, L. and {Newman}, J.~A. and {Nikakhtar}, F. and {Nikutta}, R. and {Niz}, G. and {Noriega}, H.~E.},
        title = "{Data Release 1 of the Dark Energy Spectroscopic Instrument}",
      journal = {arXiv e-prints},
         year = 2025,
        month = mar,
          eid = {arXiv:2503.14745},
        pages = {arXiv:2503.14745},
          doi = {10.48550/arXiv.2503.14745},
archivePrefix = {arXiv},
       eprint = {2503.14745},
 primaryClass = {astro-ph.CO},
       adsurl = {https://ui.adsabs.harvard.edu/abs/2025arXiv250314745D}
}

@ARTICLE{Dessart2025,
       author = {{Dessart}, Luc},
        title = "{Probing red supergiant atmospheres and winds with early-time, high-cadence, high-resolution type II supernova spectra}",
      journal = {\aap},
         year = 2025,
        month = feb,
       volume = {694},
          eid = {A132},
        pages = {A132},
          doi = {10.1051/0004-6361/202452769},
archivePrefix = {arXiv},
       eprint = {2410.20486},
 primaryClass = {astro-ph.HE},
       adsurl = {https://ui.adsabs.harvard.edu/abs/2025A&A...694A.132D}
}

@ARTICLE{Dey2019,
       author = {{Dey}, Arjun and {Schlegel}, David J. and {Lang}, Dustin and {Blum}, Robert and {Burleigh}, Kaylan and {Fan}, Xiaohui and {Findlay}, Joseph R. and {Finkbeiner}, Doug and {Herrera}, David and {Juneau}, St{\'e}phanie and {Landriau}, Martin and {Levi}, Michael and {McGreer}, Ian and {Meisner}, Aaron and {Myers}, Adam D. and {Moustakas}, John and {Nugent}, Peter and {Patej}, Anna and {Schlafly}, Edward F. and {Walker}, Alistair R. and {Valdes}, Francisco and {Weaver}, Benjamin A. and {Y{\`e}che}, Christophe and {Zou}, Hu and {Zhou}, Xu and {Abareshi}, Behzad and {Abbott}, T.~M.~C. and {Abolfathi}, Bela and {Aguilera}, C. and {Alam}, Shadab and {Allen}, Lori and {Alvarez}, A. and {Annis}, James and {Ansarinejad}, Behzad and {Aubert}, Marie and {Beechert}, Jacqueline and {Bell}, Eric F. and {BenZvi}, Segev Y. and {Beutler}, Florian and {Bielby}, Richard M. and {Bolton}, Adam S. and {Brice{\~n}o}, C{\'e}sar and {Buckley-Geer}, Elizabeth J. and {Butler}, Karen and {Calamida}, Annalisa and {Carlberg}, Raymond G. and {Carter}, Paul and {Casas}, Ricard and {Castander}, Francisco J. and {Choi}, Yumi and {Comparat}, Johan and {Cukanovaite}, Elena and {Delubac}, Timoth{\'e}e and {DeVries}, Kaitlin and {Dey}, Sharmila and {Dhungana}, Govinda and {Dickinson}, Mark and {Ding}, Zhejie and {Donaldson}, John B. and {Duan}, Yutong and {Duckworth}, Christopher J. and {Eftekharzadeh}, Sarah and {Eisenstein}, Daniel J. and {Etourneau}, Thomas and {Fagrelius}, Parker A. and {Farihi}, Jay and {Fitzpatrick}, Mike and {Font-Ribera}, Andreu and {Fulmer}, Leah and {G{\"a}nsicke}, Boris T. and {Gaztanaga}, Enrique and {George}, Koshy and {Gerdes}, David W. and {Gontcho}, Satya Gontcho A. and {Gorgoni}, Claudio and {Green}, Gregory and {Guy}, Julien and {Harmer}, Diane and {Hernandez}, M. and {Honscheid}, Klaus and {Huang}, Lijuan Wendy and {James}, David J. and {Jannuzi}, Buell T. and {Jiang}, Linhua and {Joyce}, Richard and {Karcher}, Armin and {Karkar}, Sonia and {Kehoe}, Robert and {Kneib}, Jean-Paul and {Kueter-Young}, Andrea and {Lan}, Ting-Wen and {Lauer}, Tod R. and {Le Guillou}, Laurent and {Le Van Suu}, Auguste and {Lee}, Jae Hyeon and {Lesser}, Michael and {Perreault Levasseur}, Laurence and {Li}, Ting S. and {Mann}, Justin L. and {Marshall}, Robert and {Mart{\'\i}nez-V{\'a}zquez}, C.~E. and {Martini}, Paul and {du Mas des Bourboux}, H{\'e}lion and {McManus}, Sean and {Meier}, Tobias Gabriel and {M{\'e}nard}, Brice and {Metcalfe}, Nigel and {Mu{\~n}oz-Guti{\'e}rrez}, Andrea and {Najita}, Joan and {Napier}, Kevin and {Narayan}, Gautham and {Newman}, Jeffrey A. and {Nie}, Jundan and {Nord}, Brian and {Norman}, Dara J. and {Olsen}, Knut A.~G. and {Paat}, Anthony and {Palanque-Delabrouille}, Nathalie and {Peng}, Xiyan and {Poppett}, Claire L. and {Poremba}, Megan R. and {Prakash}, Abhishek and {Rabinowitz}, David and {Raichoor}, Anand and {Rezaie}, Mehdi and {Robertson}, A.~N. and {Roe}, Natalie A. and {Ross}, Ashley J. and {Ross}, Nicholas P. and {Rudnick}, Gregory and {Safonova}, Sasha and {Saha}, Abhijit and {S{\'a}nchez}, F. Javier and {Savary}, Elodie and {Schweiker}, Heidi and {Scott}, Adam and {Seo}, Hee-Jong and {Shan}, Huanyuan and {Silva}, David R. and {Slepian}, Zachary and {Soto}, Christian and {Sprayberry}, David and {Staten}, Ryan and {Stillman}, Coley M. and {Stupak}, Robert J. and {Summers}, David L. and {Sien Tie}, Suk and {Tirado}, H. and {Vargas-Maga{\~n}a}, Mariana and {Vivas}, A. Katherina and {Wechsler}, Risa H. and {Williams}, Doug and {Yang}, Jinyi and {Yang}, Qian and {Yapici}, Tolga and {Zaritsky}, Dennis and {Zenteno}, A. and {Zhang}, Kai and {Zhang}, Tianmeng and {Zhou}, Rongpu and {Zhou}, Zhimin},
        title = "{Overview of the DESI Legacy Imaging Surveys}",
      journal = {\aj},
         year = 2019,
        month = may,
       volume = {157},
       number = {5},
          eid = {168},
        pages = {168},
          doi = {10.3847/1538-3881/ab089d},
archivePrefix = {arXiv},
       eprint = {1804.08657},
 primaryClass = {astro-ph.IM},
       adsurl = {https://ui.adsabs.harvard.edu/abs/2019AJ....157..168D}
}

@ARTICLE{Diaz-Garcia2016,
       author = {{D{\'\i}az-Garc{\'\i}a}, S. and {Salo}, H. and {Laurikainen}, E. and {Herrera-Endoqui}, M.},
        title = "{Characterization of galactic bars from 3.6 {\ensuremath{\mu}}m S$^{4}$G imaging}",
      journal = {\aap},
         year = 2016,
        month = mar,
       volume = {587},
          eid = {A160},
        pages = {A160},
          doi = {10.1051/0004-6361/201526161},
archivePrefix = {arXiv},
       eprint = {1509.06743},
 primaryClass = {astro-ph.GA},
       adsurl = {https://ui.adsabs.harvard.edu/abs/2016A&A...587A.160D}
}

@ARTICLE{Disney1983,
       author = {{Disney}, M. and {Phillipps}, S.},
        title = "{The visibility of galaxies as a function of central surface brightness.}",
      journal = {\mnras},
         year = 1983,
        month = dec,
       volume = {205},
        pages = {1253-1265},
          doi = {10.1093/mnras/205.4.1253},
       adsurl = {https://ui.adsabs.harvard.edu/abs/1983MNRAS.205.1253D}
}

@ARTICLE{vanDokkum2013,
       author = {{van Dokkum}, Pieter G. and {Leja}, Joel and {Nelson}, Erica June and {Patel}, Shannon and {Skelton}, Rosalind E. and {Momcheva}, Ivelina and {Brammer}, Gabriel and {Whitaker}, Katherine E. and {Lundgren}, Britt and {Fumagalli}, Mattia and {Conroy}, Charlie and {F{\"o}rster Schreiber}, Natascha and {Franx}, Marijn and {Kriek}, Mariska and {Labb{\'e}}, Ivo and {Marchesini}, Danilo and {Rix}, Hans-Walter and {van der Wel}, Arjen and {Wuyts}, Stijn},
        title = "{The Assembly of Milky-Way-like Galaxies Since z \raisebox{-0.5ex}\textasciitilde 2.5}",
      journal = {\apjl},
         year = 2013,
        month = jul,
       volume = {771},
       number = {2},
          eid = {L35},
        pages = {L35},
          doi = {10.1088/2041-8205/771/2/L35},
archivePrefix = {arXiv},
       eprint = {1304.2391},
 primaryClass = {astro-ph.CO},
       adsurl = {https://ui.adsabs.harvard.edu/abs/2013ApJ...771L..35V}
}

@ARTICLE{Dubois2012,
       author = {{Dubois}, Yohan and {Devriendt}, Julien and {Slyz}, Adrianne and {Teyssier}, Romain},
        title = "{Self-regulated growth of supermassive black holes by a dual jet-heating active galactic nucleus feedback mechanism: methods, tests and implications for cosmological simulations}",
      journal = {\mnras},
         year = 2012,
        month = mar,
       volume = {420},
       number = {3},
        pages = {2662-2683},
          doi = {10.1111/j.1365-2966.2011.20236.x},
archivePrefix = {arXiv},
       eprint = {1108.0110},
 primaryClass = {astro-ph.CO},
       adsurl = {https://ui.adsabs.harvard.edu/abs/2012MNRAS.420.2662D}
}

@ARTICLE{Dubois2013,
       author = {{Dubois}, Yohan and {Gavazzi}, Rapha{\"e}l and {Peirani}, S{\'e}bastien and {Silk}, Joseph},
        title = "{AGN-driven quenching of star formation: morphological and dynamical implications for early-type galaxies}",
      journal = {\mnras},
         year = 2013,
        month = aug,
       volume = {433},
       number = {4},
        pages = {3297-3313},
          doi = {10.1093/mnras/stt997},
archivePrefix = {arXiv},
       eprint = {1301.3092},
 primaryClass = {astro-ph.CO},
       adsurl = {https://ui.adsabs.harvard.edu/abs/2013MNRAS.433.3297D}
}

@ARTICLE{Dubois2014,
       author = {{Dubois}, Y. and {Pichon}, C. and {Welker}, C. and {Le Borgne}, D. and {Devriendt}, J. and {Laigle}, C. and {Codis}, S. and {Pogosyan}, D. and {Arnouts}, S. and {Benabed}, K. and {Bertin}, E. and {Blaizot}, J. and {Bouchet}, F. and {Cardoso}, J. -F. and {Colombi}, S. and {de Lapparent}, V. and {Desjacques}, V. and {Gavazzi}, R. and {Kassin}, S. and {Kimm}, T. and {McCracken}, H. and {Milliard}, B. and {Peirani}, S. and {Prunet}, S. and {Rouberol}, S. and {Silk}, J. and {Slyz}, A. and {Sousbie}, T. and {Teyssier}, R. and {Tresse}, L. and {Treyer}, M. and {Vibert}, D. and {Volonteri}, M.},
        title = "{Dancing in the dark: galactic properties trace spin swings along the cosmic web}",
      journal = {\mnras},
         year = 2014,
        month = oct,
       volume = {444},
       number = {2},
        pages = {1453-1468},
          doi = {10.1093/mnras/stu1227},
archivePrefix = {arXiv},
       eprint = {1402.1165},
 primaryClass = {astro-ph.CO},
       adsurl = {https://ui.adsabs.harvard.edu/abs/2014MNRAS.444.1453D}
}

@ARTICLE{Dubois2016,
       author = {{Dubois}, Yohan and {Peirani}, S{\'e}bastien and {Pichon}, Christophe and {Devriendt}, Julien and {Gavazzi}, Rapha{\"e}l and {Welker}, Charlotte and {Volonteri}, Marta},
        title = "{The HORIZON-AGN simulation: morphological diversity of galaxies promoted by AGN feedback}",
      journal = {\mnras},
         year = 2016,
        month = dec,
       volume = {463},
       number = {4},
        pages = {3948-3964},
          doi = {10.1093/mnras/stw2265},
archivePrefix = {arXiv},
       eprint = {1606.03086},
 primaryClass = {astro-ph.GA},
       adsurl = {https://ui.adsabs.harvard.edu/abs/2016MNRAS.463.3948D}
}

@ARTICLE{Ellison2011,
       author = {{Ellison}, Sara L. and {Patton}, David R. and {Mendel}, J. Trevor and {Scudder}, Jillian M.},
        title = "{Galaxy pairs in the Sloan Digital Sky Survey - IV. Interactions trigger active galactic nuclei}",
      journal = {\mnras},
         year = 2011,
        month = dec,
       volume = {418},
       number = {3},
        pages = {2043-2053},
          doi = {10.1111/j.1365-2966.2011.19624.x},
archivePrefix = {arXiv},
       eprint = {1108.2711},
 primaryClass = {astro-ph.CO},
       adsurl = {https://ui.adsabs.harvard.edu/abs/2011MNRAS.418.2043E}
}

@ARTICLE{Ellison2019,
       author = {{Ellison}, Sara L. and {Viswanathan}, Akshara and {Patton}, David R. and {Bottrell}, Connor and {McConnachie}, Alan W. and {Gwyn}, Stephen and {Cuillandre}, Jean-Charles},
        title = "{A definitive merger-AGN connection at z {\ensuremath{\sim}} 0 with CFIS: mergers have an excess of AGN and AGN hosts are more frequently disturbed}",
      journal = {\mnras},
         year = 2019,
        month = aug,
       volume = {487},
       number = {2},
        pages = {2491-2504},
          doi = {10.1093/mnras/stz1431},
archivePrefix = {arXiv},
       eprint = {1905.08830},
 primaryClass = {astro-ph.GA},
       adsurl = {https://ui.adsabs.harvard.edu/abs/2019MNRAS.487.2491E}
}

@ARTICLE{Elmegreen2008,
       author = {{Elmegreen}, Bruce G. and {Bournaud}, Fr{\'e}d{\'e}ric and {Elmegreen}, Debra Meloy},
        title = "{Bulge Formation by the Coalescence of Giant Clumps in Primordial Disk Galaxies}",
      journal = {\apj},
         year = 2008,
        month = nov,
       volume = {688},
       number = {1},
        pages = {67-77},
          doi = {10.1086/592190},
archivePrefix = {arXiv},
       eprint = {0808.0716},
 primaryClass = {astro-ph},
       adsurl = {https://ui.adsabs.harvard.edu/abs/2008ApJ...688...67E}
}

@ARTICLE{Elvis2002,
       author = {{Elvis}, M. and {Risaliti}, G. and {Zamorani}, G.},
        title = "{Most Supermassive Black Holes Must Be Rapidly Rotating}",
      journal = {\apjl},
         year = 2002,
        month = feb,
       volume = {565},
       number = {2},
        pages = {L75-L77},
          doi = {10.1086/339197},
archivePrefix = {arXiv},
       eprint = {astro-ph/0112413},
 primaryClass = {astro-ph},
       adsurl = {https://ui.adsabs.harvard.edu/abs/2002ApJ...565L..75E}
}

@ARTICLE{Erwin2018,
       author = {{Erwin}, Peter},
        title = "{The dependence of bar frequency on galaxy mass, colour, and gas content - and angular resolution - in the local universe}",
      journal = {\mnras},
         year = 2018,
        month = mar,
       volume = {474},
       number = {4},
        pages = {5372-5392},
          doi = {10.1093/mnras/stx3117},
archivePrefix = {arXiv},
       eprint = {1711.04867},
 primaryClass = {astro-ph.GA},
       adsurl = {https://ui.adsabs.harvard.edu/abs/2018MNRAS.474.5372E}
}

@ARTICLE{Eskridge2000,
       author = {{Eskridge}, Paul B. and {Frogel}, Jay A. and {Pogge}, Richard W. and {Quillen}, Alice C. and {Davies}, Roger L. and {DePoy}, D.~L. and {Houdashelt}, Mark L. and {Kuchinski}, Leslie E. and {Ram{\'\i}rez}, Solange V. and {Sellgren}, K. and {Terndrup}, Donald M. and {Tiede}, Glenn P.},
        title = "{The Frequency of Barred Spiral Galaxies in the Near-Infrared}",
      journal = {\aj},
         year = 2000,
        month = feb,
       volume = {119},
       number = {2},
        pages = {536-544},
          doi = {10.1086/301203},
archivePrefix = {arXiv},
       eprint = {astro-ph/9910479},
 primaryClass = {astro-ph},
       adsurl = {https://ui.adsabs.harvard.edu/abs/2000AJ....119..536E}
}

@ARTICLE{Fahey2025,
       author = {{Fahey}, Matthew J. and {Garland}, Izzy L. and {Simmons}, Brooke D. and {Keel}, William C. and {Shanahan}, Jesse and {Coil}, Alison and {Glikman}, Eilat and {Lintott}, Chris J. and {Masters}, Karen L. and {Moran}, Ed and {Smethurst}, Rebecca J. and {G{\'e}ron}, Tobias and {Thorne}, Matthew R.},
        title = "{Structural decomposition of merger-free galaxies hosting luminous AGNs}",
      journal = {\mnras},
         year = 2025,
        month = mar,
       volume = {537},
       number = {4},
        pages = {3511-3524},
          doi = {10.1093/mnras/staf239},
archivePrefix = {arXiv},
       eprint = {2410.22404},
 primaryClass = {astro-ph.GA},
       adsurl = {https://ui.adsabs.harvard.edu/abs/2025MNRAS.537.3511F}
}

@ARTICLE{Ferrarese2000,
       author = {{Ferrarese}, Laura and {Merritt}, David},
        title = "{A Fundamental Relation between Supermassive Black Holes and Their Host Galaxies}",
      journal = {\apjl},
         year = 2000,
        month = aug,
       volume = {539},
       number = {1},
        pages = {L9-L12},
          doi = {10.1086/312838},
archivePrefix = {arXiv},
       eprint = {astro-ph/0006053},
 primaryClass = {astro-ph},
       adsurl = {https://ui.adsabs.harvard.edu/abs/2000ApJ...539L...9F}
}

@ARTICLE{Ferreira2022,
       author = {{Ferreira}, Leonardo and {Adams}, Nathan and {Conselice}, Christopher J. and {Sazonova}, Elizaveta and {Austin}, Duncan and {Caruana}, Joseph and {Ferrari}, Fabricio and {Verma}, Aprajita and {Trussler}, James and {Broadhurst}, Tom and {Diego}, Jose and {Frye}, Brenda L. and {Pascale}, Massimo and {Wilkins}, Stephen M. and {Windhorst}, Rogier A. and {Zitrin}, Adi},
        title = "{Panic! at the Disks: First Rest-frame Optical Observations of Galaxy Structure at z > 3 with JWST in the SMACS 0723 Field}",
      journal = {\apjl},
         year = 2022,
        month = oct,
       volume = {938},
       number = {1},
          eid = {L2},
        pages = {L2},
          doi = {10.3847/2041-8213/ac947c},
archivePrefix = {arXiv},
       eprint = {2207.09428},
 primaryClass = {astro-ph.GA},
       adsurl = {https://ui.adsabs.harvard.edu/abs/2022ApJ...938L...2F}
}

@article{Ferrers1877,
  author = {Ferrers, N. M.},
  title  = {On the potentials of certain ellipsoids (and related results)},
  journal= {Quarterly Journal of Pure and Applied Mathematics},
  volume = {14},
  pages  = {1--22},
  year   = {1877}
}

@ARTICLE{Fisher2008,
       author = {{Fisher}, David B. and {Drory}, Niv},
        title = "{The Structure of Classical Bulges and Pseudobulges: the Link Between Pseudobulges and S{\'E}RSIC Index}",
      journal = {\aj},
         year = 2008,
        month = aug,
       volume = {136},
       number = {2},
        pages = {773-839},
          doi = {10.1088/0004-6256/136/2/773},
archivePrefix = {arXiv},
       eprint = {0805.4206},
 primaryClass = {astro-ph},
       adsurl = {https://ui.adsabs.harvard.edu/abs/2008AJ....136..773F}
}

@ARTICLE{Flaugher2015,
       author = {{Flaugher}, B. and {Diehl}, H.~T. and {Honscheid}, K. and {Abbott}, T.~M.~C. and {Alvarez}, O. and {Angstadt}, R. and {Annis}, J.~T. and {Antonik}, M. and {Ballester}, O. and {Beaufore}, L. and {Bernstein}, G.~M. and {Bernstein}, R.~A. and {Bigelow}, B. and {Bonati}, M. and {Boprie}, D. and {Brooks}, D. and {Buckley-Geer}, E.~J. and {Campa}, J. and {Cardiel-Sas}, L. and {Castander}, F.~J. and {Castilla}, J. and {Cease}, H. and {Cela-Ruiz}, J.~M. and {Chappa}, S. and {Chi}, E. and {Cooper}, C. and {da Costa}, L.~N. and {Dede}, E. and {Derylo}, G. and {DePoy}, D.~L. and {de Vicente}, J. and {Doel}, P. and {Drlica-Wagner}, A. and {Eiting}, J. and {Elliott}, A.~E. and {Emes}, J. and {Estrada}, J. and {Fausti Neto}, A. and {Finley}, D.~A. and {Flores}, R. and {Frieman}, J. and {Gerdes}, D. and {Gladders}, M.~D. and {Gregory}, B. and {Gutierrez}, G.~R. and {Hao}, J. and {Holland}, S.~E. and {Holm}, S. and {Huffman}, D. and {Jackson}, C. and {James}, D.~J. and {Jonas}, M. and {Karcher}, A. and {Karliner}, I. and {Kent}, S. and {Kessler}, R. and {Kozlovsky}, M. and {Kron}, R.~G. and {Kubik}, D. and {Kuehn}, K. and {Kuhlmann}, S. and {Kuk}, K. and {Lahav}, O. and {Lathrop}, A. and {Lee}, J. and {Levi}, M.~E. and {Lewis}, P. and {Li}, T.~S. and {Mandrichenko}, I. and {Marshall}, J.~L. and {Martinez}, G. and {Merritt}, K.~W. and {Miquel}, R. and {Mu{\~n}oz}, F. and {Neilsen}, E.~H. and {Nichol}, R.~C. and {Nord}, B. and {Ogando}, R. and {Olsen}, J. and {Palaio}, N. and {Patton}, K. and {Peoples}, J. and {Plazas}, A.~A. and {Rauch}, J. and {Reil}, K. and {Rheault}, J. -P. and {Roe}, N.~A. and {Rogers}, H. and {Roodman}, A. and {Sanchez}, E. and {Scarpine}, V. and {Schindler}, R.~H. and {Schmidt}, R. and {Schmitt}, R. and {Schubnell}, M. and {Schultz}, K. and {Schurter}, P. and {Scott}, L. and {Serrano}, S. and {Shaw}, T.~M. and {Smith}, R.~C. and {Soares-Santos}, M. and {Stefanik}, A. and {Stuermer}, W. and {Suchyta}, E. and {Sypniewski}, A. and {Tarle}, G. and {Thaler}, J. and {Tighe}, R. and {Tran}, C. and {Tucker}, D. and {Walker}, A.~R. and {Wang}, G. and {Watson}, M. and {Weaverdyck}, C. and {Wester}, W. and {Woods}, R. and {Yanny}, B. and {DES Collaboration}},
        title = "{The Dark Energy Camera}",
      journal = {\aj},
         year = 2015,
        month = nov,
       volume = {150},
       number = {5},
          eid = {150},
        pages = {150},
          doi = {10.1088/0004-6256/150/5/150},
archivePrefix = {arXiv},
       eprint = {1504.02900},
 primaryClass = {astro-ph.IM},
       adsurl = {https://ui.adsabs.harvard.edu/abs/2015AJ....150..150F}
}

@ARTICLE{Fritz2006,
       author = {{Fritz}, J. and {Franceschini}, A. and {Hatziminaoglou}, E.},
        title = "{Revisiting the infrared spectra of active galactic nuclei with a new torus emission model}",
      journal = {\mnras},
         year = 2006,
        month = mar,
       volume = {366},
       number = {3},
        pages = {767-786},
          doi = {10.1111/j.1365-2966.2006.09866.x},
archivePrefix = {arXiv},
       eprint = {astro-ph/0511428},
 primaryClass = {astro-ph},
       adsurl = {https://ui.adsabs.harvard.edu/abs/2006MNRAS.366..767F}
}

@ARTICLE{Furtak2024,
       author = {{Furtak}, Lukas J. and {Labb{\'e}}, Ivo and {Zitrin}, Adi and {Greene}, Jenny E. and {Dayal}, Pratika and {Chemerynska}, Iryna and {Kokorev}, Vasily and {Miller}, Tim B. and {Goulding}, Andy D. and {de Graaff}, Anna and {Bezanson}, Rachel and {Brammer}, Gabriel B. and {Cutler}, Sam E. and {Leja}, Joel and {Pan}, Richard and {Price}, Sedona H. and {Wang}, Bingjie and {Weaver}, John R. and {Whitaker}, Katherine E. and {Atek}, Hakim and {Bogd{\'a}n}, {\'A}kos and {Charlot}, St{\'e}phane and {Curtis-Lake}, Emma and {van Dokkum}, Pieter and {Endsley}, Ryan and {Feldmann}, Robert and {Fudamoto}, Yoshinobu and {Fujimoto}, Seiji and {Glazebrook}, Karl and {Juneau}, St{\'e}phanie and {Marchesini}, Danilo and {Maseda}, Micheal V. and {Nelson}, Erica and {Oesch}, Pascal A. and {Plat}, Ad{\`e}le and {Setton}, David J. and {Stark}, Daniel P. and {Williams}, Christina C.},
        title = "{A high black-hole-to-host mass ratio in a lensed AGN in the early Universe}",
      journal = {\nat},
         year = 2024,
        month = apr,
       volume = {628},
       number = {8006},
        pages = {57-61},
          doi = {10.1038/s41586-024-07184-8},
archivePrefix = {arXiv},
       eprint = {2308.05735},
 primaryClass = {astro-ph.GA},
       adsurl = {https://ui.adsabs.harvard.edu/abs/2024Natur.628...57F}
}

@ARTICLE{Gabor2009,
       author = {{Gabor}, J.~M. and {Impey}, C.~D. and {Jahnke}, K. and {Simmons}, B.~D. and {Trump}, J.~R. and {Koekemoer}, A.~M. and {Brusa}, M. and {Cappelluti}, N. and {Schinnerer}, E. and {Smol{\v{c}}i{\'c}}, V. and {Salvato}, M. and {Rhodes}, J.~D. and {Mobasher}, B. and {Capak}, P. and {Massey}, R. and {Leauthaud}, A. and {Scoville}, N.},
        title = "{Active Galactic Nucleus Host Galaxy Morphologies in COSMOS}",
      journal = {\apj},
         year = 2009,
        month = jan,
       volume = {691},
       number = {1},
        pages = {705-722},
          doi = {10.1088/0004-637X/691/1/705},
archivePrefix = {arXiv},
       eprint = {0809.0309},
 primaryClass = {astro-ph},
       adsurl = {https://ui.adsabs.harvard.edu/abs/2009ApJ...691..705G}
}

@ARTICLE{Gadotti2008,
       author = {{Gadotti}, Dimitri Alexei},
        title = "{Image decomposition of barred galaxies and AGN hosts}",
      journal = {\mnras},
         year = 2008,
        month = feb,
       volume = {384},
       number = {1},
        pages = {420-439},
          doi = {10.1111/j.1365-2966.2007.12723.x},
archivePrefix = {arXiv},
       eprint = {0708.3870},
 primaryClass = {astro-ph},
       adsurl = {https://ui.adsabs.harvard.edu/abs/2008MNRAS.384..420G}
}

@ARTICLE{Gadotti2020,
       author = {{Gadotti}, Dimitri A. and {Bittner}, Adrian and {Falc{\'o}n-Barroso}, Jes{\'u}s and {M{\'e}ndez-Abreu}, Jairo and {Kim}, Taehyun and {Fragkoudi}, Francesca and {de Lorenzo-C{\'a}ceres}, Adriana and {Leaman}, Ryan and {Neumann}, Justus and {Querejeta}, Miguel and {S{\'a}nchez-Bl{\'a}zquez}, Patricia and {Martig}, Marie and {Mart{\'\i}n-Navarro}, Ignacio and {P{\'e}rez}, Isabel and {Seidel}, Marja K. and {van de Ven}, Glenn},
        title = "{Kinematic signatures of nuclear discs and bar-driven secular evolution in nearby galaxies of the MUSE TIMER project}",
      journal = {\aap},
         year = 2020,
        month = nov,
       volume = {643},
          eid = {A14},
        pages = {A14},
          doi = {10.1051/0004-6361/202038448},
archivePrefix = {arXiv},
       eprint = {2009.01852},
 primaryClass = {astro-ph.GA},
       adsurl = {https://ui.adsabs.harvard.edu/abs/2020A&A...643A..14G}
}

@ARTICLE{Galloway2015,
       author = {{Galloway}, Melanie A. and {Willett}, Kyle W. and {Fortson}, Lucy F. and {Cardamone}, Carolin N. and {Schawinski}, Kevin and {Cheung}, Edmond and {Lintott}, Chris J. and {Masters}, Karen L. and {Melvin}, Thomas and {Simmons}, Brooke D.},
        title = "{Galaxy Zoo: the effect of bar-driven fuelling on the presence of an active galactic nucleus in disc galaxies}",
      journal = {\mnras},
         year = 2015,
        month = apr,
       volume = {448},
       number = {4},
        pages = {3442-3454},
          doi = {10.1093/mnras/stv235},
archivePrefix = {arXiv},
       eprint = {1502.01033},
 primaryClass = {astro-ph.GA},
       adsurl = {https://ui.adsabs.harvard.edu/abs/2015MNRAS.448.3442G}
}

@ARTICLE{Garland2023,
       author = {{Garland}, Izzy L. and {Fahey}, Matthew J. and {Simmons}, Brooke D. and {Smethurst}, Rebecca J. and {Lintott}, Chris J. and {Shanahan}, Jesse and {Silcock}, Maddie S. and {Smith}, Joshua and {Keel}, William C. and {Coil}, Alison and {G{\'e}ron}, Tobias and {Kruk}, Sandor and {Masters}, Karen L. and {O'Ryan}, David and {Thorne}, Matthew R. and {Wiersema}, Klaas},
        title = "{The most luminous, merger-free AGNs show only marginal correlation with bar presence}",
      journal = {\mnras},
         year = 2023,
        month = jun,
       volume = {522},
       number = {1},
        pages = {211-225},
          doi = {10.1093/mnras/stad966},
archivePrefix = {arXiv},
       eprint = {2304.01260},
 primaryClass = {astro-ph.GA},
       adsurl = {https://ui.adsabs.harvard.edu/abs/2023MNRAS.522..211G}
}

@ARTICLE{Garland2024,
       author = {{Garland}, Izzy L. and {Walmsley}, Mike and {Silcock}, Maddie S. and {Potts}, Leah M. and {Smith}, Josh and {Simmons}, Brooke D. and {Lintott}, Chris J. and {Smethurst}, Rebecca J. and {Dawson}, James M. and {Keel}, William C. and {Kruk}, Sandor and {Mantha}, Kameswara Bharadwaj and {Masters}, Karen L. and {O'Ryan}, David and {Popp}, J{\"u}rgen J. and {Thorne}, Matthew R.},
        title = "{Galaxy Zoo DESI: large-scale bars as a secular mechanism for triggering AGNs}",
      journal = {\mnras},
         year = 2024,
        month = aug,
       volume = {532},
       number = {2},
        pages = {2320-2330},
          doi = {10.1093/mnras/stae1620},
archivePrefix = {arXiv},
       eprint = {2406.20096},
 primaryClass = {astro-ph.GA},
       adsurl = {https://ui.adsabs.harvard.edu/abs/2024MNRAS.532.2320G}
}

@ARTICLE{Gaskell1988,
       author = {{Gaskell}, C. Martin},
        title = "{Direct Evidence for Gravitational Domination of the Motion of Gas within One Light-Week of the Central Object in NGC 4151 and the Determination of the Mass of the Probable Black Hole}",
      journal = {\apj},
         year = 1988,
        month = feb,
       volume = {325},
        pages = {114},
          doi = {10.1086/165986},
       adsurl = {https://ui.adsabs.harvard.edu/abs/1988ApJ...325..114G}
}

@ARTICLE{Geron2021,
       author = {{G{\'e}ron}, Tobias and {Smethurst}, R.~J. and {Lintott}, Chris and {Kruk}, Sandor and {Masters}, Karen L. and {Simmons}, Brooke and {Stark}, David V.},
        title = "{Galaxy zoo: stronger bars facilitate quenching in star-forming galaxies}",
      journal = {\mnras},
         year = 2021,
        month = nov,
       volume = {507},
       number = {3},
        pages = {4389-4408},
          doi = {10.1093/mnras/stab2064},
archivePrefix = {arXiv},
       eprint = {2107.06913},
 primaryClass = {astro-ph.GA},
       adsurl = {https://ui.adsabs.harvard.edu/abs/2021MNRAS.507.4389G}
}

@ARTICLE{Geron2023,
       author = {{G{\'e}ron}, Tobias and {Smethurst}, Rebecca J. and {Lintott}, Chris and {Kruk}, Sandor and {Masters}, Karen L. and {Simmons}, Brooke and {Mantha}, Kameswara Bharadwaj and {Walmsley}, Mike and {Garma-Oehmichen}, L. and {Drory}, Niv and {Lane}, Richard R.},
        title = "{Galaxy Zoo: kinematics of strongly and weakly barred galaxies}",
      journal = {\mnras},
         year = 2023,
        month = may,
       volume = {521},
       number = {2},
        pages = {1775-1793},
          doi = {10.1093/mnras/stad501},
archivePrefix = {arXiv},
       eprint = {2302.05464},
 primaryClass = {astro-ph.GA},
       adsurl = {https://ui.adsabs.harvard.edu/abs/2023MNRAS.521.1775G}
}

@ARTICLE{Goulding2017,
       author = {{Goulding}, A.~D. and {Matthaey}, E. and {Greene}, J.~E. and {Hickox}, R.~C. and {Alexander}, D.~M. and {Forman}, W.~R. and {Jones}, C. and {Lehmer}, B.~D. and {Griffis}, S. and {Kanek}, S. and {Oulmakki}, M.},
        title = "{Galaxy-scale Bars in Late-type Sloan Digital Sky Survey Galaxies Do Not Influence the Average Accretion Rates of Supermassive Black Holes}",
      journal = {\apj},
         year = 2017,
        month = jul,
       volume = {843},
       number = {2},
          eid = {135},
        pages = {135},
          doi = {10.3847/1538-4357/aa755b},
archivePrefix = {arXiv},
       eprint = {1705.08895},
 primaryClass = {astro-ph.GA},
       adsurl = {https://ui.adsabs.harvard.edu/abs/2017ApJ...843..135G}
}

@ARTICLE{Goulding2023,
       author = {{Goulding}, Andy D. and {Greene}, Jenny E. and {Setton}, David J. and {Labbe}, Ivo and {Bezanson}, Rachel and {Miller}, Tim B. and {Atek}, Hakim and {Bogd{\'a}n}, {\'A}kos and {Brammer}, Gabriel and {Chemerynska}, Iryna and {Cutler}, Sam E. and {Dayal}, Pratika and {Fudamoto}, Yoshinobu and {Fujimoto}, Seiji and {Furtak}, Lukas J. and {Kokorev}, Vasily and {Khullar}, Gourav and {Leja}, Joel and {Marchesini}, Danilo and {Natarajan}, Priyamvada and {Nelson}, Erica and {Oesch}, Pascal A. and {Pan}, Richard and {Papovich}, Casey and {Price}, Sedona H. and {van Dokkum}, Pieter and {Wang}, Bingjie and {Weaver}, John R. and {Whitaker}, Katherine E. and {Zitrin}, Adi},
        title = "{UNCOVER: The Growth of the First Massive Black Holes from JWST/NIRSpec-Spectroscopic Redshift Confirmation of an X-Ray Luminous AGN at z = 10.1}",
      journal = {\apjl},
         year = 2023,
        month = sep,
       volume = {955},
       number = {1},
          eid = {L24},
        pages = {L24},
          doi = {10.3847/2041-8213/acf7c5},
archivePrefix = {arXiv},
       eprint = {2308.02750},
 primaryClass = {astro-ph.GA},
       adsurl = {https://ui.adsabs.harvard.edu/abs/2023ApJ...955L..24G}
}

@ARTICLE{GRAVITY2024,
       author = {{GRAVITY Collaboration} and {Amorim}, A. and {Bourdarot}, G. and {Brandner}, W. and {Cao}, Y. and {Cl{\'e}net}, Y. and {Davies}, R. and {de Zeeuw}, P.~T. and {Dexter}, J. and {Drescher}, A. and {Eckart}, A. and {Eisenhauer}, F. and {Fabricius}, M. and {Feuchtgruber}, H. and {F{\"o}rster Schreiber}, N.~M. and {Garcia}, P.~J.~V. and {Genzel}, R. and {Gillessen}, S. and {Gratadour}, D. and {H{\"o}nig}, S. and {Kishimoto}, M. and {Lacour}, S. and {Lutz}, D. and {Millour}, F. and {Netzer}, H. and {Ott}, T. and {Paumard}, T. and {Perraut}, K. and {Perrin}, G. and {Peterson}, B.~M. and {Petrucci}, P.~O. and {Pfuhl}, O. and {Prieto}, M.~A. and {Rabien}, S. and {Rouan}, D. and {Santos}, D.~J.~D. and {Shangguan}, J. and {Shimizu}, T. and {Sternberg}, A. and {Straubmeier}, C. and {Sturm}, E. and {Tacconi}, L.~J. and {Tristram}, K.~R.~W. and {Widmann}, F. and {Woillez}, J.},
        title = "{The size-luminosity relation of local active galactic nuclei from interferometric observations of the broad-line region}",
      journal = {\aap},
         year = 2024,
        month = apr,
       volume = {684},
          eid = {A167},
        pages = {A167},
          doi = {10.1051/0004-6361/202348167},
archivePrefix = {arXiv},
       eprint = {2401.07676},
 primaryClass = {astro-ph.GA},
       adsurl = {https://ui.adsabs.harvard.edu/abs/2024A&A...684A.167G}
}

@ARTICLE{Greene2005,
       author = {{Greene}, Jenny E. and {Ho}, Luis C.},
        title = "{Estimating Black Hole Masses in Active Galaxies Using the H{\ensuremath{\alpha}} Emission Line}",
      journal = {\apj},
         year = 2005,
        month = sep,
       volume = {630},
       number = {1},
        pages = {122-129},
          doi = {10.1086/431897},
archivePrefix = {arXiv},
       eprint = {astro-ph/0508335},
 primaryClass = {astro-ph},
       adsurl = {https://ui.adsabs.harvard.edu/abs/2005ApJ...630..122G}
}

@ARTICLE{Greene2007,
       author = {{Greene}, Jenny E. and {Ho}, Luis C.},
        title = "{The Mass Function of Active Black Holes in the Local Universe}",
      journal = {\apj},
         year = 2007,
        month = sep,
       volume = {667},
       number = {1},
        pages = {131-148},
          doi = {10.1086/520497},
archivePrefix = {arXiv},
       eprint = {0705.0020},
 primaryClass = {astro-ph},
       adsurl = {https://ui.adsabs.harvard.edu/abs/2007ApJ...667..131G}
}

@ARTICLE{Greene2020,
       author = {{Greene}, Jenny E. and {Strader}, Jay and {Ho}, Luis C.},
        title = "{Intermediate-Mass Black Holes}",
      journal = {\araa},
         year = 2020,
        month = aug,
       volume = {58},
        pages = {257-312},
          doi = {10.1146/annurev-astro-032620-021835},
archivePrefix = {arXiv},
       eprint = {1911.09678},
 primaryClass = {astro-ph.GA},
       adsurl = {https://ui.adsabs.harvard.edu/abs/2020ARA&A..58..257G}
}

@ARTICLE{Grier2013,
       author = {{Grier}, C.~J. and {Martini}, P. and {Watson}, L.~C. and {Peterson}, B.~M. and {Bentz}, M.~C. and {Dasyra}, K.~M. and {Dietrich}, M. and {Ferrarese}, L. and {Pogge}, R.~W. and {Zu}, Y.},
        title = "{Stellar Velocity Dispersion Measurements in High-luminosity Quasar Hosts and Implications for the AGN Black Hole Mass Scale}",
      journal = {\apj},
         year = 2013,
        month = aug,
       volume = {773},
       number = {2},
          eid = {90},
        pages = {90},
          doi = {10.1088/0004-637X/773/2/90},
archivePrefix = {arXiv},
       eprint = {1305.2447},
 primaryClass = {astro-ph.CO},
       adsurl = {https://ui.adsabs.harvard.edu/abs/2013ApJ...773...90G}
}

@ARTICLE{Grier2017,
       author = {{Grier}, C.~J. and {Pancoast}, A. and {Barth}, A.~J. and {Fausnaugh}, M.~M. and {Brewer}, B.~J. and {Treu}, T. and {Peterson}, B.~M.},
        title = "{The Structure of the Broad-line Region in Active Galactic Nuclei. II. Dynamical Modeling of Data From the AGN10 Reverberation Mapping Campaign}",
      journal = {\apj},
         year = 2017,
        month = nov,
       volume = {849},
       number = {2},
          eid = {146},
        pages = {146},
          doi = {10.3847/1538-4357/aa901b},
archivePrefix = {arXiv},
       eprint = {1705.02346},
 primaryClass = {astro-ph.GA},
       adsurl = {https://ui.adsabs.harvard.edu/abs/2017ApJ...849..146G}
}

@ARTICLE{Gutierrez2017,
       author = {{Guti{\'e}rrez}, Claudia P. and {Anderson}, Joseph P. and {Hamuy}, Mario and {Morrell}, Nidia and {Gonz{\'a}lez-Gaitan}, Santiago and {Stritzinger}, Maximilian D. and {Phillips}, Mark M. and {Galbany}, Lluis and {Folatelli}, Gast{\'o}n and {Dessart}, Luc and {Contreras}, Carlos and {Della Valle}, Massimo and {Freedman}, Wendy L. and {Hsiao}, Eric Y. and {Krisciunas}, Kevin and {Madore}, Barry F. and {Maza}, Jos{\'e} and {Suntzeff}, Nicholas B. and {Prieto}, Jose Luis and {Gonz{\'a}lez}, Luis and {Cappellaro}, Enrico and {Navarrete}, Mauricio and {Pizzella}, Alessandro and {Ruiz}, Maria T. and {Smith}, R. Chris and {Turatto}, Massimo},
        title = "{Type II Supernova Spectral Diversity. I. Observations, Sample Characterization, and Spectral Line Evolution}",
      journal = {\apj},
         year = 2017,
        month = nov,
       volume = {850},
       number = {1},
          eid = {89},
        pages = {89},
          doi = {10.3847/1538-4357/aa8f52},
archivePrefix = {arXiv},
       eprint = {1709.02487},
 primaryClass = {astro-ph.HE},
       adsurl = {https://ui.adsabs.harvard.edu/abs/2017ApJ...850...89G}
}

@ARTICLE{Habouzit2021,
       author = {{Habouzit}, M{\'e}lanie and {Li}, Yuan and {Somerville}, Rachel S. and {Genel}, Shy and {Pillepich}, Annalisa and {Volonteri}, Marta and {Dav{\'e}}, Romeel and {Rosas-Guevara}, Yetli and {McAlpine}, Stuart and {Peirani}, S{\'e}bastien and {Hernquist}, Lars and {Angl{\'e}s-Alc{\'a}zar}, Daniel and {Reines}, Amy and {Bower}, Richard and {Dubois}, Yohan and {Nelson}, Dylan and {Pichon}, Christophe and {Vogelsberger}, Mark},
        title = "{Supermassive black holes in cosmological simulations I: M$_{BH}$ - M$_{{\ensuremath{\star}}}$ relation and black hole mass function}",
      journal = {\mnras},
         year = 2021,
        month = may,
       volume = {503},
       number = {2},
        pages = {1940-1975},
          doi = {10.1093/mnras/stab496},
archivePrefix = {arXiv},
       eprint = {2006.10094},
 primaryClass = {astro-ph.GA},
       adsurl = {https://ui.adsabs.harvard.edu/abs/2021MNRAS.503.1940H}
}

@ARTICLE{Harikane2023,
       author = {{Harikane}, Yuichi and {Zhang}, Yechi and {Nakajima}, Kimihiko and {Ouchi}, Masami and {Isobe}, Yuki and {Ono}, Yoshiaki and {Hatano}, Shun and {Xu}, Yi and {Umeda}, Hiroya},
        title = "{A JWST/NIRSpec First Census of Broad-line AGNs at z = 4-7: Detection of 10 Faint AGNs with M $_{BH}$ {}10$^{6}$-{}10$^{8}$ M $_{{\ensuremath{\odot}}}$ and Their Host Galaxy Properties}",
      journal = {\apj},
         year = 2023,
        month = dec,
       volume = {959},
       number = {1},
          eid = {39},
        pages = {39},
          doi = {10.3847/1538-4357/ad029e},
archivePrefix = {arXiv},
       eprint = {2303.11946},
 primaryClass = {astro-ph.GA},
       adsurl = {https://ui.adsabs.harvard.edu/abs/2023ApJ...959...39H}
}

@ARTICLE{Haring2004,
       author = {{H{\"a}ring}, Nadine and {Rix}, Hans-Walter},
        title = "{On the Black Hole Mass-Bulge Mass Relation}",
      journal = {\apjl},
         year = 2004,
        month = apr,
       volume = {604},
       number = {2},
        pages = {L89-L92},
          doi = {10.1086/383567},
archivePrefix = {arXiv},
       eprint = {astro-ph/0402376},
 primaryClass = {astro-ph},
       adsurl = {https://ui.adsabs.harvard.edu/abs/2004ApJ...604L..89H}
}

@ARTICLE{Haslbauer2022,
       author = {{Haslbauer}, Moritz and {Banik}, Indranil and {Kroupa}, Pavel and {Wittenburg}, Nils and {Javanmardi}, Behnam},
        title = "{The High Fraction of Thin Disk Galaxies Continues to Challenge {\ensuremath{\Lambda}}CDM Cosmology}",
      journal = {\apj},
         year = 2022,
        month = feb,
       volume = {925},
       number = {2},
          eid = {183},
        pages = {183},
          doi = {10.3847/1538-4357/ac46ac},
archivePrefix = {arXiv},
       eprint = {2202.01221},
 primaryClass = {astro-ph.GA},
       adsurl = {https://ui.adsabs.harvard.edu/abs/2022ApJ...925..183H}
}

@ARTICLE{Haussler2013,
       author = {{H{\"a}u{\ss}ler}, Boris and {Bamford}, Steven P. and {Vika}, Marina and {Rojas}, Alex L. and {Barden}, Marco and {Kelvin}, Lee S. and {Alpaslan}, Mehmet and {Robotham}, Aaron S.~G. and {Driver}, Simon P. and {Baldry}, I.~K. and {Brough}, Sarah and {Hopkins}, Andrew M. and {Liske}, Jochen and {Nichol}, Robert C. and {Popescu}, Cristina C. and {Tuffs}, Richard J.},
        title = "{MegaMorph - multiwavelength measurement of galaxy structure: complete S{\'e}rsic profile information from modern surveys}",
      journal = {\mnras},
         year = 2013,
        month = mar,
       volume = {430},
       number = {1},
        pages = {330-369},
          doi = {10.1093/mnras/sts633},
archivePrefix = {arXiv},
       eprint = {1212.3332},
 primaryClass = {astro-ph.CO},
       adsurl = {https://ui.adsabs.harvard.edu/abs/2013MNRAS.430..330H}
}

@ARTICLE{Heckman2004,
       author = {{Heckman}, Timothy M. and {Kauffmann}, Guinevere and {Brinchmann}, Jarle and {Charlot}, St{\'e}phane and {Tremonti}, Christy and {White}, Simon D.~M.},
        title = "{Present-Day Growth of Black Holes and Bulges: The Sloan Digital Sky Survey Perspective}",
      journal = {\apj},
         year = 2004,
        month = sep,
       volume = {613},
       number = {1},
        pages = {109-118},
          doi = {10.1086/422872},
archivePrefix = {arXiv},
       eprint = {astro-ph/0406218},
 primaryClass = {astro-ph},
       adsurl = {https://ui.adsabs.harvard.edu/abs/2004ApJ...613..109H}
}

@ARTICLE{Heckman2014,
       author = {{Heckman}, Timothy M. and {Best}, Philip N.},
        title = "{The Coevolution of Galaxies and Supermassive Black Holes: Insights from Surveys of the Contemporary Universe}",
      journal = {\araa},
         year = 2014,
        month = aug,
       volume = {52},
        pages = {589-660},
          doi = {10.1146/annurev-astro-081913-035722},
archivePrefix = {arXiv},
       eprint = {1403.4620},
 primaryClass = {astro-ph.GA},
       adsurl = {https://ui.adsabs.harvard.edu/abs/2014ARA&A..52..589H}
}

@ARTICLE{Hopkins2010,
       author = {{Hopkins}, Philip F. and {Bundy}, Kevin and {Croton}, Darren and {Hernquist}, Lars and {Keres}, Dusan and {Khochfar}, Sadegh and {Stewart}, Kyle and {Wetzel}, Andrew and {Younger}, Joshua D.},
        title = "{Mergers and Bulge Formation in {\ensuremath{\Lambda}}CDM: Which Mergers Matter?}",
      journal = {\apj},
         year = 2010,
        month = may,
       volume = {715},
       number = {1},
        pages = {202-229},
          doi = {10.1088/0004-637X/715/1/202},
archivePrefix = {arXiv},
       eprint = {0906.5357},
 primaryClass = {astro-ph.CO},
       adsurl = {https://ui.adsabs.harvard.edu/abs/2010ApJ...715..202H}
}

@ARTICLE{Hopkins2010a,
       author = {{Hopkins}, Philip F. and {Quataert}, Eliot},
        title = "{How do massive black holes get their gas?}",
      journal = {\mnras},
         year = 2010,
        month = sep,
       volume = {407},
       number = {3},
        pages = {1529-1564},
          doi = {10.1111/j.1365-2966.2010.17064.x},
archivePrefix = {arXiv},
       eprint = {0912.3257},
 primaryClass = {astro-ph.CO},
       adsurl = {https://ui.adsabs.harvard.edu/abs/2010MNRAS.407.1529H}
}

@ARTICLE{Huertas-Company2024,
       author = {{Huertas-Company}, M. and {Iyer}, K.~G. and {Angeloudi}, E. and {Bagley}, M.~B. and {Finkelstein}, S.~L. and {Kartaltepe}, J. and {McGrath}, E.~J. and {Sarmiento}, R. and {Vega-Ferrero}, J. and {Arrabal Haro}, P. and {Behroozi}, P. and {Buitrago}, F. and {Cheng}, Y. and {Costantin}, L. and {Dekel}, A. and {Dickinson}, M. and {Elbaz}, D. and {Grogin}, N.~A. and {Hathi}, N.~P. and {Holwerda}, B.~W. and {Koekemoer}, A.~M. and {Lucas}, R.~A. and {Papovich}, C. and {P{\'e}rez-Gonz{\'a}lez}, P.~G. and {Pirzkal}, N. and {Seill{\'e}}, L.-M. and {de la Vega}, A. and {Wuyts}, S. and {Yang}, G. and {Yung}, L.~Y.~A.},
        title = "{Galaxy morphology from z {\ensuremath{\sim}} 6 through the lens of JWST}",
      journal = {\aap},
         year = 2024,
        month = may,
       volume = {685},
          eid = {A48},
        pages = {A48},
          doi = {10.1051/0004-6361/202346800},
archivePrefix = {arXiv},
       eprint = {2305.02478},
 primaryClass = {astro-ph.GA},
       adsurl = {https://ui.adsabs.harvard.edu/abs/2024A&A...685A..48H}
}

@ARTICLE{Hutchinson-Smith2026,
       author = {{Hutchinson-Smith}, Tenley and {Simmons}, Brooke D. and {Masters}, Karen L. and {Coil}, Alison and {Garland}, Izzy and {G{\'e}ron}, Tobias and {Kruk}, Sandor and {Lintott}, Chris and {Smethurst}, Rebecca and {Tapia}, Amauri and {Willett}, Kyle and {Baeten}, Elisabeth and {Beer}, Sylvia and {Peck}, Michael L. and {Wilcox}, Julianne},
        title = "{Galaxy Zoo Bar Lengths: A Catalogue of Measurements from Hubble Space Telescope Images and the Evolution of Galactic Bar Structure at z < 1}",
      journal = {arXiv e-prints},
         year = 2026,
        month = apr,
          eid = {arXiv:2604.27100},
        pages = {arXiv:2604.27100},
          doi = {10.48550/arXiv.2604.27100},
archivePrefix = {arXiv},
       eprint = {2604.27100},
 primaryClass = {astro-ph.GA},
       adsurl = {https://ui.adsabs.harvard.edu/abs/2026arXiv260427100H}
}

@ARTICLE{Ilbert2010,
       author = {{Ilbert}, O. and {Salvato}, M. and {Le Floc'h}, E. and {Aussel}, H. and {Capak}, P. and {McCracken}, H.~J. and {Mobasher}, B. and {Kartaltepe}, J. and {Scoville}, N. and {Sanders}, D.~B. and {Arnouts}, S. and {Bundy}, K. and {Cassata}, P. and {Kneib}, J.-P. and {Koekemoer}, A. and {Le F{\`e}vre}, O. and {Lilly}, S. and {Surace}, J. and {Taniguchi}, Y. and {Tasca}, L. and {Thompson}, D. and {Tresse}, L. and {Zamojski}, M. and {Zamorani}, G. and {Zucca}, E.},
        title = "{Galaxy Stellar Mass Assembly Between 0.2 < z < 2 from the S-COSMOS Survey}",
      journal = {\apj},
         year = 2010,
        month = feb,
       volume = {709},
       number = {2},
        pages = {644-663},
          doi = {10.1088/0004-637X/709/2/644},
archivePrefix = {arXiv},
       eprint = {0903.0102},
 primaryClass = {astro-ph.CO},
       adsurl = {https://ui.adsabs.harvard.edu/abs/2010ApJ...709..644I}
}

@ARTICLE{Inoue2011,
       author = {{Inoue}, Akio K.},
        title = "{Rest-frame ultraviolet-to-optical spectral characteristics of extremely metal-poor and metal-free galaxies}",
      journal = {\mnras},
         year = 2011,
        month = aug,
       volume = {415},
       number = {3},
        pages = {2920-2931},
          doi = {10.1111/j.1365-2966.2011.18906.x},
archivePrefix = {arXiv},
       eprint = {1102.5150},
 primaryClass = {astro-ph.CO},
       adsurl = {https://ui.adsabs.harvard.edu/abs/2011MNRAS.415.2920I}
}

@INPROCEEDINGS{deJong1996,
       author = {{de Jong}, R.~S.},
        title = "{Colour Gradients in the Optical and Near-IR}",
    booktitle = {Spiral Galaxies in the Near-IR},
         year = 1996,
       editor = {{Minniti}, Dante and {Rix}, Hans-Walter},
        month = jan,
        pages = {43},
          doi = {10.1007/978-3-540-49739-4_6},
archivePrefix = {arXiv},
       eprint = {astro-ph/9509001},
 primaryClass = {astro-ph},
       adsurl = {https://ui.adsabs.harvard.edu/abs/1996sgni.conf...43D}
}

@ARTICLE{Kannappan2004,
       author = {{Kannappan}, Sheila J.},
        title = "{Linking Gas Fractions to Bimodalities in Galaxy Properties}",
      journal = {\apjl},
         year = 2004,
        month = aug,
       volume = {611},
       number = {2},
        pages = {L89-L92},
          doi = {10.1086/423785},
archivePrefix = {arXiv},
       eprint = {astro-ph/0405136},
 primaryClass = {astro-ph},
       adsurl = {https://ui.adsabs.harvard.edu/abs/2004ApJ...611L..89K}
}

@ARTICLE{Kartaltepe2023,
       author = {{Kartaltepe}, Jeyhan S. and {Rose}, Caitlin and {Vanderhoof}, Brittany N. and {McGrath}, Elizabeth J. and {Costantin}, Luca and {Cox}, Isabella G. and {Yung}, L.~Y. Aaron and {Kocevski}, Dale D. and {Wuyts}, Stijn and {Ferguson}, Henry C. and {Bagley}, Micaela B. and {Finkelstein}, Steven L. and {Amor{\'\i}n}, Ricardo O. and {Andrews}, Brett H. and {Arrabal Haro}, Pablo and {Backhaus}, Bren E. and {Behroozi}, Peter and {Bisigello}, Laura and {Calabr{\`o}}, Antonello and {Casey}, Caitlin M. and {Coogan}, Rosemary T. and {Cooper}, M.~C. and {Croton}, Darren and {de la Vega}, Alexander and {Dickinson}, Mark and {Fontana}, Adriano and {Franco}, Maximilien and {Grazian}, Andrea and {Grogin}, Norman A. and {Hathi}, Nimish P. and {Holwerda}, Benne W. and {Huertas-Company}, Marc and {Iyer}, Kartheik G. and {Jogee}, Shardha and {Jung}, Intae and {Kewley}, Lisa J. and {Kirkpatrick}, Allison and {Koekemoer}, Anton M. and {Liu}, James and {Lotz}, Jennifer M. and {Lucas}, Ray A. and {Newman}, Jeffrey A. and {Pacifici}, Camilla and {Pandya}, Viraj and {Papovich}, Casey and {Pentericci}, Laura and {P{\'e}rez-Gonz{\'a}lez}, Pablo G. and {Petersen}, Jayse and {Pirzkal}, Nor and {Rafelski}, Marc and {Ravindranath}, Swara and {Simons}, Raymond C. and {Snyder}, Gregory F. and {Somerville}, Rachel S. and {Stanway}, Elizabeth R. and {Straughn}, Amber N. and {Tacchella}, Sandro and {Trump}, Jonathan R. and {Vega-Ferrero}, Jes{\'u}s and {Wilkins}, Stephen M. and {Yang}, Guang and {Zavala}, Jorge A.},
        title = "{CEERS Key Paper. III. The Diversity of Galaxy Structure and Morphology at z = 3-9 with JWST}",
      journal = {\apjl},
         year = 2023,
        month = mar,
       volume = {946},
       number = {1},
          eid = {L15},
        pages = {L15},
          doi = {10.3847/2041-8213/acad01},
archivePrefix = {arXiv},
       eprint = {2210.14713},
 primaryClass = {astro-ph.GA},
       adsurl = {https://ui.adsabs.harvard.edu/abs/2023ApJ...946L..15K}
}

@ARTICLE{Kauffmann2003,
       author = {{Kauffmann}, Guinevere and {Heckman}, Timothy M. and {White}, Simon D.~M. and {Charlot}, St{\'e}phane and {Tremonti}, Christy and {Peng}, Eric W. and {Seibert}, Mark and {Brinkmann}, Jon and {Nichol}, Robert C. and {SubbaRao}, Mark and {York}, Don},
        title = "{The dependence of star formation history and internal structure on stellar mass for {}10$^{5}$ low-redshift galaxies}",
      journal = {\mnras},
         year = 2003,
        month = may,
       volume = {341},
       number = {1},
        pages = {54-69},
          doi = {10.1046/j.1365-8711.2003.06292.x},
archivePrefix = {arXiv},
       eprint = {astro-ph/0205070},
 primaryClass = {astro-ph},
       adsurl = {https://ui.adsabs.harvard.edu/abs/2003MNRAS.341...54K}
}

@ARTICLE{Kauffmann2004,
       author = {{Kauffmann}, Guinevere and {White}, Simon D.~M. and {Heckman}, Timothy M. and {M{\'e}nard}, Brice and {Brinchmann}, Jarle and {Charlot}, St{\'e}phane and {Tremonti}, Christy and {Brinkmann}, Jon},
        title = "{The environmental dependence of the relations between stellar mass, structure, star formation and nuclear activity in galaxies}",
      journal = {\mnras},
         year = 2004,
        month = sep,
       volume = {353},
       number = {3},
        pages = {713-731},
          doi = {10.1111/j.1365-2966.2004.08117.x},
archivePrefix = {arXiv},
       eprint = {astro-ph/0402030},
 primaryClass = {astro-ph},
       adsurl = {https://ui.adsabs.harvard.edu/abs/2004MNRAS.353..713K}
}

@ARTICLE{Kaviraj2017,
       author = {{Kaviraj}, S. and {Laigle}, C. and {Kimm}, T. and {Devriendt}, J.~E.~G. and {Dubois}, Y. and {Pichon}, C. and {Slyz}, A. and {Chisari}, E. and {Peirani}, S.},
        title = "{The Horizon-AGN simulation: evolution of galaxy properties over cosmic time}",
      journal = {\mnras},
         year = 2017,
        month = jun,
       volume = {467},
       number = {4},
        pages = {4739-4752},
          doi = {10.1093/mnras/stx126},
archivePrefix = {arXiv},
       eprint = {1605.09379},
 primaryClass = {astro-ph.GA},
       adsurl = {https://ui.adsabs.harvard.edu/abs/2017MNRAS.467.4739K}
}

@ARTICLE{Kelly2007,
       author = {{Kelly}, Brandon C.},
        title = "{Some Aspects of Measurement Error in Linear Regression of Astronomical Data}",
      journal = {\apj},
         year = 2007,
        month = aug,
       volume = {665},
       number = {2},
        pages = {1489-1506},
          doi = {10.1086/519947},
archivePrefix = {arXiv},
       eprint = {0705.2774},
 primaryClass = {astro-ph},
       adsurl = {https://ui.adsabs.harvard.edu/abs/2007ApJ...665.1489K}
}

@ARTICLE{Keres2005,
       author = {{Kere{\v{s}}}, Du{\v{s}}an and {Katz}, Neal and {Weinberg}, David H. and {Dav{\'e}}, Romeel},
        title = "{How do galaxies get their gas?}",
      journal = {\mnras},
         year = 2005,
        month = oct,
       volume = {363},
       number = {1},
        pages = {2-28},
          doi = {10.1111/j.1365-2966.2005.09451.x},
archivePrefix = {arXiv},
       eprint = {astro-ph/0407095},
 primaryClass = {astro-ph},
       adsurl = {https://ui.adsabs.harvard.edu/abs/2005MNRAS.363....2K}
}

@ARTICLE{Kewley2006,
       author = {{Kewley}, Lisa J. and {Groves}, Brent and {Kauffmann}, Guinevere and {Heckman}, Tim},
        title = "{The host galaxies and classification of active galactic nuclei}",
      journal = {\mnras},
         year = 2006,
        month = nov,
       volume = {372},
       number = {3},
        pages = {961-976},
          doi = {10.1111/j.1365-2966.2006.10859.x},
archivePrefix = {arXiv},
       eprint = {astro-ph/0605681},
 primaryClass = {astro-ph},
       adsurl = {https://ui.adsabs.harvard.edu/abs/2006MNRAS.372..961K}
}

@ARTICLE{Kim2008a,
       author = {{Kim}, Minjin and {Ho}, Luis C. and {Peng}, Chien Y. and {Barth}, Aaron J. and {Im}, Myungshin},
        title = "{Decomposition of the Host Galaxies of Active Galactic Nuclei Using Hubble Space Telescope Images}",
      journal = {\apjs},
         year = 2008,
        month = dec,
       volume = {179},
       number = {2},
        pages = {283-305},
          doi = {10.1086/591796},
archivePrefix = {arXiv},
       eprint = {0807.1334},
 primaryClass = {astro-ph},
       adsurl = {https://ui.adsabs.harvard.edu/abs/2008ApJS..179..283K}
}

@ARTICLE{Kim2014,
       author = {{Kim}, Yonghwi and {Kim}, Woong-Tae},
        title = "{Gaseous spiral structure and mass drift in spiral galaxies}",
      journal = {\mnras},
         year = 2014,
        month = may,
       volume = {440},
       number = {1},
        pages = {208-224},
          doi = {10.1093/mnras/stu276},
archivePrefix = {arXiv},
       eprint = {1402.2291},
 primaryClass = {astro-ph.GA},
       adsurl = {https://ui.adsabs.harvard.edu/abs/2014MNRAS.440..208K}
}

@ARTICLE{Kim2015,
       author = {{Kim}, Taehyun and {Sheth}, Kartik and {Gadotti}, Dimitri A. and {Lee}, Myung Gyoon and {Zaritsky}, Dennis and {Elmegreen}, Bruce G. and {Athanassoula}, E. and {Bosma}, Albert and {Holwerda}, Benne and {Ho}, Luis C. and {Comer{\'o}n}, S{\'e}bastien and {Knapen}, Johan H. and {Hinz}, Joannah L. and {Mu{\~n}oz-Mateos}, Juan-Carlos and {Erroz-Ferrer}, Santiago and {Buta}, Ronald J. and {Kim}, Minjin and {Laurikainen}, Eija and {Salo}, Heikki and {Madore}, Barry F. and {Laine}, Jarkko and {Men{\'e}ndez-Delmestre}, Kar{\'\i}n and {Regan}, Michael W. and {de Swardt}, Bonita and {Gil de Paz}, Armando and {Seibert}, Mark and {Mizusawa}, Trisha},
        title = "{The Mass Profile and Shape of Bars in the Spitzer Survey of Stellar Structure in Galaxies (S$^{4}$G): Search for an Age Indicator for Bars}",
      journal = {\apj},
         year = 2015,
        month = jan,
       volume = {799},
       number = {1},
          eid = {99},
        pages = {99},
          doi = {10.1088/0004-637X/799/1/99},
archivePrefix = {arXiv},
       eprint = {1411.4650},
 primaryClass = {astro-ph.GA},
       adsurl = {https://ui.adsabs.harvard.edu/abs/2015ApJ...799...99K}
}

@ARTICLE{Kokorev2023,
       author = {{Kokorev}, Vasily and {Fujimoto}, Seiji and {Labbe}, Ivo and {Greene}, Jenny E. and {Bezanson}, Rachel and {Dayal}, Pratika and {Nelson}, Erica J. and {Atek}, Hakim and {Brammer}, Gabriel and {Caputi}, Karina I. and {Chemerynska}, Iryna and {Cutler}, Sam E. and {Feldmann}, Robert and {Fudamoto}, Yoshinobu and {Furtak}, Lukas J. and {Goulding}, Andy D. and {de Graaff}, Anna and {Leja}, Joel and {Marchesini}, Danilo and {Miller}, Tim B. and {Nanayakkara}, Themiya and {Oesch}, Pascal A. and {Pan}, Richard and {Price}, Sedona H. and {Setton}, David J. and {Smit}, Renske and {Stefanon}, Mauro and {Wang}, Bingjie and {Weaver}, John R. and {Whitaker}, Katherine E. and {Williams}, Christina C. and {Zitrin}, Adi},
        title = "{UNCOVER: A NIRSpec Identification of a Broad-line AGN at z = 8.50}",
      journal = {\apjl},
         year = 2023,
        month = nov,
       volume = {957},
       number = {1},
          eid = {L7},
        pages = {L7},
          doi = {10.3847/2041-8213/ad037a},
archivePrefix = {arXiv},
       eprint = {2308.11610},
 primaryClass = {astro-ph.GA},
       adsurl = {https://ui.adsabs.harvard.edu/abs/2023ApJ...957L...7K}
}

@INPROCEEDINGS{Kormendy2001,
       author = {{Kormendy}, John and {Gebhardt}, Karl},
        title = "{Supermassive black holes in galactic nuclei}",
    booktitle = {20th Texas Symposium on relativistic astrophysics},
         year = 2001,
       editor = {{Wheeler}, J. Craig and {Martel}, Hugo},
       series = {American Institute of Physics Conference Series},
       volume = {586},
        month = oct,
    publisher = {AIP},
        pages = {363-381},
          doi = {10.1063/1.1419581},
archivePrefix = {arXiv},
       eprint = {astro-ph/0105230},
 primaryClass = {astro-ph},
       adsurl = {https://ui.adsabs.harvard.edu/abs/2001AIPC..586..363K}
}

@ARTICLE{Kormendy2004,
       author = {{Kormendy}, John and {Kennicutt}, Jr., Robert C.},
        title = "{Secular Evolution and the Formation of Pseudobulges in Disk Galaxies}",
      journal = {\araa},
         year = 2004,
        month = sep,
       volume = {42},
       number = {1},
        pages = {603-683},
          doi = {10.1146/annurev.astro.42.053102.134024},
archivePrefix = {arXiv},
       eprint = {astro-ph/0407343},
 primaryClass = {astro-ph},
       adsurl = {https://ui.adsabs.harvard.edu/abs/2004ARA&A..42..603K}
}

@ARTICLE{Kraljic2012,
       author = {{Kraljic}, Katarina and {Bournaud}, Fr{\'e}d{\'e}ric and {Martig}, Marie},
        title = "{The Two-phase Formation History of Spiral Galaxies Traced by the Cosmic Evolution of the Bar Fraction}",
      journal = {\apj},
         year = 2012,
        month = sep,
       volume = {757},
       number = {1},
          eid = {60},
        pages = {60},
          doi = {10.1088/0004-637X/757/1/60},
archivePrefix = {arXiv},
       eprint = {1207.0351},
 primaryClass = {astro-ph.GA},
       adsurl = {https://ui.adsabs.harvard.edu/abs/2012ApJ...757...60K}
}

@ARTICLE{Kruk2018,
       author = {{Kruk}, Sandor J. and {Lintott}, Chris J. and {Bamford}, Steven P. and {Masters}, Karen L. and {Simmons}, Brooke D. and {H{\"a}u{\ss}ler}, Boris and {Cardamone}, Carolin N. and {Hart}, Ross E. and {Kelvin}, Lee and {Schawinski}, Kevin and {Smethurst}, Rebecca J. and {Vika}, Marina},
        title = "{Galaxy Zoo: secular evolution of barred galaxies from structural decomposition of multiband images}",
      journal = {\mnras},
         year = 2018,
        month = feb,
       volume = {473},
       number = {4},
        pages = {4731-4753},
          doi = {10.1093/mnras/stx2605},
archivePrefix = {arXiv},
       eprint = {1710.00093},
 primaryClass = {astro-ph.GA},
       adsurl = {https://ui.adsabs.harvard.edu/abs/2018MNRAS.473.4731K}
}

@ARTICLE{Kubryk2013,
       author = {{Kubryk}, M. and {Prantzos}, N. and {Athanassoula}, E.},
        title = "{Radial migration in a bar-dominated disc galaxy - I. Impact on chemical evolution}",
      journal = {\mnras},
         year = 2013,
        month = dec,
       volume = {436},
       number = {2},
        pages = {1479-1491},
          doi = {10.1093/mnras/stt1667},
       adsurl = {https://ui.adsabs.harvard.edu/abs/2013MNRAS.436.1479K}
}

@ARTICLE{LaMarca2025,
       author = {{La Marca}, A. and {Nardone}, M.~T. and {Wang}, L. and {Margalef-Bentabol}, B. and {Kruk}, S. and {Trager}, S.~C.},
        title = "{Galactic bars and active galactic nucleus fuelling in the second half of cosmic history}",
      journal = {arXiv e-prints},
         year = 2025,
        month = oct,
          eid = {arXiv:2510.23522},
        pages = {arXiv:2510.23522},
          doi = {10.48550/arXiv.2510.23522},
archivePrefix = {arXiv},
       eprint = {2510.23522},
 primaryClass = {astro-ph.GA},
       adsurl = {https://ui.adsabs.harvard.edu/abs/2025arXiv251023522L}
}

@ARTICLE{Land2008,
       author = {{Land}, Kate and {Slosar}, An{\v{z}}e and {Lintott}, Chris and {Andreescu}, Dan and {Bamford}, Steven and {Murray}, Phil and {Nichol}, Robert and {Raddick}, M. Jordan and {Schawinski}, Kevin and {Szalay}, Alex and {Thomas}, Daniel and {Vandenberg}, Jan},
        title = "{Galaxy Zoo: the large-scale spin statistics of spiral galaxies in the Sloan Digital Sky Survey}",
      journal = {\mnras},
         year = 2008,
        month = aug,
       volume = {388},
       number = {4},
        pages = {1686-1692},
          doi = {10.1111/j.1365-2966.2008.13490.x},
archivePrefix = {arXiv},
       eprint = {0803.3247},
 primaryClass = {astro-ph},
       adsurl = {https://ui.adsabs.harvard.edu/abs/2008MNRAS.388.1686L}
}

@ARTICLE{Lin2013,
       author = {{Lin}, Lien-Hsuan and {Wang}, Hsiang-Hsu and {Hsieh}, Pei-Ying and {Taam}, Ronald E. and {Yang}, Chao-Chin and {Yen}, David C.~C.},
        title = "{Hydrodynamical Simulations of the Barred Spiral Galaxy NGC 1097}",
      journal = {\apj},
         year = 2013,
        month = jul,
       volume = {771},
       number = {1},
          eid = {8},
        pages = {8},
          doi = {10.1088/0004-637X/771/1/8},
archivePrefix = {arXiv},
       eprint = {1305.1963},
 primaryClass = {astro-ph.CO},
       adsurl = {https://ui.adsabs.harvard.edu/abs/2013ApJ...771....8L}
}

@ARTICLE{Liu2019,
       author = {{Liu}, He-Yang and {Liu}, Wen-Juan and {Dong}, Xiao-Bo and {Zhou}, Hongyan and {Wang}, Tinggui and {Lu}, Honglin and {Yuan}, Weimin},
        title = "{A Comprehensive and Uniform Sample of Broad-line Active Galactic Nuclei from the SDSS DR7}",
      journal = {\apjs},
         year = 2019,
        month = aug,
       volume = {243},
       number = {2},
          eid = {21},
        pages = {21},
          doi = {10.3847/1538-4365/ab298b},
archivePrefix = {arXiv},
       eprint = {1906.05597},
 primaryClass = {astro-ph.GA},
       adsurl = {https://ui.adsabs.harvard.edu/abs/2019ApJS..243...21L}
}

@ARTICLE{Liu2024,
       author = {{Liu}, H.~T. and {Feng}, Hai-Cheng and {Li}, Sha-Sha and {Bai}, J.~M. and {Li}, H.~Z.},
        title = "{Measuring the Virial Factor in SDSS DR7 Active Galactic Nuclei with Redshifted H{\ensuremath{\beta}} and H{\ensuremath{\alpha}} Broad Emission Lines}",
      journal = {\apj},
         year = 2024,
        month = mar,
       volume = {963},
       number = {1},
          eid = {30},
        pages = {30},
          doi = {10.3847/1538-4357/ad1ab8},
       adsurl = {https://ui.adsabs.harvard.edu/abs/2024ApJ...963...30L}
}

@ARTICLE{Lopez-Coba2020,
       author = {{L{\'o}pez-Cob{\'a}}, Carlos and {S{\'a}nchez}, Sebasti{\'a}n F. and {Anderson}, Joseph P. and {Cruz-Gonz{\'a}lez}, Irene and {Galbany}, Llu{\'\i}s and {Ruiz-Lara}, Tom{\'a}s and {Barrera-Ballesteros}, Jorge K. and {Prieto}, Jos{\'e} L. and {Kuncarayakti}, Hanindyo},
        title = "{The AMUSING++ Nearby Galaxy Compilation. I. Full Sample Characterization and Galactic-scale Outflow Selection}",
      journal = {\aj},
         year = 2020,
        month = apr,
       volume = {159},
       number = {4},
          eid = {167},
        pages = {167},
          doi = {10.3847/1538-3881/ab7848},
archivePrefix = {arXiv},
       eprint = {2002.09328},
 primaryClass = {astro-ph.GA},
       adsurl = {https://ui.adsabs.harvard.edu/abs/2020AJ....159..167L}
}

@ARTICLE{Maciejewski2002,
       author = {{Maciejewski}, Witold and {Teuben}, Peter J. and {Sparke}, Linda S. and {Stone}, James M.},
        title = "{Gas inflow in barred galaxies - effects of secondary bars}",
      journal = {\mnras},
         year = 2002,
        month = jan,
       volume = {329},
       number = {3},
        pages = {502-512},
          doi = {10.1046/j.1365-8711.2002.04957.x},
archivePrefix = {arXiv},
       eprint = {astro-ph/0109431},
 primaryClass = {astro-ph},
       adsurl = {https://ui.adsabs.harvard.edu/abs/2002MNRAS.329..502M}
}

@ARTICLE{Maciejewski2004,
       author = {{Maciejewski}, Witold},
        title = "{Nuclear spirals in galaxies: gas response to an asymmetric potential - II. Hydrodynamical models}",
      journal = {\mnras},
         year = 2004,
        month = nov,
       volume = {354},
       number = {3},
        pages = {892-904},
          doi = {10.1111/j.1365-2966.2004.08254.x},
archivePrefix = {arXiv},
       eprint = {astro-ph/0408100},
 primaryClass = {astro-ph},
       adsurl = {https://ui.adsabs.harvard.edu/abs/2004MNRAS.354..892M}
}

@ARTICLE{Magorrian1998,
       author = {{Magorrian}, John and {Tremaine}, Scott and {Richstone}, Douglas and {Bender}, Ralf and {Bower}, Gary and {Dressler}, Alan and {Faber}, S.~M. and {Gebhardt}, Karl and {Green}, Richard and {Grillmair}, Carl and {Kormendy}, John and {Lauer}, Tod},
        title = "{The Demography of Massive Dark Objects in Galaxy Centers}",
      journal = {\aj},
         year = 1998,
        month = jun,
       volume = {115},
       number = {6},
        pages = {2285-2305},
          doi = {10.1086/300353},
archivePrefix = {arXiv},
       eprint = {astro-ph/9708072},
 primaryClass = {astro-ph},
       adsurl = {https://ui.adsabs.harvard.edu/abs/1998AJ....115.2285M}
}

@ARTICLE{Maiolino2024,
       author = {{Maiolino}, Roberto and {Scholtz}, Jan and {Curtis-Lake}, Emma and {Carniani}, Stefano and {Baker}, William and {de Graaff}, Anna and {Tacchella}, Sandro and {{\"U}bler}, Hannah and {D'Eugenio}, Francesco and {Witstok}, Joris and {Curti}, Mirko and {Arribas}, Santiago and {Bunker}, Andrew J. and {Charlot}, St{\'e}phane and {Chevallard}, Jacopo and {Eisenstein}, Daniel J. and {Egami}, Eiichi and {Ji}, Zhiyuan and {Jones}, Gareth C. and {Lyu}, Jianwei and {Rawle}, Tim and {Robertson}, Brant and {Rujopakarn}, Wiphu and {Perna}, Michele and {Sun}, Fengwu and {Venturi}, Giacomo and {Williams}, Christina C. and {Willott}, Chris},
        title = "{JADES: The diverse population of infant black holes at 4 < z < 11: Merging, tiny, poor, but mighty}",
      journal = {\aap},
         year = 2024,
        month = nov,
       volume = {691},
          eid = {A145},
        pages = {A145},
          doi = {10.1051/0004-6361/202347640},
archivePrefix = {arXiv},
       eprint = {2308.01230},
 primaryClass = {astro-ph.GA},
       adsurl = {https://ui.adsabs.harvard.edu/abs/2024A&A...691A.145M}
}

@ARTICLE{Malkan2026,
       author = {{Malkan}, Matthew A. and {Jensen}, Lisbeth D. and {Hao}, Lei},
        title = "{The Narrow Emission Lines of Seyfert 1 Galaxies: Comparisons with a Large SDSS Sample}",
      journal = {\apj},
         year = 2026,
        month = feb,
       volume = {998},
       number = {1},
          eid = {165},
        pages = {165},
          doi = {10.3847/1538-4357/ae232e},
archivePrefix = {arXiv},
       eprint = {2512.10186},
 primaryClass = {astro-ph.GA},
       adsurl = {https://ui.adsabs.harvard.edu/abs/2026ApJ...998..165M}
}

@ARTICLE{Marconi2003,
       author = {{Marconi}, Alessandro and {Hunt}, Leslie K.},
        title = "{The Relation between Black Hole Mass, Bulge Mass, and Near-Infrared Luminosity}",
      journal = {\apjl},
         year = 2003,
        month = may,
       volume = {589},
       number = {1},
        pages = {L21-L24},
          doi = {10.1086/375804},
archivePrefix = {arXiv},
       eprint = {astro-ph/0304274},
 primaryClass = {astro-ph},
       adsurl = {https://ui.adsabs.harvard.edu/abs/2003ApJ...589L..21M}
}

@ARTICLE{Marinova2007,
       author = {{Marinova}, Irina and {Jogee}, Shardha},
        title = "{Characterizing Bars at z \raisebox{-0.5ex}\textasciitilde 0 in the Optical and NIR: Implications for the Evolution of Barred Disks with Redshift}",
      journal = {\apj},
         year = 2007,
        month = apr,
       volume = {659},
       number = {2},
        pages = {1176-1197},
          doi = {10.1086/512355},
archivePrefix = {arXiv},
       eprint = {astro-ph/0608039},
 primaryClass = {astro-ph},
       adsurl = {https://ui.adsabs.harvard.edu/abs/2007ApJ...659.1176M}
}

@ARTICLE{Martin2018,
       author = {{Martin}, G. and {Kaviraj}, S. and {Volonteri}, M. and {Simmons}, B.~D. and {Devriendt}, J.~E.~G. and {Lintott}, C.~J. and {Smethurst}, R.~J. and {Dubois}, Y. and {Pichon}, C.},
        title = "{Normal black holes in bulge-less galaxies: the largely quiescent, merger-free growth of black holes over cosmic time}",
      journal = {\mnras},
         year = 2018,
        month = may,
       volume = {476},
       number = {2},
        pages = {2801-2812},
          doi = {10.1093/mnras/sty324},
archivePrefix = {arXiv},
       eprint = {1801.09699},
 primaryClass = {astro-ph.GA},
       adsurl = {https://ui.adsabs.harvard.edu/abs/2018MNRAS.476.2801M}
}

@ARTICLE{Martin2018b,
       author = {{Martin}, G. and {Kaviraj}, S. and {Devriendt}, J.~E.~G. and {Dubois}, Y. and {Pichon}, C.},
        title = "{The role of mergers in driving morphological transformation over cosmic time}",
      journal = {\mnras},
         year = 2018,
        month = oct,
       volume = {480},
       number = {2},
        pages = {2266-2283},
          doi = {10.1093/mnras/sty1936},
archivePrefix = {arXiv},
       eprint = {1807.08761},
 primaryClass = {astro-ph.GA},
       adsurl = {https://ui.adsabs.harvard.edu/abs/2018MNRAS.480.2266M}
}

@article{Massey1951,
author = {{Massey}, Frank J. Jr.},
title = {The Kolmogorov-Smirnov Test for Goodness of Fit},
journal = {Journal of the American Statistical Association},
volume = {46},
number = {253},
pages = {68--78},
year = {1951},
publisher = {Taylor \& Francis},
doi = {10.1080/01621459.1951.10500769},
URL = {https://www.tandfonline.com/doi/abs/10.1080/01621459.1951.10500769},
eprint = {https://www.tandfonline.com/doi/pdf/10.1080/01621459.1951.10500769}
}

@ARTICLE{Masters2011,
       author = {{Masters}, Karen L. and {Nichol}, Robert C. and {Hoyle}, Ben and {Lintott}, Chris and {Bamford}, Steven P. and {Edmondson}, Edward M. and {Fortson}, Lucy and {Keel}, William C. and {Schawinski}, Kevin and {Smith}, Arfon M. and {Thomas}, Daniel},
        title = "{Galaxy Zoo: bars in disc galaxies}",
      journal = {\mnras},
         year = 2011,
        month = mar,
       volume = {411},
       number = {3},
        pages = {2026-2034},
          doi = {10.1111/j.1365-2966.2010.17834.x},
archivePrefix = {arXiv},
       eprint = {1003.0449},
 primaryClass = {astro-ph.CO},
       adsurl = {https://ui.adsabs.harvard.edu/abs/2011MNRAS.411.2026M}
}

@ARTICLE{McAlpine2020,
       author = {{McAlpine}, Stuart and {Harrison}, Chris M. and {Rosario}, David J. and {Alexander}, David M. and {Ellison}, Sara L. and {Johansson}, Peter H. and {Patton}, David R.},
        title = "{Galaxy mergers in EAGLE do not induce a significant amount of black hole growth yet do increase the rate of luminous AGN}",
      journal = {\mnras},
         year = 2020,
        month = jun,
       volume = {494},
       number = {4},
        pages = {5713-5733},
          doi = {10.1093/mnras/staa1123},
archivePrefix = {arXiv},
       eprint = {2002.00959},
 primaryClass = {astro-ph.GA},
       adsurl = {https://ui.adsabs.harvard.edu/abs/2020MNRAS.494.5713M}
}

@ARTICLE{McCarthy2010,
       author = {{McCarthy}, I.~G. and {Schaye}, J. and {Ponman}, T.~J. and {Bower}, R.~G. and {Booth}, C.~M. and {Dalla Vecchia}, C. and {Crain}, R.~A. and {Springel}, V. and {Theuns}, T. and {Wiersma}, R.~P.~C.},
        title = "{The case for AGN feedback in galaxy groups}",
      journal = {\mnras},
         year = 2010,
        month = aug,
       volume = {406},
       number = {2},
        pages = {822-839},
          doi = {10.1111/j.1365-2966.2010.16750.x},
archivePrefix = {arXiv},
       eprint = {0911.2641},
 primaryClass = {astro-ph.CO},
       adsurl = {https://ui.adsabs.harvard.edu/abs/2010MNRAS.406..822M}
}

@ARTICLE{McConnell2013,
       author = {{McConnell}, Nicholas J. and {Ma}, Chung-Pei},
        title = "{Revisiting the Scaling Relations of Black Hole Masses and Host Galaxy Properties}",
      journal = {\apj},
         year = 2013,
        month = feb,
       volume = {764},
       number = {2},
          eid = {184},
        pages = {184},
          doi = {10.1088/0004-637X/764/2/184},
archivePrefix = {arXiv},
       eprint = {1211.2816},
 primaryClass = {astro-ph.CO},
       adsurl = {https://ui.adsabs.harvard.edu/abs/2013ApJ...764..184M}
}

@ARTICLE{McLure2002,
       author = {{McLure}, R.~J. and {Dunlop}, J.~S.},
        title = "{On the black hole-bulge mass relation in active and inactive galaxies}",
      journal = {\mnras},
         year = 2002,
        month = apr,
       volume = {331},
       number = {3},
        pages = {795-804},
          doi = {10.1046/j.1365-8711.2002.05236.x},
archivePrefix = {arXiv},
       eprint = {astro-ph/0108417},
 primaryClass = {astro-ph},
       adsurl = {https://ui.adsabs.harvard.edu/abs/2002MNRAS.331..795M}
}

@ARTICLE{Menendez-Delmestre2007,
       author = {{Men{\'e}ndez-Delmestre}, Kar{\'\i}n and {Sheth}, Kartik and {Schinnerer}, Eva and {Jarrett}, Thomas H. and {Scoville}, Nick Z.},
        title = "{A Near-Infrared Study of 2MASS Bars in Local Galaxies: An Anchor for High-Redshift Studies}",
      journal = {\apj},
         year = 2007,
        month = mar,
       volume = {657},
       number = {2},
        pages = {790-804},
          doi = {10.1086/511025},
archivePrefix = {arXiv},
       eprint = {astro-ph/0611540},
 primaryClass = {astro-ph},
       adsurl = {https://ui.adsabs.harvard.edu/abs/2007ApJ...657..790M}
}

@ARTICLE{Merloni2010,
       author = {{Merloni}, A. and {Bongiorno}, A. and {Bolzonella}, M. and {Brusa}, M. and {Civano}, F. and {Comastri}, A. and {Elvis}, M. and {Fiore}, F. and {Gilli}, R. and {Hao}, H. and {Jahnke}, K. and {Koekemoer}, A.~M. and {Lusso}, E. and {Mainieri}, V. and {Mignoli}, M. and {Miyaji}, T. and {Renzini}, A. and {Salvato}, M. and {Silverman}, J. and {Trump}, J. and {Vignali}, C. and {Zamorani}, G. and {Capak}, P. and {Lilly}, S.~J. and {Sanders}, D. and {Taniguchi}, Y. and {Bardelli}, S. and {Carollo}, C.~M. and {Caputi}, K. and {Contini}, T. and {Coppa}, G. and {Cucciati}, O. and {de la Torre}, S. and {de Ravel}, L. and {Franzetti}, P. and {Garilli}, B. and {Hasinger}, G. and {Impey}, C. and {Iovino}, A. and {Iwasawa}, K. and {Kampczyk}, P. and {Kneib}, J. -P. and {Knobel}, C. and {Kova{\v{c}}}, K. and {Lamareille}, F. and {Le Borgne}, J. -F. and {Le Brun}, V. and {Le F{\`e}vre}, O. and {Maier}, C. and {Pello}, R. and {Peng}, Y. and {Perez Montero}, E. and {Ricciardelli}, E. and {Scodeggio}, M. and {Tanaka}, M. and {Tasca}, L.~A.~M. and {Tresse}, L. and {Vergani}, D. and {Zucca}, E.},
        title = "{On the Cosmic Evolution of the Scaling Relations Between Black Holes and Their Host Galaxies: Broad-Line Active Galactic Nuclei in the zCOSMOS Survey}",
      journal = {\apj},
         year = 2010,
        month = jan,
       volume = {708},
       number = {1},
        pages = {137-157},
          doi = {10.1088/0004-637X/708/1/137},
archivePrefix = {arXiv},
       eprint = {0910.4970},
 primaryClass = {astro-ph.CO},
       adsurl = {https://ui.adsabs.harvard.edu/abs/2010ApJ...708..137M}
}

@ARTICLE{Moffat1969,
       author = {{Moffat}, A.~F.~J.},
        title = "{A Theoretical Investigation of Focal Stellar Images in the Photographic Emulsion and Application to Photographic Photometry}",
      journal = {\aap},
         year = 1969,
        month = dec,
       volume = {3},
        pages = {455},
       adsurl = {https://ui.adsabs.harvard.edu/abs/1969A&A.....3..455M}
}

@misc{Moustakas2023,
  author = {Moustakas, John and Buhler, Jeremy and Scholte, Dirk and Dey, Biprateep and Khederlarian, Ashod},
  title = {FastSpecFit: Fast spectral synthesis and emission-line fitting of DESI spectra},
  year = {2023},
  month = {August},
  note = {Astrophysics Source Code Library, record ascl:2308.005},
  url = {https://ui.adsabs.harvard.edu/abs/2023ascl.soft08005M}
}

@ARTICLE{Moffett2015,
       author = {{Moffett}, Amanda J. and {Kannappan}, Sheila J. and {Berlind}, Andreas A. and {Eckert}, Kathleen D. and {Stark}, David V. and {Hendel}, David and {Norris}, Mark A. and {Grogin}, Norman A.},
        title = "{ECO and RESOLVE: Galaxy Disk Growth in Environmental Context}",
      journal = {\apj},
         year = 2015,
        month = oct,
       volume = {812},
       number = {2},
          eid = {89},
        pages = {89},
          doi = {10.1088/0004-637X/812/2/89},
archivePrefix = {arXiv},
       eprint = {1508.00948},
 primaryClass = {astro-ph.GA},
       adsurl = {https://ui.adsabs.harvard.edu/abs/2015ApJ...812...89M}
}

@ARTICLE{Moffett2016,
       author = {{Moffett}, Amanda J. and {Ingarfield}, Stephen A. and {Driver}, Simon P. and {Robotham}, Aaron S.~G. and {Kelvin}, Lee S. and {Lange}, Rebecca and {Me{\v{s}}tri{\'c}}, Uro{\v{s}} and {Alpaslan}, Mehmet and {Baldry}, Ivan K. and {Bland-Hawthorn}, Joss and {Brough}, Sarah and {Cluver}, Michelle E. and {Davies}, Luke J.~M. and {Holwerda}, Benne W. and {Hopkins}, Andrew M. and {Kafle}, Prajwal R. and {Kennedy}, Rebecca and {Norberg}, Peder and {Taylor}, Edward N.},
        title = "{Galaxy And Mass Assembly (GAMA): the stellar mass budget by galaxy type}",
      journal = {\mnras},
         year = 2016,
        month = apr,
       volume = {457},
       number = {2},
        pages = {1308-1319},
          doi = {10.1093/mnras/stv2883},
archivePrefix = {arXiv},
       eprint = {1512.02342},
 primaryClass = {astro-ph.GA},
       adsurl = {https://ui.adsabs.harvard.edu/abs/2016MNRAS.457.1308M}
}

@ARTICLE{Oh2012,
       author = {{Oh}, Sree and {Oh}, Kyuseok and {Yi}, Sukyoung K.},
        title = "{Bar Effects on Central Star Formation and Active Galactic Nucleus Activity}",
      journal = {\apjs},
         year = 2012,
        month = jan,
       volume = {198},
       number = {1},
          eid = {4},
        pages = {4},
          doi = {10.1088/0067-0049/198/1/4},
archivePrefix = {arXiv},
       eprint = {1111.3623},
 primaryClass = {astro-ph.GA},
       adsurl = {https://ui.adsabs.harvard.edu/abs/2012ApJS..198....4O}
}

@ARTICLE{Oh2015,
       author = {{Oh}, Kyuseok and {Yi}, Sukyoung K. and {Schawinski}, Kevin and {Koss}, Michael and {Trakhtenbrot}, Benny and {Soto}, Kurt},
        title = "{A New Catalog of Type 1 AGNs and its Implications on the AGN Unified Model}",
      journal = {\apjs},
         year = 2015,
        month = jul,
       volume = {219},
       number = {1},
          eid = {1},
        pages = {1},
          doi = {10.1088/0067-0049/219/1/1},
archivePrefix = {arXiv},
       eprint = {1504.07247},
 primaryClass = {astro-ph.GA},
       adsurl = {https://ui.adsabs.harvard.edu/abs/2015ApJS..219....1O}
}

@ARTICLE{Onken2004,
       author = {{Onken}, Christopher A. and {Ferrarese}, Laura and {Merritt}, David and {Peterson}, Bradley M. and {Pogge}, Richard W. and {Vestergaard}, Marianne and {Wandel}, Amri},
        title = "{Supermassive Black Holes in Active Galactic Nuclei. II. Calibration of the Black Hole Mass-Velocity Dispersion Relationship for Active Galactic Nuclei}",
      journal = {\apj},
         year = 2004,
        month = nov,
       volume = {615},
       number = {2},
        pages = {645-651},
          doi = {10.1086/424655},
archivePrefix = {arXiv},
       eprint = {astro-ph/0407297},
 primaryClass = {astro-ph},
       adsurl = {https://ui.adsabs.harvard.edu/abs/2004ApJ...615..645O}
}

@ARTICLE{Peng2002,
       author = {{Peng}, Chien Y. and {Ho}, Luis C. and {Impey}, Chris D. and {Rix}, Hans-Walter},
        title = "{Detailed Structural Decomposition of Galaxy Images}",
      journal = {\aj},
         year = 2002,
        month = jul,
       volume = {124},
       number = {1},
        pages = {266-293},
          doi = {10.1086/340952},
archivePrefix = {arXiv},
       eprint = {astro-ph/0204182},
 primaryClass = {astro-ph},
       adsurl = {https://ui.adsabs.harvard.edu/abs/2002AJ....124..266P}
}

@ARTICLE{Peng2010,
       author = {{Peng}, Chien Y. and {Ho}, Luis C. and {Impey}, Chris D. and {Rix}, Hans-Walter},
        title = "{Detailed Decomposition of Galaxy Images. II. Beyond Axisymmetric Models}",
      journal = {\aj},
         year = 2010,
        month = jun,
       volume = {139},
       number = {6},
        pages = {2097-2129},
          doi = {10.1088/0004-6256/139/6/2097},
archivePrefix = {arXiv},
       eprint = {0912.0731},
 primaryClass = {astro-ph.CO},
       adsurl = {https://ui.adsabs.harvard.edu/abs/2010AJ....139.2097P}
}

@INPROCEEDINGS{Peterson2010,
       author = {{Peterson}, Bradley M.},
        title = "{Toward Precision Measurement of Central Black Hole Masses}",
    booktitle = {Co-Evolution of Central Black Holes and Galaxies},
         year = 2010,
       editor = {{Peterson}, Bradley M. and {Somerville}, Rachel S. and {Storchi-Bergmann}, Thaisa},
       series = {IAU Symposium},
       volume = {267},
        month = may,
        pages = {151-160},
          doi = {10.1017/S1743921310006095},
archivePrefix = {arXiv},
       eprint = {1001.3675},
 primaryClass = {astro-ph.GA},
       adsurl = {https://ui.adsabs.harvard.edu/abs/2010IAUS..267..151P}
}

@ARTICLE{Planck2020,
       author = {{Planck Collaboration} and {Aghanim}, N. and {Akrami}, Y. and {Ashdown}, M. and {Aumont}, J. and {Baccigalupi}, C. and {Ballardini}, M. and {Banday}, A.~J. and {Barreiro}, R.~B. and {Bartolo}, N. and {Basak}, S. and {Battye}, R. and {Benabed}, K. and {Bernard}, J.-P. and {Bersanelli}, M. and {Bielewicz}, P. and {Bock}, J.~J. and {Bond}, J.~R. and {Borrill}, J. and {Bouchet}, F.~R. and {Boulanger}, F. and {Bucher}, M. and {Burigana}, C. and {Butler}, R.~C. and {Calabrese}, E. and {Cardoso}, J.-F. and {Carron}, J. and {Challinor}, A. and {Chiang}, H.~C. and {Chluba}, J. and {Colombo}, L.~P.~L. and {Combet}, C. and {Contreras}, D. and {Crill}, B.~P. and {Cuttaia}, F. and {de Bernardis}, P. and {de Zotti}, G. and {Delabrouille}, J. and {Delouis}, J.-M. and {Di Valentino}, E. and {Diego}, J.~M. and {Dor{\'e}}, O. and {Douspis}, M. and {Ducout}, A. and {Dupac}, X. and {Dusini}, S. and {Efstathiou}, G. and {Elsner}, F. and {En{\ss}lin}, T.~A. and {Eriksen}, H.~K. and {Fantaye}, Y. and {Farhang}, M. and {Fergusson}, J. and {Fernandez-Cobos}, R. and {Finelli}, F. and {Forastieri}, F. and {Frailis}, M. and {Fraisse}, A.~A. and {Franceschi}, E. and {Frolov}, A. and {Galeotta}, S. and {Galli}, S. and {Ganga}, K. and {G{\'e}nova-Santos}, R.~T. and {Gerbino}, M. and {Ghosh}, T. and {Gonz{\'a}lez-Nuevo}, J. and {G{\'o}rski}, K.~M. and {Gratton}, S. and {Gruppuso}, A. and {Gudmundsson}, J.~E. and {Hamann}, J. and {Handley}, W. and {Hansen}, F.~K. and {Herranz}, D. and {Hildebrandt}, S.~R. and {Hivon}, E. and {Huang}, Z. and {Jaffe}, A.~H. and {Jones}, W.~C. and {Karakci}, A. and {Keih{\"a}nen}, E. and {Keskitalo}, R. and {Kiiveri}, K. and {Kim}, J. and {Kisner}, T.~S. and {Knox}, L. and {Krachmalnicoff}, N. and {Kunz}, M. and {Kurki-Suonio}, H. and {Lagache}, G. and {Lamarre}, J.-M. and {Lasenby}, A. and {Lattanzi}, M. and {Lawrence}, C.~R. and {Le Jeune}, M. and {Lemos}, P. and {Lesgourgues}, J. and {Levrier}, F. and {Lewis}, A. and {Liguori}, M. and {Lilje}, P.~B. and {Lilley}, M. and {Lindholm}, V. and {L{\'o}pez-Caniego}, M. and {Lubin}, P.~M. and {Ma}, Y.-Z. and {Mac{\'\i}as-P{\'e}rez}, J.~F. and {Maggio}, G. and {Maino}, D. and {Mandolesi}, N. and {Mangilli}, A. and {Marcos-Caballero}, A. and {Maris}, M. and {Martin}, P.~G. and {Martinelli}, M. and {Mart{\'\i}nez-Gonz{\'a}lez}, E. and {Matarrese}, S. and {Mauri}, N. and {McEwen}, J.~D. and {Meinhold}, P.~R. and {Melchiorri}, A. and {Mennella}, A. and {Migliaccio}, M. and {Millea}, M. and {Mitra}, S. and {Miville-Desch{\^e}nes}, M.-A. and {Molinari}, D. and {Montier}, L. and {Morgante}, G. and {Moss}, A. and {Natoli}, P. and {N{\o}rgaard-Nielsen}, H.~U. and {Pagano}, L. and {Paoletti}, D. and {Partridge}, B. and {Patanchon}, G. and {Peiris}, H.~V. and {Perrotta}, F. and {Pettorino}, V. and {Piacentini}, F. and {Polastri}, L. and {Polenta}, G. and {Puget}, J.-L. and {Rachen}, J.~P. and {Reinecke}, M. and {Remazeilles}, M. and {Renzi}, A. and {Rocha}, G. and {Rosset}, C. and {Roudier}, G. and {Rubi{\~n}o-Mart{\'\i}n}, J.~A. and {Ruiz-Granados}, B. and {Salvati}, L. and {Sandri}, M. and {Savelainen}, M. and {Scott}, D. and {Shellard}, E.~P.~S. and {Sirignano}, C. and {Sirri}, G. and {Spencer}, L.~D. and {Sunyaev}, R. and {Suur-Uski}, A.-S. and {Tauber}, J.~A. and {Tavagnacco}, D. and {Tenti}, M. and {Toffolatti}, L. and {Tomasi}, M. and {Trombetti}, T. and {Valenziano}, L. and {Valiviita}, J. and {Van Tent}, B. and {Vibert}, L. and {Vielva}, P. and {Villa}, F. and {Vittorio}, N. and {Wandelt}, B.~D. and {Wehus}, I.~K. and {White}, M. and {White}, S.~D.~M. and {Zacchei}, A. and {Zonca}, A.},
        title = "{Planck 2018 results. VI. Cosmological parameters}",
      journal = {\aap},
         year = 2020,
        month = sep,
       volume = {641},
          eid = {A6},
        pages = {A6},
          doi = {10.1051/0004-6361/201833910},
archivePrefix = {arXiv},
       eprint = {1807.06209},
 primaryClass = {astro-ph.CO},
       adsurl = {https://ui.adsabs.harvard.edu/abs/2020A&A...641A...6P}
}

@ARTICLE{Pucha2025,
       author = {{Pucha}, Ragadeepika and {Juneau}, S. and {Dey}, Arjun and {Siudek}, M. and {Mezcua}, M. and {Moustakas}, J. and {BenZvi}, S. and {Hainline}, K. and {Hviding}, R. and {Mao}, Yao-Yuan and {Alexander}, D.~M. and {Alfarsy}, R. and {Circosta}, C. and {Guo}, Wei-Jian and {Manwadkar}, V. and {Martini}, P. and {Weaver}, B.~A. and {Aguilar}, J. and {Ahlen}, S. and {Bianchi}, D. and {Brooks}, D. and {Canning}, R. and {Claybaugh}, T. and {Dawson}, K. and {de la Macorra}, A. and {Dey}, Biprateep and {Doel}, P. and {Font-Ribera}, A. and {Forero-Romero}, J.~E. and {Gazta{\~n}aga}, E. and {Gontcho A Gontcho}, S. and {Gutierrez}, G. and {Honscheid}, K. and {Kehoe}, R. and {Koposov}, S.~E. and {Lambert}, A. and {Landriau}, M. and {Le Guillou}, L. and {Meisner}, A. and {Miquel}, R. and {Prada}, F. and {Rossi}, G. and {Sanchez}, E. and {Schlegel}, D. and {Schubnell}, M. and {Seo}, H. and {Sprayberry}, D. and {Tarl{\'e}}, G. and {Zou}, H.},
        title = "{Tripling the Census of Dwarf AGN Candidates Using DESI Early Data}",
      journal = {\apj},
         year = 2025,
        month = mar,
       volume = {982},
       number = {1},
          eid = {10},
        pages = {10},
          doi = {10.3847/1538-4357/adb1dd},
archivePrefix = {arXiv},
       eprint = {2411.00091},
 primaryClass = {astro-ph.GA},
       adsurl = {https://ui.adsabs.harvard.edu/abs/2025ApJ...982...10P}
}

@ARTICLE{Pucha2026,
       author = {{Pucha}, Ragadeepika and {Juneau}, S. and {Mezcua}, M. and {Dey}, Arjun and {Mao}, Y.-Y. and {Alexander}, D.~M. and {Circosta}, C. and {Fawcett}, V.~A. and {Guo}, Wei-Jian and {Moustakas}, J. and {Panda}, S. and {Siudek}, M. and {Yu}, Z. and {Aguilar}, J. and {Ahlen}, S. and {Bianchi}, D. and {Brooks}, D. and {Claybaugh}, T. and {Dawson}, K.~S. and {de la Macorra}, A. and {Doel}, P. and {Ferraro}, S. and {Font-Ribera}, A. and {Forero-Romero}, J.~E. and {Gazta{\~n}aga}, E. and {Gontcho}, Satya Gontcho A and {Gutierrez}, G. and {Hahn}, C. and {Honscheid}, K. and {Joyce}, R. and {Kehoe}, R. and {Kisner}, T. and {Kremin}, A. and {Landriau}, M. and {Le Guillou}, L. and {Manera}, M. and {Meisner}, A. and {Miquel}, R. and {Nadathur}, S. and {Percival}, W.~J. and {Prada}, F. and {P{\'e}rez-R{\`a}fols}, I. and {Rossi}, G. and {Sanchez}, E. and {Schlegel}, D. and {Schubnell}, M. and {Silber}, J. and {Sprayberry}, D. and {Tarl{\'e}}, G. and {Weaver}, B.~A. and {Zou}, H.},
        title = "{A New Record Census of Dwarf AGN and a Bimodal $M_{\rm BH}$-$M_{\star}$ Scaling Relation with DESI DR1}",
      journal = {arXiv e-prints},
         year = 2026,
        month = jun,
          eid = {arXiv:2606.02699},
        pages = {arXiv:2606.02699},
archivePrefix = {arXiv},
       eprint = {2606.02699},
 primaryClass = {astro-ph.GA},
       adsurl = {https://ui.adsabs.harvard.edu/abs/2026arXiv260602699P}
}

@ARTICLE{Regan2004,
       author = {{Regan}, Michael W. and {Teuben}, Peter J.},
        title = "{Bar-driven Mass Inflow: How Bar Characteristics Affect the Inflow}",
      journal = {\apj},
         year = 2004,
        month = jan,
       volume = {600},
       number = {2},
        pages = {595-612},
          doi = {10.1086/380116},
       adsurl = {https://ui.adsabs.harvard.edu/abs/2004ApJ...600..595R}
}


\bsp	
\label{lastpage}
\end{document}